\documentclass[twocolumn,aps,showpacs,prc,superscriptaddress]{revtex4}

\usepackage{epstopdf} 
\usepackage{sidecap}
\usepackage{ulem}
\usepackage{epsfig}
\usepackage{amsmath,amssymb,amsthm}
\usepackage{graphicx}
\usepackage{bm}
\usepackage{color}

\usepackage{relsize}
\usepackage[utf8]{inputenc}
\usepackage{slashed}

\usepackage{subcaption}
\usepackage{ragged2e}
\DeclareCaptionJustification{justified}{\justifying}
\usepackage[capitalize]{cleveref}

\DeclareMathAlphabet{\mathpzc}{OT1}{pzc}{m}{it}

\DeclareMathOperator{\Tr}{Tr}
\begin{document}

\renewcommand{\textfraction}{0.00}
\def\mean#1{\left< #1 \right>}


\newcommand{\vAi}{{\cal A}_{i_1\cdots i_n}}
\newcommand{\vAim}{{\cal A}_{i_1\cdots i_{n-1}}}
\newcommand{\vAbi}{\bar{\cal A}^{i_1\cdots i_n}}
\newcommand{\vAbim}{\bar{\cal A}^{i_1\cdots i_{n-1}}}
\newcommand{\htS}{\hat{S}}
\newcommand{\htR}{\hat{R}}
\newcommand{\htB}{\hat{B}}
\newcommand{\htD}{\hat{D}}
\newcommand{\htV}{\hat{V}}
\newcommand{\cT}{{\cal T}}
\newcommand{\cM}{{\cal M}}
\newcommand{\cMs}{{\cal M}^*}
\newcommand{\vk}{\vec{\mathbf{k}}}
\newcommand{\bk}{\bm{k}}
\newcommand{\kt}{\bm{k}_\perp}
\newcommand{\kp}{k_\perp}
\newcommand{\km}{k_\mathrm{max}}
\newcommand{\vl}{\vec{\mathbf{l}}}
\newcommand{\bl}{\bm{l}}
\newcommand{\bK}{\bm{K}}
\newcommand{\bb}{\bm{b}}
\newcommand{\qm}{q_\mathrm{max}}
\newcommand{\vp}{\vec{\mathbf{p}}}
\newcommand{\bp}{\bm{p}}
\newcommand{\vq}{\vec{\mathbf{q}}}
\newcommand{\bq}{\bm{q}}
\newcommand{\qt}{\bm{q}_\perp}
\newcommand{\qp}{q_\perp}
\newcommand{\bQ}{\bm{Q}}
\newcommand{\vx}{\vec{\mathbf{x}}}
\newcommand{\bx}{\bm{x}}
\newcommand{\tr}{{{\rm Tr\,}}}
\newcommand{\bc}{\textcolor{blue}}

\newcommand{\beq}{\begin{equation}}
\newcommand{\eeq}[1]{\label{#1} \end{equation}}
\newcommand{\ee}{\end{equation}}
\newcommand{\bea}{\begin{eqnarray}}
\newcommand{\eea}{\end{eqnarray}}
\newcommand{\beqar}{\begin{eqnarray}}
\newcommand{\eeqar}[1]{\label{#1}\end{eqnarray}}

\newcommand{\half}{{\textstyle\frac{1}{2}}}
\newcommand{\ben}{\begin{enumerate}}
\newcommand{\een}{\end{enumerate}}
\newcommand{\bit}{\begin{itemize}}
\newcommand{\eit}{\end{itemize}}
\newcommand{\ec}{\end{center}}
\newcommand{\bra}[1]{\langle {#1}|}
\newcommand{\ket}[1]{|{#1}\rangle}
\newcommand{\norm}[2]{\langle{#1}|{#2}\rangle}
\newcommand{\brac}[3]{\langle{#1}|{#2}|{#3}\rangle}
\newcommand{\hilb}{{\cal H}}
\newcommand{\pleft}{\stackrel{\leftarrow}{\partial}}
\newcommand{\pright}{\stackrel{\rightarrow}{\partial}}

\newcommand{\squeezeup}{\vspace{-2.5mm}}


\title{Calculating hard probe radiative energy loss beyond soft-gluon approximation: how valid is the approximation?}

\date{\today}

\author{Bojana Blagojevic}
\author{Magdalena Djordjevic}
\affiliation{Institute of Physics Belgrade, University of Belgrade, Belgrade, Serbia}
\author{Marko Djordjevic}
\affiliation{Faculty of Biology, Institute of Physiology and Biochemistry, University of Belgrade, Belgrade, Serbia}

\begin{abstract}
The soft-gluon approximation, which implies that radiated gluon carries away a small fraction of initial parton's energy, is a commonly used assumption in calculating radiative energy loss of high momentum partons traversing QGP created at RHIC and LHC.  While soft-gluon approximation is convenient, different theoretical approaches reported significant radiative energy loss of high $p_{\perp}$ partons, thereby questioning its validity. To address this issue, we relaxed the soft-gluon approximation within DGLV formalism. The obtained analytical expressions are quite distinct compared to the soft-gluon case. However, numerical results for the first order in opacity fractional energy loss lead to small differences in predictions for the two cases. The difference in the predicted number of radiated gluons is also small. Moreover, the effect on these two variables has an opposite sign, which when combined results in almost overlapping suppression predictions. Therefore, our results imply that, contrary to the commonly held doubts, the soft-gluon approximation in practice works surprisingly well in DGLV formalism. Finally, we also discuss generalizing this relaxation in the dynamical QCD medium, which suggests a more general applicability of the conclusions obtained here.
\end{abstract}
\maketitle
\section{\label{sec:Intro}Introduction}

One of the main assumptions in the radiative energy loss calculations of energetic parton (in the further text referred to as jet)  traversing the QGP medium, is the soft-gluon approximation which assumes that radiated gluon carries away a small portion of initial jet energy,  
i.e. $x=\omega/E \ll 1$, where $E$ is the energy of initial jet and $\omega$ is the radiated gluon energy.

Such assumption was widely used in various energy loss models: {\it i)} in multiple soft scattering based ASW model~\citep{ASW1, ASW2, ASW3};  {\it ii)} BDMPS~\citep{BDMPS1, BDMPS2} and BDMPS-Z~\citep{Z,Z1};  {\it iii)} in opacity expansion based GLV model~\citep{GLV1,GLV} and {\it iv)} in multi-gluon evolution based HT approach~\citep{HT,HT1}, etc. These various energy loss models predict a significant medium induced radiative energy loss, questioning the validity of the soft-gluon approximation. To address this issue, a finite $x$ (or large $x$ limit) was introduced in some of these models~\citep{ASW_bsg,HT_bsg} or their extensions~\citep{Vitev_beyond_g}. However, introduction of finite $x$ lead to different conclusions on the importance of relaxing the soft-gluon approximation, which was assessed from relatively small~\citep{Vitev_beyond_g}, but noticeable, to moderately large~\citep{HT_bsg}.

The soft-gluon approximation was also used in the development of our dynamical energy loss formalism~\citep{DynEL1,DynEL,RunnC}, specifically in its radiative energy loss component. This formalism was comprehensively tested against angular averaged nuclear modification factor $R_{AA}$~\citep{STE2,RAA} data, where we obtained robust agreement for wide range of probes~\citep{RunnC,CRHIC}, centralities~\citep{NCLHC} and beam energies~\citep{CRHIC,HFLHC}, including clear predictions for future experiments~\citep{51,massTom}. This might strongly suggest that our energy loss formalism can well explain the jet-medium interactions in QGP, making this formalism suitable for the tomography of QCD medium.

However, the soft-gluon approximation obviously breaks-down for: {\it i)} intermediate momentum ranges ($5 < p_{\perp} < 10$ GeV) where the experimental data are most abundant and with the smallest error-bars, and {\it ii)} gluon energy loss, since due to the color factor of $9/4$ gluons lose significantly more energy compared to quark jets, therefore questioning the reliability of our formalism in such cases. Due to this, and for precise predictions, it became necessary  to relax the soft-gluon approximation, and consequently test its validity in dynamical energy loss formalism.

This paper presents our first step toward this goal. Since the dynamical energy loss is computationally very demanding, we will, in this study, start with relaxing this approximation on its simpler predecessor, i.e. DGLV~\citep{DGLVstatic} formalism. Within this, we will concentrate on gluon jets, since, due to their color factor, the soft-gluon approximation has the largest impact for this type of partons. For the gluon jets, we perform the radiative energy loss calculation, to the first order in the number of scattering centers (opacity), where we consider that the radiation of one gluon is induced by one collisional interaction with the medium.

Our calculation is done within the pQCD approach for a finite size, optically thin QCD medium and since it is technically demanding – it will be divided in several steps: {\it i)} First, the calculation will be done in the simplest case of massless gluons in the system of static scattering centers~\citep{GW} within GLV, {\it ii)} Then it will be extended towards the gluons with the effective mass~\citep{mg}, which presents expansion of DGLV~\citep{DGLVstatic} toward larger loss of jet energy via radiated gluon, and {\it iii)} Finally, we will discuss the impact of finite $x$ on the radiative energy loss, when dynamical medium~\citep{RunnC} (i.e. a recoil with the medium constituents) is accounted.

In that manner we will assess the validity of the soft-gluon assumption for gluon jets, and this will also provide an insight into whether or not a finite $x$ has to be implemented in quark-jet radiative energy loss calculations within our formalism. Namely, if the relaxation of the soft-gluon approximation only slightly modifies gluon-jet radiative energy loss, then even smaller modification  would be expected in quark-jet case, thus making this relaxation redundant. Otherwise, if the effect of a finite $x$ appears to be a significant in gluon-jet case, then the relaxation in quark-jet case may also be required, and would represent an important future task.

Secondly, as stated above, the relaxation of the soft-gluon approximation is needed in order to extend the applicability of our model~\citep{RunnC} towards intermediate momentum region. Thus, the another benefit of this relaxation would be to extend the $p_{\perp}$ range in which our predictions are valid.

The sections are organized as follows: In section II, we provide the theoretical framework. In section III,  we outline the computation of the zeroth order in opacity gluon-jet radiative energy loss in static QCD medium, beyond soft-gluon approximation, in the cases of both massless and massive gluons. For $x \ll 1$ the results from~\citep{GLV, DGLVstatic} are reproduced.

Section IV contains concise description of relaxing the soft-gluon approximation in calculating the first order in opacity radiative energy loss for massless gluon jet in static QCD medium. In a limit of very small $x$ result from~\citep{GLV} is recovered.

In section V we explain the computation of the first order in opacity gluon-jet energy loss in static QCD medium, with effective gluon mass~\citep{mg} included, and beyond soft-gluon approximation. This presents an extension of the calculations from~\citep{DGLVstatic} toward finite $x$,
so that results from~\citep{DGLVstatic} can be recovered in $x \ll 1$ limit. The detailed calculations corresponding to sections III - V are presented in the Appendices~\ref{sec:M0}-\ref{sec:Emg}.

In section VI we outline the numerical estimates based on our beyond soft-gluon calculations for gluon jet and the comparison with our previous results from~\citep{DGLVstatic}, i.e. the results with soft-gluon approximation. Particularly, we investigate the effect of finite $x$ on gluon-jet fractional radiative energy loss, number of radiated gluons, fractional differential radiative energy loss (intensity spectrum), single gluon radiation spectrum and gluon suppression~\citep{Bjorken}.
Conclusions and outlook are presented in section VII.

\section{Theoretical framework}
  In this work, we concentrate on relaxing soft-gluon approximation in calculating the first order in opacity radiative energy loss of high $p_{\perp}$ eikonal gluon jets  within (GLV) DGLV~\citep{DGLVstatic} formalism. That is, we assume that high $p_{\perp}$ gluon jet is produced inside a "thin" finite QGP medium at some initial point ($t_0, z_0, {\mathbf{x}}_0$), and that the medium is composed of static scattering centers~\citep{GW}. Therefore, we model the interactions in QGP assuming a static (Debye) colored-screened Yukawa potential, 
 whose Fourier and color structure acquires the following form (\citep{GLV,DGLVstatic,GW}):
\begin{align}~\label{Yukawa_rad}
V_n=V(q_n)e^{i{q_n}{x_n}}= {} & 2 {\pi} {\delta(q^0_n)} v(\vec{\bf{q}}_n) e^{-i{\vec{\bf{q}}_n} \cdot {\vec{\bf{x}}_n}}  \nonumber \\
 \times {} & T_{a_n}(R)\otimes T_{a_n}(n),
\end{align}
\begin{eqnarray}
 v(\vec{\bf{q}}_n)=\frac{4\pi\alpha_s}{{\vec{\bf{q}}_n}^2+\mu^2},
\label{vrad}
\end{eqnarray}
where $x_n$ denotes time-space coordinate of the $n^{th}$ scattering center, $\mu$ is Debye screening mass, $\alpha_s=g^2_s/{4\pi}$ is strong coupling constant, while $T_{a_n}(R)$ and $T_{a_n}(n)$ denote the generators in $SU(N_c=3)$ color representation of gluon jet and target (scattering center), respectively.

For consistency with~\citep{GLV, DGLVstatic}, we use the same notation for 4D vectors (e.g. momenta), which is described
 in detail in Appendix~\ref{sec:Notation} and  proceed throughout using Light-cone coordinates. The same Appendix contains algebra manipulation and identities for $SU(N_c)$ generators, as well as the Feynman rules, used in these calculations.

 The approximations that we assume throughout the paper are stated in Appendix~\ref{sec:Assumptions}.

The small transverse momentum transfer elastic cross section for interaction between gluon jet and target parton in GW approach~\citep{GW,GLV1} is given by:
\begin{align}~\label{ECS}
\frac{d\sigma_{el}}{d^2{\mathbf{q}}_1} =\frac{C_2(G)C_2(T)}{d_G} \frac{|v(0,{\mathbf{q}}_1)|^2}{(2 \pi)^2},
\end{align}
 where ${\mathbf{q}}_1$ corresponds to transverse momentum of exchanged gluon, $C_2(G)$ represents Casimir operator in  adjoint representation $(G)$ of gluons $SU(N_c=3)$  with dimension $d_G=8$, whereas $C_2(T)$ denotes Casimir operator in target (T) representation.

Since this formalism assumes optically "thin" plasma, the final results are expanded in powers of opacity, which is defined as the mean number of collisions in the medium: $L/\lambda = N \sigma_{el}/A_{\perp}$~\citep{GLV}, where $L$ is the thickness of the QCD medium, $\lambda$ is a mean free path, while $N$ denotes the number of scatterers (targets) in transverse area $A_{\perp}$. Note that, we restrict our calculations to the first order in opacity, which is shown to be the dominant term (\citep{MGFprvi,prvi1}).

\section{\label{treci} Zeroth order radiative energy loss} 

 To gradually introduce technically involving beyond soft-gluon calculations, we first concentrate on massless gluons traversing static QCD medium. 

 We start with $M_0$ Feynman diagram, which corresponds to the source $J$ that produces off-shell gluon with momentum $p+k$, that further, without interactions with QCD medium, radiates on-shell gluon with momentum $k$ and emerges with momentum $p$. We will further refer to these two outgoing gluons as the radiated ($k$) and the final ($p$) gluon. Note that, both in this and consecutive sections that involve interactions with one and two scattering centers, we consistently assume that initial jet propagates along the longitudinal $z$ axis. The detailed calculation of $M_0$ for finite $x$ in massless case is presented in Appendix~\ref{sec:M0}, with all assumptions listed in Appendix~\ref{sec:Assumptions}.

 We also assume that gluons are transversely polarized particles and although we work in covariant gauge, we can choose any polarization vector for the external on-shell gluons~\citep{Vitev_beyond_g}, so in accordance with~\citep{DGLVstatic,GLV,Vitev_beyond_g} we choose   $n^{\mu}=[0, 2, {\bf{0}}]$ (i.e. $\epsilon(k)\cdot k=0$, $\epsilon(k)\cdot n=0$ and $\epsilon(p)\cdot p=0$, $\epsilon(p)\cdot n=0$). Likewise, we assume that the source
has also the physical polarization as real gluons~\citep{Vitev_beyond_g} (i.e. $\epsilon(p+k)\cdot (p+k)=0$, $\epsilon(p+k)\cdot n=0$). Thus, for massless gluon's momenta we have:
 \begin{align}~\label{pk}
 & p+k=[E^+,E^-,{\bf{0}}], \quad  k =[xE^+,\frac{{\bf{k}}^2}{xE^+},{\bf{k}}], \nonumber \\
 & p = [(1-x)E^+,\frac{{\bf{p}}^2}{(1-x)E^+},{\bf{p}}],
\end{align}
 where  $E^+=p^0+k^0+p_z+k_z$, $E^-=p^0+k^0-p_z-k_z$ and due to 4-momentum conservation:
 \begin{align}~\label{zoi}
  &{\bf{p}}+ {\bf{k}}=0.
  \end{align}
The polarization vectors read:
\begin{align}~\label{kin4M0rad}
& \epsilon_i(k)=[0,
\frac{2{\bf{\boldsymbol{\epsilon}}}_i \cdot {\bf{k}}}{xE^+},  {\boldsymbol{\epsilon}}_i], \qquad \epsilon_i(p)=[0, \frac{2{\bf{\boldsymbol{\epsilon}}}_i \cdot {\bf{p}}}{(1-x)E^+},{\boldsymbol{\epsilon}}_i],\nonumber \\
& \epsilon_i(p+k)=[0,0,{\boldsymbol{\epsilon}}_i],
\end{align}
where $i=1, 2$, and we also make use of Eq.~\eqref{zoi}. So, the amplitude that gluon jet, produced at $x_0$ inside QCD medium, radiates a gluon of color $c$ without final state interactions reads:
\begin{align}~\label{M0_rad}
M_{0} = {} & J_a(p+k)e^{i(p+k)x_0} (-2ig_s)(1-x+x^2)\frac{{\mathbf{\boldsymbol{\epsilon}}}\cdot{\mathbf{k}}}{{\mathbf{k}}^2} (T^c)_{da}.
\end{align}
The radiation spectrum is obtained when Eq.~\eqref{M0_rad} is substituted in:
\def\mean#1{\left< #1 \right>}
\begin{align}~\label{E0}
d^3N^{(0)}_g d^3N_J \approx \Tr\mean{|M_0|^2} \frac{d^3\vec{\mathbf{p}}}{(2\pi)^3 2p^0} \frac{d^3\vec{\mathbf{k}}}{(2\pi)^3 2\omega},
\end{align}
where $\omega = k_0$, and where $d^3N_J$ reads:
 \begin{align}~\label{dnj2t}
 d^3N_J = {} & d_G {|J(p+k)|^2} \frac{d^3\vec{\mathbf{p}}_J}{(2\pi)^3 2E_J}.
\end{align}
Here $E_J=E=p_0 +k_0$ and ${\vec{\mathbf{p}}}_J$ denotes energy and 3D momentum of the initial gluon jet, respectively. Note that $E$ retains the same expression for other diagrams as well. The jet part can be decoupled by using the equality:
\begin{align}~\label{cvt}
 \frac{d^3\vec{\mathbf{p}}}{(2\pi)^3 2p^0} \frac{d^3\vec{\mathbf{k}}}{(2\pi)^3 2\omega} = {} & \frac{d^3\vec{\mathbf{p}}_J}{(2\pi)^3 2E_J} \frac{dx d^2{\mathbf{k}}}{(2\pi)^3 2x(1-x)},
\end{align}
\newline
which is obtained by substituting $p_z, k_z \rightarrow p^J_z, xE$. Finally, energy spectrum acquires the form:
  \begin{align}~\label{Zero1}
\frac{x d^3N^{(0)}_g}{dx d{\mathbf{k}}^2} = {} & \frac{\alpha_s}{\pi} \frac{C_2(G)}{{\mathbf{k}}^2} \frac{(1-x+x^2)^2}{1-x},
\end{align}
which recovers the well-known Altarrelli-Parisi~\citep{Altarelli} result. 

We now briefly concentrate on generating result in finite temperature QCD medium, since in~\citep{mg}, it was shown that gluons in finite temperature QGP can be approximated as a massive transverse plasmons with mass  $m_g=\mu / \sqrt{2}$, where $\mu$ is the Debye mass. In this case, $M_0$ amplitude becomes:
\begin{align}~\label{Zero_mg}
M_{0}= {} & J_a(p+k)e^{i(p+k)x_0} (-2ig_s)(1-x+x^2)  \nonumber \\
 \times {} & \frac{{\mathbf{\boldsymbol{\epsilon}}}\cdot{\mathbf{k}}}{{\mathbf{k}}^2 + m^2_g (1-x+x^2)} (T^c)_{da},
\end{align}
leading to:
 \begin{align}~\label{Zero2}
\frac{x d^3N^{(0)}_g}{dx d{\mathbf{k}}^2} = {} & \frac{\alpha_s}{\pi} \frac{C_2(G) \; {\mathbf{k}}^2}{({\mathbf{k}}^2 + m^2_g (1-x+x^2))^2}  \nonumber \\
 \times {} & \frac{(1-x+x^2)^2}{1-x}.
\end{align}

\section{First order radiative energy loss in massless case}

In accordance with~\citep{DGLVstatic}, we compute the first order in opacity radiative energy loss of gluon jet for finite $x$ starting from the expression:
\def\mean#1{\left< #1 \right>}
\begin{widetext}
\begin{align}~\label{E1rad}
d^3N^{(1)}_g d^3N_J = \Big( \frac{1}{d_T} \Tr\mean{|M_1|^2}  + \frac{2}{d_T} Re \Tr\mean{M_2 M^*_0} \Big) \frac{d^3\vec{\mathbf{p}}}{(2\pi)^3 2p^0} \frac{d^3\vec{\mathbf{k}}}{(2\pi)^3 2\omega}, 
\end{align}
\end{widetext}
where $M_0$ corresponds to the diagram without final state interactions with QCD medium, introduced in previous section, $M_1$ is the sum of all diagrams with one scattering center, $M_2$ is the sum of all diagrams with two scattering centers in the contact-limit case, while $d_T$ denotes the dimension of the target color representation (for pure gluon medium $d_T=8$). In obtaining the expression for differential energy loss, we again incorporate~\cref{dnj2t,cvt} in Eq.~\eqref{E1rad}.

The assumption that initial jet propagates along $z$-axis, takes the following form in the two cases stated below:
\begin{enumerate}
\item{One interaction with QCD medium ($M_1$)}:
\begin{align}~\label{kin1M1rad}
& p+k-q_1=[E^+ - q_{1z},E^- +q_{1z},\bf{0}],
\end{align}
 where $p+k-q_1$ corresponds to the initial jet, while $k$ and $p$ retain the same expressions as in Eq.~\eqref{pk}, with the distinction that now ${\mathbf{p}} \neq -{\mathbf{k}}$, since due to 4-momentum conservation, the following relation holds:
\begin{align}~\label{timpulsiM1rad}
{\mathbf{q}}_1 = {\mathbf{p + k}};
\end{align}
The rest of the notation is the same as in Eq.~\eqref{pk}.
\item{Two interactions with QCD medium ($M_2$):}
\begin{align}~\label{kin1M2rad}
 p+k-q_1 - q_2=[E^+ - q_{1z} - q_{2z},E^- +q_{1z}+q_{2z},\bf{0}],
\end{align}
where $p+k-q_1 - q_2$ corresponds to the initial jet and $q_i=[q_{iz}, -q_{iz}, {\bf{q}}_i]$ to exchanged gluons, $i=1,2$ with $q^0_i=0$, while $p$, $k$ retain the same expressions as in Eq.~\eqref{pk}. Also, due to 4-momentum conservation, the following relation between gluon transverse momenta holds:
\begin{align}~\label{timpulsiM2rad}
{\mathbf{p + k}} = {\mathbf{q}}_1 + {\mathbf{q}}_2,
\end{align}
which in the contact-limit case (when ${\mathbf{q}}_1 + {\mathbf{q}}_2=0$) reduces to ${\mathbf{p + k}} = 0$.
 \end{enumerate}

Note that Eq.~\eqref{timpulsiM1rad} has to be satisfied for $M_1$ diagrams in order to claim that initial jet propagates along $z$-axis, i.e. for $M_1$ diagrams ${\mathbf{p + k}}$ is different from $0$. This is an important distinction between the calculations presented in our study, and the calculations done within SCET formalism (see e.g.~\citep{Vitev_beyond_g}), where ${\mathbf{p + k}} = 0$ was used in calculation of {\it both} $M_1$ and $M_2$ diagrams, though the assumption that initial jet propagates along $z$-axis was used in that study as well.

 The transverse polarization vectors $\epsilon_i(k)$ and $\epsilon_i(p)$ for both: $M_1$ and $M_2$ amplitudes are given by the same expression as in the previous section (with an addition that in $M_1$ case: ${\mathbf{p}} \neq -{\mathbf{k}}$, as discussed above), while $\epsilon$ for initial jets consistently has the same form as in Eq.~\eqref{kin4M0rad}, i.e. $\epsilon_i(p+k-q_1)=[0,0,{\boldsymbol{\epsilon}}_i]$ for $M_1$ amplitudes, and $\epsilon_i(p+k-q_1-q_2)=[0,0,{\boldsymbol{\epsilon}}_i]$ for $M_2$ amplitudes.

 The detailed calculation of the remaining 10 Feynman diagrams, under the approximations stated in Appendix~\ref{sec:Assumptions}, contributing to the first order in opacity radiative energy loss, is given in Appendices~\ref{sec:M1}-\ref{sec:M210}, whereas thorough derivation of the single gluon radiation spectrum  beyond soft-gluon approximation in massless case is given in Appendix~\ref{sec:E} and reads (energy loss expression can be straightforwardly extracted by using $dE^{(1)}/dx \equiv \omega dN^{(1)}_g/dx \approx xE dN^{(1)}_g/dx$):
\begin{widetext}
\begin{align}~\label{dE_dx_massless}
\frac{dN^{(1)}_g}{dx^{}}={} & \frac{C_2(G) \alpha_s}{\pi} \frac{L}{\lambda} \frac{(1-x+x^2)^2}{x(1-x)}\int{\frac{d^2{\mathbf{q}}_1}{\pi} \frac{\mu^2}{({\mathbf{q}}_1^2 + \mu^2)^2}}
\int{d{\mathbf{k}}^2}  \nonumber \\
 \times {} & \Big\{  \frac{({\mathbf{k}}- {\mathbf{q}}_1)^2}{(\frac{4x(1-x)E}{L})^2 +({\mathbf{k}}- {\mathbf{q}}_1)^4} \Big( 2 -\frac{{\mathbf{k}} \cdot ({\mathbf{k}}- {\mathbf{q}}_1)}{{\mathbf{k}}^2} - \frac{({\mathbf{k}}- {\mathbf{q}}_1) \cdot ({\mathbf{k}}- x{\mathbf{q}}_1)}{({\mathbf{k}}- x{\mathbf{q}}_1)^2} \Big)  \nonumber \\
 + {} & \frac{{\mathbf{k}}^2}{(\frac{4x(1-x)E}{L})^2 + {\mathbf{k}}^4} \Big( 1 - \frac{{\mathbf{k}} \cdot ({\mathbf{k}}- x{\mathbf{q}}_1)}{({\mathbf{k}}- x{\mathbf{q}}_1)^2} \Big) + \Big( \frac{1}{({\mathbf{k}}- x{\mathbf{q}}_1)^2} - \frac{1}{{\mathbf{k}}^2} \Big) \Big\},
\end{align}
\end{widetext}
where we assumed a simple exponential distribution $\frac{2}{L} e^{\frac{-2\Delta z}{L}}$ of longitudinal distance between the gluon-jet production site and target rescattering site, emerging as $(\frac{4x(1-x)E}{L})^2$
in the denominators of the integrand. Beside facilitating the calculations, this assumption is in accordance with~\citep{GLV1,DGLVstatic,GLVexp,MGFprvi}, which allows direct comparison of our results with the corresponding (GLV) DGLV results. Specifically, as the calculations from this paper present a generalization of the previous GLV (DGLV) obtaining toward finite $x$, in the soft-gluon limit they should recover GLV (DGLV) results. To this end, note that, Eq.~\eqref{dE_dx_massless} reduces to massless case of Eq.~(11) from~\citep{DGLVstatic} in the $x\rightarrow 0$ limit, as expected. 

Additionally, we tested the robustness of all results from this manuscript, by taking into account the alternative (opposite) assumption of uniform distribution of scattering centers (as done in~\citep{DynEL1,DynEL}). The same results with the respect to the importance of soft-gluon approximation are obtained, i.e. the conclusions presented in this manuscript are robust to the presumed longitudinal distance distribution (formulas and data shown in Appendix~\ref{sec:Uniform}).

 It is straightforward to show that our result is symmetric under the exchange of radiated ($k$) and final ($p$) gluon, as expected beyond soft-gluon approximation, and due to inability to distinguish between these two identical gluons.


\section{Gluon radiative energy loss in finite temperature QCD medium}

 Next, we note that in ultra-relativistic heavy ion collisions, finite temperature QCD medium is created, that modifies the gluon self energies, and can consequently significantly influence the radiative energy loss results. It is therefore essential to include finite temperature effects in gluon radiative energy loss calculations beyond soft-gluon approximation, which is the main goal of this section. To address this issue, we note that in~\citep{mg}, it was shown that gluons can be approximated as massive transverse plasmons with effective mass $m_g$ (for gluons with the hard momenta $k \gtrsim T$) equal to its asymptotic value.
The assumption of initial jet propagating along $z$-axis, for massive case, leads to the following form of momenta, in the three cases stated below:
\begin{enumerate}
\item{No interaction with QCD medium ($M_0$)}:
\begin{align}~\label{kin1M0radmg}
& p+k=[E^+ ,E^-,{\bf{0}}],\quad
 k =[xE^+,\frac{{\bf{k}}^2 +m^2_g}{xE^+},{\bf{k}}], \nonumber \\
&  p = [(1-x)E^+,\frac{{\bf{p}}^2 +m^2_g}{(1-x)E^+},{\bf{p}}],
\end{align}
where Eq.~\eqref{zoi} holds;
\item{One interaction with QCD medium ($M_1$)}:

 $k$ and $p$ retain the same expressions as in Eq.~\eqref{kin1M0radmg}, with addition that (as in the previous section) Eq.~\eqref{timpulsiM1rad} holds due to conservation of 4-momentum, while initial jet has the momentum of the same form as in Eq.~\eqref{kin1M1rad}.
\item{Two interactions with QCD medium ($M_2$)}:

 $p$, $k$ have the same expressions as in Eq.~\eqref{kin1M0radmg}. Also, due to 4-momentum conservation Eq.~\eqref{timpulsiM2rad} holds and
in the contact-limit case reduces to ${\mathbf{p + k}} = 0$, while initial  jet momentum retains the same form as in Eq.~\eqref{kin1M2rad}.
\end{enumerate}

The transverse polarization vectors remain the same as in the massless case.

We retain all approximations from the previous section, which are reviewed in Appendix~\ref{sec:Assumptions}, and recalculate the same 11 diagrams from Appendices~\ref{sec:M0}-\ref{sec:M210}, also beyond soft-gluon approximation. The overview of all intermediate results is contained in Appendix~\ref{sec:Emg}. Thus, Eq.~\eqref{dE_dx_massless} in the massive case acquires more complex form given by:
\begin{widetext}
\begin{align}~\label{dN_dxmassive}
\frac{dN^{(1)}_{g}}{dx^{}}={} & \frac{C_2(G) \alpha_s}{\pi} \frac{L}{\lambda} \frac{(1-x+x^2)^2}{x(1-x)}\int{\frac{d^2{\mathbf{q}}_1}{\pi} \frac{\mu^2}{({\mathbf{q}}^2_1 + \mu^2)^2}}
\int{d{\mathbf{k}}^2} \nonumber \\
 \times {} & \Big\{  \frac{({\mathbf{k}}-{\mathbf{q}}_1)^2 + \chi}{(\frac{4x(1-x)E}{L})^2 +(({\mathbf{k}}-{\mathbf{q}}_1)^2+ \chi)^2} \Big(2\frac{({\mathbf{k}}-{\mathbf{q}}_1)^2}{({\mathbf{k}}-{\mathbf{q}}_1)^2 + \chi} -\frac{{\mathbf{k}} \cdot ({\mathbf{k}}-{\mathbf{q}}_1)}{{\mathbf{k}}^2 + \chi} - \frac{({\mathbf{k}}-{\mathbf{q}}_1) \cdot ({\mathbf{k}}-x{\mathbf{q}}_1)}{({\mathbf{k}}-x{\mathbf{q}}_1)^2 + \chi} \Big)  \nonumber \\
  + {} & \frac{{\mathbf{k}}^2 + \chi}{(\frac{4x(1-x)E}{L})^2 +({\mathbf{k}}^2+ \chi)^2}
 \Big(\frac{{\mathbf{k}}^2}{{\mathbf{k}}^2+ \chi} - \frac{{\mathbf{k}} \cdot ({\mathbf{k}}-x{\mathbf{q}}_1) }{({\mathbf{k}}-x{\mathbf{q}}_1)^2 + \chi} \Big)  + \Big( \frac{({\mathbf{k}}-x{\mathbf{q}}_1)^2}{(({\mathbf{k}}-x{\mathbf{q}}_1)^2 + \chi)^2}
  - \frac{{\mathbf{k}}^2}{({\mathbf{k}}^2+ \chi)^2} \Big) \Big\},
\end{align}
\end{widetext}
where $\chi=m^2_g(1-x+x^2)$. It can easily be verified that, in the soft-gluon limit, we recover Eq.~(11) from~\citep{DGLVstatic} (note that for gluon jet $M \equiv m_g$, so that the term $M^2x^2$ from ~\citep{DGLVstatic} should be neglected), and that in the massless  limit Eq.~\eqref{dN_dxmassive} reduces to Eq.~\eqref{dE_dx_massless}.

To our knowledge, this result presents the first introduction of effective gluon mass beyond-soft-gluon-approximation radiative energy loss. Additionally, we again verified that single gluon radiation spectrum is symmetric to substitution of $p$ and $k$ gluons, as necessary (see the previous section and Appendix~\ref{sec:Emg}). Furthermore, note that the analytical form of Eq.~\eqref{dN_dxmassive} is quite different from the corresponding expression with the soft-gluon approximation (Eq.~(11) from~\citep{DGLVstatic}). In the next section, we will evaluate the extent of numerical differences to which these two different analytical expressions lead.

In particular, we are interested in what is the effect of finite $x$ on gluon fractional radiative energy loss ($\frac{\Delta E^{(1)}}{E}$), number of radiated gluons ($N^{(1)}_g$) and on the suppression ($R_{AA}$). We accordingly note that $\frac{dE^{(1)}}{dx}\equiv \omega \frac{dN^{(1)}_g}{dx} \approx xE \frac{dN^{(1)}_g}{dx}$ from which we can further straightforwardly numerically evaluate  $\frac{\Delta E^{(1)}}{E}$, as well as the number of radiated gluons ($N^{(1)}_g$).

\section{\label{sec:Num}Numerical results}

 \begin{figure*}
\includegraphics[scale=0.5]{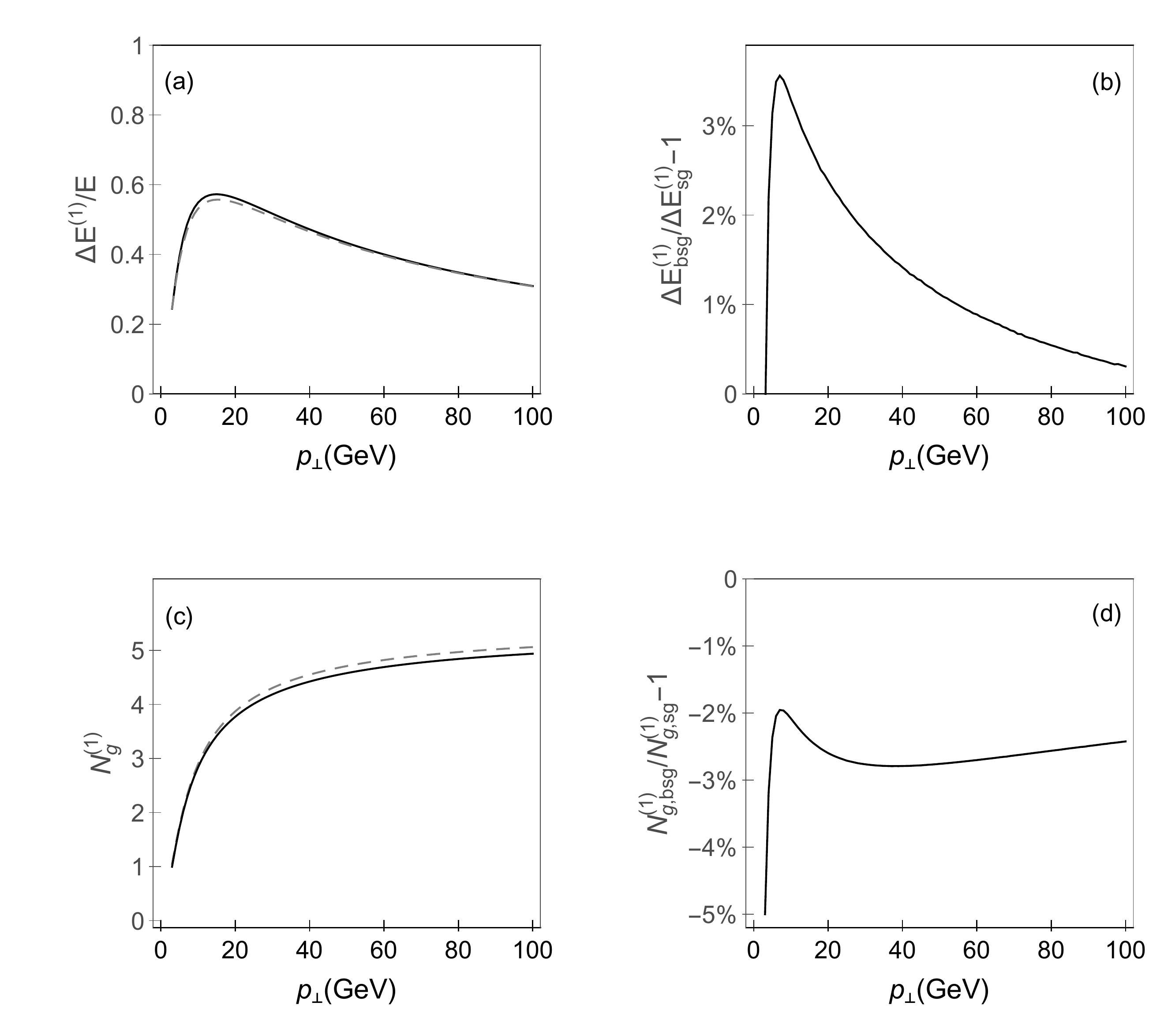}
\caption{\small The effect of relaxing the soft-gluon approximation on integrated variables to the $1^{st}$ order in opacity of DGLV formalism, as a function of $p_{\perp}$. 
(a) Comparison of gluon's fractional radiative energy loss without (the solid curve) and  with (the dashed curve) soft-gluon approximation. 
 (b) The relative change of the radiative energy loss when the soft-gluon approximation is relaxed with respect to the soft-gluon limit. 
  (c) Comparison of number of radiated gluons without (the solid curve) and with (the dashed curve) soft-gluon approximation. 
  (d) A percentage of radiated gluon number change when soft-gluon approximation is relaxed.}
\label{1}
\end{figure*}

\begin{figure*}
\includegraphics[scale=0.5]{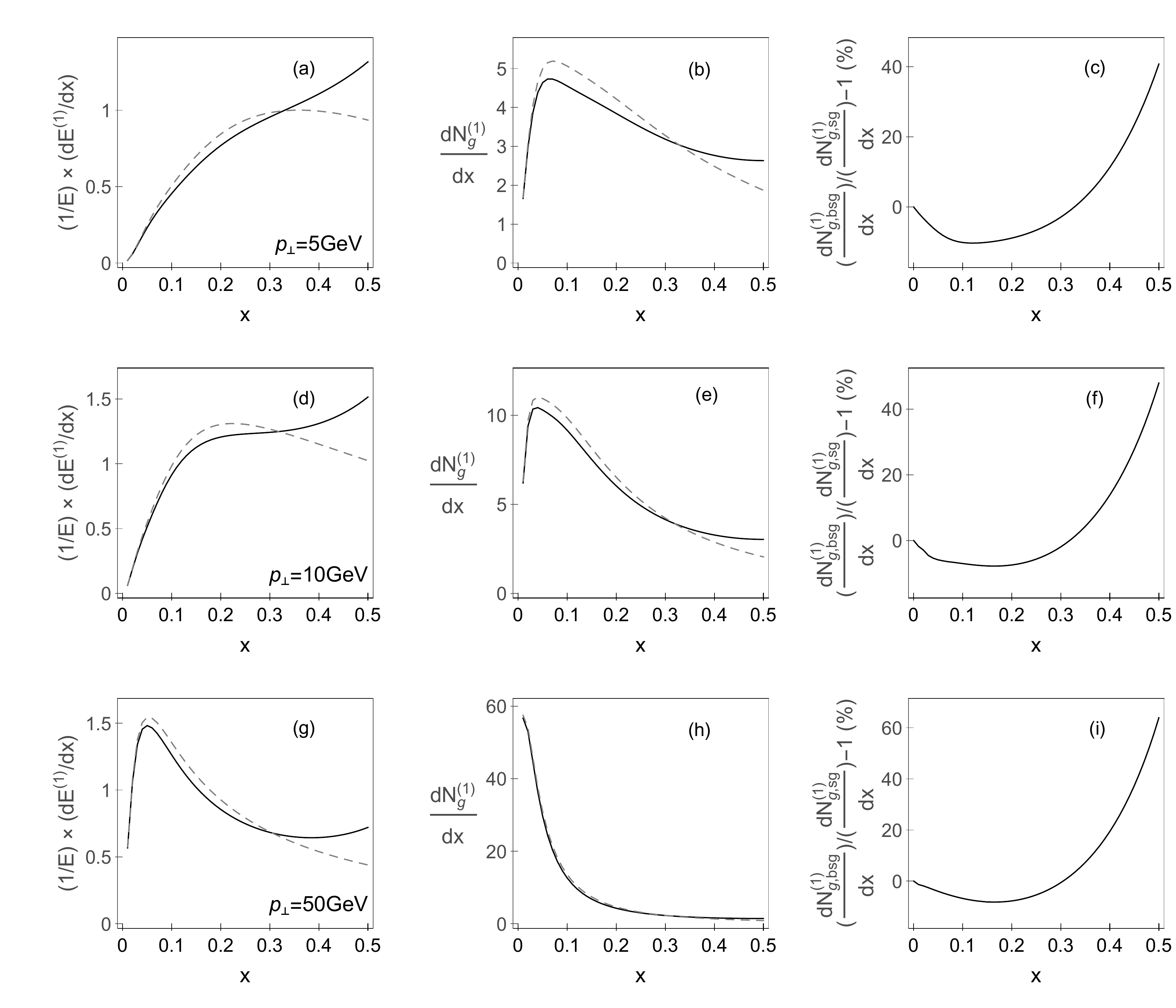}
\caption{\small The effect of relaxing the soft-gluon approximation on differential variables to the $1^{st}$ order in opacity of DGLV formalism, as a function of $x$. The comparison of: {\it i)} fractional differential gluon radiative energy loss ($(1/E) \times (dE^{(1)}/dx)$); {\it ii)} single gluon radiation (spectrum) distribution in momentum fraction ($dN^{(1)}_{g}/dx$) between {\it bsg} (the solid curve) and {\it sg} (the dashed curve) case, for different values of initial jet transverse momenta (5 GeV, 10 GeV, 50 GeV, as indicated in panels) is shown in the first ((a), (d) and (g)) and second ((b), (e) and (h)) column, respectively. The relative change of the single gluon radiation spectrum with respect to soft-gluon limit is shown in (c), (f) and (i). }
\label{2}
\end{figure*}

\begin{figure}
\includegraphics[scale=0.8]{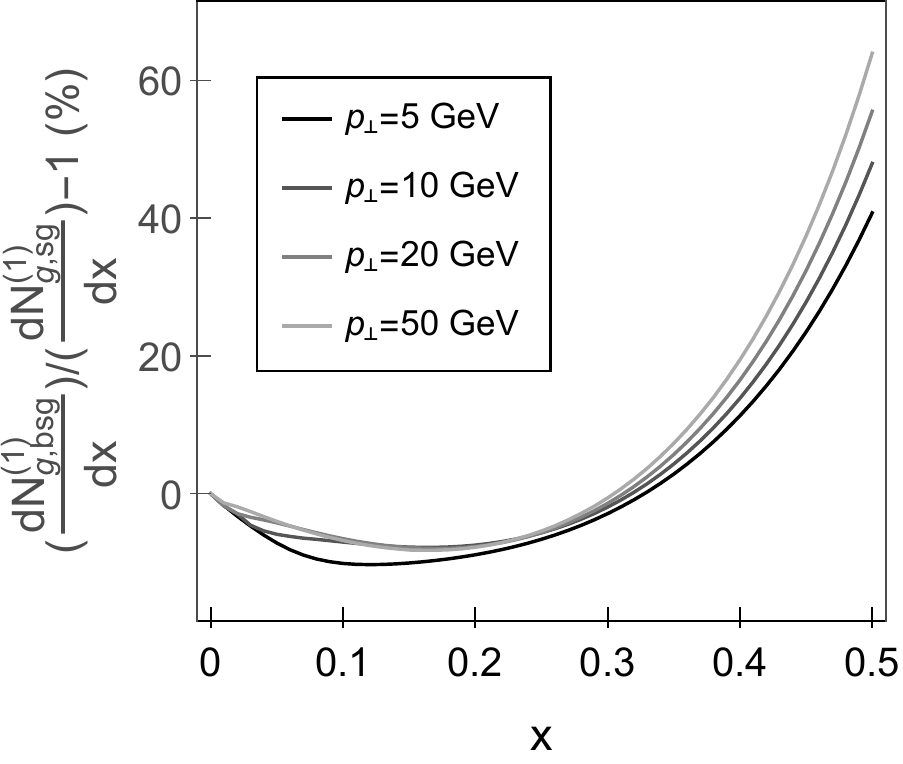}
\caption{\small The effect of relaxing the soft-gluon approximation on $dN^{(1)}_{g}/dx$ for different $p_{\perp}$ values. The relative change of the single gluon radiation spectrum with respect to soft-gluon case, calculated to the $1^{st}$ order in opacity of DGLV formalism, for different values of initial $p_{\perp}$ (as indicated in the legend) is depicted as a function of $x$. The curves fade as transverse momentum increases. }
\label{3}
\end{figure}
\begin{figure*}
\includegraphics[scale=0.5]{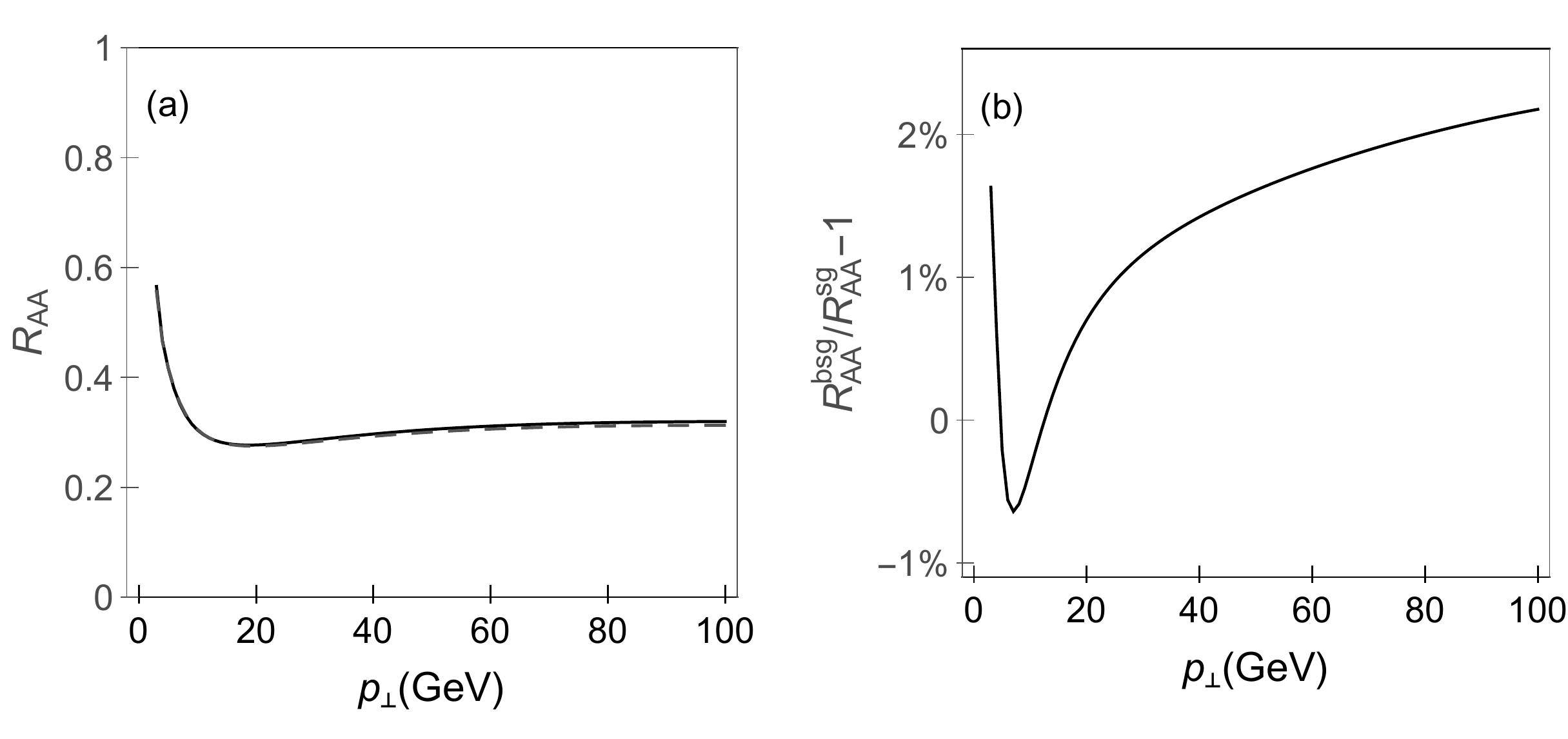}
\caption{\small The effect of relaxing the soft-gluon approximation on gluon nuclear modification factor $R_{AA}$ versus $p_{\perp}$. (a) The suppression of gluon jet beyond soft-gluon approximation (the solid curve) is compared to soft-gluon $R_{AA}$ (the dashed curve) as a function of transverse momentum. 
(b) Quantification of the effect and its expression in percentage.}
\label{4}
\end{figure*}
\begin{figure}
\includegraphics[scale=0.9]{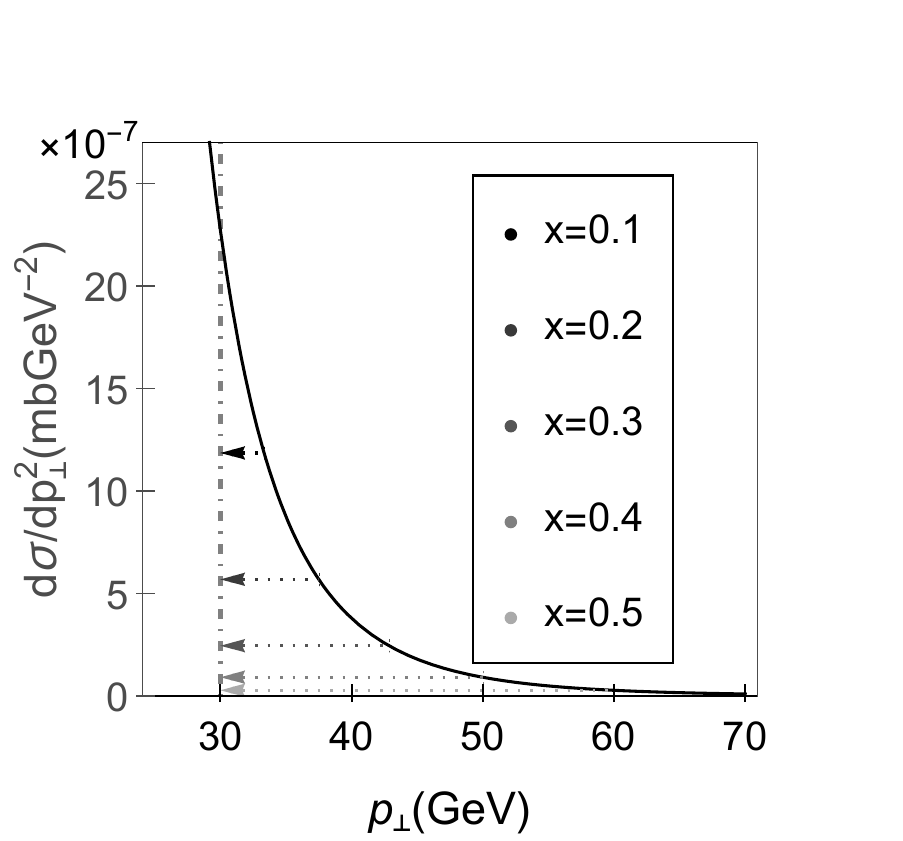}
\caption[]{\small The role of initial gluon distribution in constraining relevant $x$ region. The solid black curve represents initial gluon distribution as a function of $p_{\perp}$ at the LHC~\citep{init,initg}. The dot-dashed gray line marks the final gluon transverse momentum, while dotted arrows link parent gluons, that lost momentum fraction equal to $x$,  with their corresponding initial transverse momenta. The arrows fade as $x$ increases, as indicated in the legend.}
\label{Ilustracija}
\end{figure}
We next assess how the relaxation of soft-gluon approximation modifies gluon-jet energy loss to the $1^{st}$ order in opacity. We consequently compare the predictions based on the results derived in this paper, with the one obtained in the soft-gluon limit from~\citep{DGLVstatic} (applied to gluons) - the comparison is done for gluons with effective mass $m_g = \mu/ \sqrt{2}$, where $\mu= \sqrt{4 \pi \alpha_s (1+n_f/6)} T$ and $n_f = 3$ is the number of the effective light-quark flavors. For all figures, we use the following set of parameters: constant $\alpha_s=\frac{g^2_s}{4 \pi} =0.3$, $L=5$ fm, $\lambda=1$ fm and $T=300$ MeV, to mimic standard LHC conditions.

 The 
  comparison of the fractional radiative energy loss $\frac{\Delta E^{(1)}}{E}$, for calculations beyond the soft-gluon approximation, and with the soft-gluon approximation, as a function of initial jet transverse momentum ($p_{\perp}$) is shown in Fig.~\ref{1} (a); note that in this manuscript we concentrate on
mid-rapidity jets in relativistic heavy-ion collisions, where gluons energy is approximately equal to their transverse  momentum, due to negligible effective gluon mass compared to the transverse momentum. More specifically, the curve corresponding to beyond soft-gluon approximation ({\it bsg}) case  is obtained from  Eq.~\eqref{dN_dxmassive} multiplied by $xE$ and integrated over $x$, while the curve corresponding to soft-gluon approximation ({\it sg}) case  is obtained by numerically integrating Eq.~(11) from~\citep{DGLVstatic}. These two curves almost overlap, even converge towards one another at higher $p_{\perp}$. Note that, the upper limit of x integration is equal to 1/2 instead of 1, in order to avoid double counting.
 The upper limits of integration for $|{\mathbf{k}}|$ and $|{\mathbf{q}}_1|$, determined kinematically, are $2x(1-x)E$ and $\sqrt{4ET}$, respectively~\citep{DGLVstatic}.

The 
 comparison of number of radiated gluons in {\it bsg} and {\it sg} cases is presented in Fig.~\ref{1} (c).
These two curves also nearly overlap, with a slight disagreement at higher $p_{\perp}$.

 Quantitative assessment of relaxing the soft-gluon approximation on these two variables can be observed 
 in Figs.~\ref{1} (b) and (d). We see that finite values of $x$ slightly increase fractional radiative energy loss by maximum of $\sim 3\% $ up to $p_{\perp} \approx 10$ GeV compared to {\it sg} case. Afterwards, the difference between {\it bsg} and {\it  sg}  $\frac{\Delta E^{(1)}}{E}$ steeply decreases towards $0\%$. Additionally, finite $x$  also decreases number of radiated gluons for a small amount (up to $5\%$) compared to {\it sg} case for very low transverse momenta. Further the relative difference reaches a peak of $- 2 \%$ also at $p_{\perp} \approx 10$ GeV, and for higher transverse momenta remains nearly constant somewhat below $- 2\%$. Consequently, the overall conclusion from Fig.~\ref{1} is that the effect on both variables is small and with opposite signs.

 The effect of finite $x$ value is further assessed on the fractional differential radiative energy loss ($\frac{1}{E}\frac{dE^{(1)}}{dx} = x\frac{dN^{(1)}_{g}}{dx}$), and on single gluon radiation spectrum ($\frac{dN^{(1)}_{g}}{dx}$) and it's relative change. These effects are shown as a function of $x$ in Fig.~\ref{2}, for different values of initial jet transverse momentum $p_{\perp}$; {\it bsg} curves for $\frac{1}{E}\frac{dE^{(1)}}{dx}$ are obtained from Eq.~\eqref{dN_dxmassive} multiplied by $x$, whereas {\it sg} curves correspond  to Eq.~(11) in~\citep{DGLVstatic}. From Fig.~\ref{2}, we observe a small difference between {\it bsg} and {\it sg} results for $x \lesssim 0.3$ (roughly up to $0.4$), i.e. for smaller $x$, as expected. We also recognize $x\approx0.3$ as a "cross-over" value, below which fractional differential radiative energy loss and single gluon radiation spectrum are somewhat lower in {\it bsg} compared to {\it sg} case, and above which the opposite is true. At high value of $x$, i.e. $0.4<x\leq0.5$, the differences between our {\it bsg} fractional differential radiative energy loss 
 and previously obtained {\it sg}~\citep{DGLVstatic} ascend to notable values ($\sim 50\%$) and increase with increasing $p_{\perp}$.

To investigate the effect of relaxing the soft-gluon approximation on the single gluon radiation spectrum in more detail, the third column is added in Fig.~\ref{2}, i.e. Figs.~\ref{2} (c), (f) and (i) (see also Fig.~\ref{3}), showing relative change of $\frac{dN^{(1)}_{g}}{dx}$. This quantitative estimation (difference smaller than $10 \%$ for $x \lesssim 0.4$) is in agreement with the previous discussion. In particular, at higher $x$ values, there is a notably larger spectra in {\it bsg} compared to {\it sg} case, and this difference enhances (up to 60\% at $p_{\perp}=50$ GeV) with increasing $p_{\perp}$. Nevertheless, for both variables: ($\frac{1}{E}\frac{dE^{(1)}}{dx}$ and $\frac{dN^{(1)}_{g}}{dx}$) {\it bsg} and {\it sg} cases lead to similar results for $x \lesssim 0.4$.

 The effect of relaxing the soft-gluon approximation on  single gluon radiation  spectrum for different transverse momentum values of initial gluon jet is further addressed in Fig.~\ref{3}. We observe that a notable, that is,  tenfold increase of $p_{\perp}$ leads to a modest increase (less than 25\%) of $\frac{dN^{(1)}_{g}}{dx}$ in {\it bsg} compared to {\it sg} case. Note that the same dependence is obtained for $(\frac{1}{E}\frac{dE^{(1)}_{bsg}}{dx})/(\frac{1}{E}\frac{dE^{(1)}_{sg}}{dx}) -1$ (since $\frac{1}{E}\frac{dE^{(1)}}{dx} = x \frac{dN^{(1)}_{g}}{dx}$, so that $x$ cancels when taking the relative ratio).  Therefore, we conclude that the relaxation of the soft-gluon approximation has nearly the same effect on $\frac{dN^{(1)}_{g}}{dx}$ and $\frac{1}{E}\frac{dE^{(1)}}{dx}$ (across the whole $x$ region) independently on $p_{\perp}$ of the initial jet.

Although we showed that relaxing the soft-gluon approximation has small numerical impact on both integrated ($\frac{\Delta E^{(1)}}{E}$, $N^{(1)}_{g}$, across the whole $x$ region) and differential ($\frac{1}{E}\frac{dE^{(1)}}{dx}$, $\frac{dN^{(1)}_{g}}{dx}$, up to $x\approx0.4$) variables, the difference between {\it bsg} and {\it sg} cases can go up to $10 \%$ (and with different signs), and moreover can be quite large for $x > 0.4$. This, therefore, leads to a question, how the relaxation of the soft-gluon approximation affects predictions for measured observables, such as the angular averaged nuclear modification factor $R_{AA}$~\citep{STE2,RAA}. Comparing $R_{AA}$ with and without soft-gluon approximation allows assessing how adequate is this approximation  in obtaining reliable numerical predictions.

To that end, we next concentrate on generating the predictions for bare gluon $R_{AA}$, based only on radiative energy loss, with and without soft-gluon approximation.  $R_{AA}$ is defined as the ratio of the quenched $A + A$ spectrum to the $p + p$ spectrum, scaled by the number of binary collisions $N_{bin}$:
\begin{eqnarray}
 R_{AA}(p_{\perp}) = \frac{dN_{AA}/dp_{\perp}}{N_{bin} dN_{pp}/dp_{\perp}}.
\label{supp}
\end{eqnarray}
In order to obtain gluon quenched spectra, we use generic pQCD convolution~\citep{conv}:
\begin{eqnarray}
\frac{E_f d^3 \sigma(g)}{dp^3_f}=\frac{E_i d^3 \sigma(g)}{dp^3_i}\otimes P(E_i \rightarrow E_f),
\label{konv}
\end{eqnarray}
where $\frac{E_i d^3 \sigma(g)}{dp^3_i}$ denotes initial gluon spectrum, which is computed according to~\citep{init,initg}, while $P(E_i \rightarrow E_f)$ denotes radiative energy loss probability, which includes multi-gluon~\citep{MGFprvi} and path-length~\citep{conv} fluctuations. In accordance with~\citep{MGFprvi}, the multi-gluon fluctuations  are introduced under the assumption that the fluctuations of the gluon number are uncorrelated, and therefore presented via Poisson distribution. Specifically, the energy loss probability takes into account that the jet, during its propagation through QGP, can independently radiate number of gluons (for more details on the implementation procedure, please see ref.~\citep{MGFprvi}).

Regarding the path-length fluctuations, we take into account that jets can be produced anywhere in the nuclei overlapping area, can go in any direction, and consequently travel different distances (and lose different amounts of energy) in QGP.  
The path length probability is calculated according to the procedure described in~\citep{RAA}, where one assumes the Glauber model~\citep{Glauber} for the collision geometry, with implementation of Woods-Saxon nuclear density~\citep{WS}.

Note that we omitted fragmentation and decay functions, because we are considering the parton's quenching, as we are primarily interested in how the relaxation of the soft-gluon approximation in energy loss affects $R_{AA}$. Thereupon, we will also investigate how the initial gluon distribution influences $R_{AA}$.

Therefore, 
Fig.~\ref{4} (a) compares $R_{AA}$ predictions with and without soft-gluon approximation accounted, while the percentage change arising from relaxing the approximation is given by 
Fig.~\ref{4} (b) as a function of the final $p_{\perp}$. We observe that this relaxation barely modifies $R_{AA}$, in particular the relative change drops to somewhat less than $-1 \% $ at $p_{\perp} \approx 10$ GeV and further rises to the constant value of $2\% $, with increasing $p_{\perp}$. This very good agreement (with even smaller differences compared to previously studied variables) between {\it bsg} and {\it sg} $R_{AA}$  raises questions of: {\it i)} why relaxing the soft-gluon approximation has negligible effect on $R_{AA}$ and {\it ii)} why the large discrepancy observed in~\cref{2,3} for high $x$ values does not lead to larger difference in $R_{AA}$?

 Regarding {\it i)} above, we argue that this pattern is expected, as it is well-known that in suppression calculations both $\frac{\Delta E^{(1)}}{E}$ and $N^{(1)}_g$ non-trivially affect the $R_{AA}$.  Namely, by comparing 
 Figs.~\ref{1} (b) and (d) with 
 Fig.~\ref{4} (b) we observe that relaxing the soft-gluon approximation has opposite effects on $\frac{\Delta E^{(1)}}{E}$ and $N^{(1)}_g$, while their interplay is responsible for the negligible effect on $R_{AA}$ - i.e. the effect on $R_{AA}$ is qualitatively a superposition of the effects on $\frac{\Delta E^{(1)}}{E}$ and $N^{(1)}_{g}$.

 To answer {\it ii)} above, it is convenient to recall that suppression of gluon jet (see Eq.~\eqref{konv}) depends  not only on the energy loss probability, but also on the initial gluon distribution. In order to intuitively interpret the role of the initial gluon distribution, we refer to a descriptive Fig.~\ref{Ilustracija}, which represents its dependence on initial transverse momentum.
The concept considered is the following: Some parent gluon with unknown initial momentum traverses QGP, loses its energy by gluon bremsstrahlung, and emerges with final momentum $p_{\perp}=30$ GeV. This final gluon can descend from the parent gluon with any $p_{\perp}$ higher than its own, but we restrict ourselves to 5 different initial momenta, corresponding to different fractional momentum loss $x\in\{0.1, 0.2, 0.3, 0.4, 0.5\}$. For instance, $x=0.5$ corresponds to initial gluon momentum of $30/(1-0.5)$ GeV $=60$ GeV, i.e. to the parent gluon that lost half of its momentum etc. The question is which of these 5 gluons is the most likely to be the parent one, and how is this probability correlated with $x$? From Fig.~\ref{Ilustracija} we infer that, due to the exponentially decreasing initial gluon momentum distribution, the initial gluon corresponding to $x=0.1$ has the highest probability to be the parent one, and as $x$ increases the probability sharply decreases (i.e. for $x\gtrsim 0.4$ it diminishes for 2 orders of magnitude compared to the $x=0.1$ case). Thus, based on initial distribution, the main contribution to the suppression predictions comes from $x \lesssim 0.4$ region, making this region the most relevant one for differentiating between {\it bsg} and {\it sg} $R_{AA}$. In this region, {\it bsg} and {\it sg} $\frac{dN_g^{(1)}}{dx}$ (and equivalently $\frac{1}{E} \frac{dE^{(1)}}{dx}$) curves are very similar (according to~\cref{2,3}), which intuitively explains nearly overlapping $R_{AA}$ in Fig.~\ref{4}. Also, the relevant $x$ region qualitatively resolves the issue of why the large inconsistency between these curves at higher $x$ does not affect $R_{AA}$. 

Note however that we cannot simply reject the $x > 0.4$ region in the suppression calculations, since non-negligible $\frac{dN_g^{(1)}}{dx}$ contribution to $R_{AA}$ (see 
Figs.~\ref{2} (b), (e) and (h)) comes from it.  Therefore, for  reliable suppression results, one has to take into account the entire $x$ region, while from the above analysis, we claim that only $x \leq 0.4$ region is relevant for studying the importance of soft-gluon approximation. In order to support this in more rigorous way, we compared suppressions obtained from {\it bsg} expression for the entire $x \leq 0.5$ region, with results obtained from {\it bsg} expression for $x \leq 0.4$ combined with {\it sg} expression for $ x > 0.4$. As expected from the discussion presented in the previous paragraph, we obtained that these two approaches lead to almost the same results (with negligible differences), confirming that the region above $x = 0.4$ is not relevant for the importance of soft-gluon approximation (data shown in Appendix~\ref{sec:RelRegion} for two scenarios). 

Additionally, the effect of relaxing the soft-gluon approximation on $\frac{dN^{(1)}_{g}}{dx}$ and $\frac{1}{E}\frac{dE^{(1)}}{dx}$ is practically insensitive to initial transverse momentum (see Fig.~\ref{3}), which is the reason why finite $x$ affects equivalently gluon $R_{AA}$ regardless of it's transverse momentum, as observed in Fig.~\ref{4}.

Finally, we also recalculated our finite $x$ results, when running coupling $\alpha_s(Q^2)$, as defined in~\citep{RunA}, instead of constant value $\alpha_s=0.3$, is introduced in radiative energy loss formula. The obtained predictions lead to the same conclusions as obtained above (and are consequently omitted), which supports the generality of the obtained results.

\section{\label{sec:CO} Conclusions and outlook}

The main theoretical goal of this paper was to investigate what effect  relaxing of the soft-gluon approximation has on radiative energy loss, and consequently on suppression, which depends only on initial distribution and energy loss of high-momentum parton in QGP. Particularly we chose high $p_{\perp}$ gluon, as due to the color factor of 9/4 compared to the quarks, this assumption affects gluons the most. To this end, we                                                                                                                                                                                                                                                                                                                                                              analytically calculated all Feynman diagrams  contributing to the first order in opacity radiative energy loss beyond soft-gluon approximation, first within GLV~\citep{GLV} (massless case), and later within DGLV~\citep{DGLVstatic} (massive case), formalism, and numerically predicted: fractional integrated and differential energy loss, number of radiated gluons, single gluon radiation spectrum and gluon's suppression. Unexpectedly we obtained that, although the analytic results significantly differ from the corresponding soft-gluon results, the numerical predictions are nearly indistinguishable, i.e. within few percents. We then explained that, due to exponentially decreasing initial gluon distribution, only $x\lesssim0.4$ region effectively contributes to the differences between {\it bsg} and {\it sg} integrated variable predictions. 
We also showed that negligible suppression change is due to an interplay between the finite $x$ effects on: {\it i)} fractional energy loss and {\it ii)} number of radiated gluons, that have opposite sign. The presented comparisons are done under the assumption of fixed strong coupling constant, but also tested with running coupling leading to the same conclusions. Since we showed that gluon quenching in QCD medium composed of static scattering centers is not affected by the soft-gluon assumption, quark radiative energy loss is even less likely to be notably altered, though this remains to be further tested.

This, to our knowledge, presents the first opportunity to assess the effect of relaxing the soft-gluon approximation on radiative energy loss within DGLV formalism. Some other radiative energy loss formalisms, which also imply static scatterers, generated their results on a finite $x$. However, contrary to the conclusions derived for these formalisms, where significant difference in the radiative energy loss was obtained, we found that relaxing soft-gluon approximation brings negligible change to the results. Consequently, this implies that, within DGLV formalism, there is no need to go beyond the soft-gluon approximation. Furthermore, we also obtained that the conclusions regarding the importance of the soft-gluon approximation are robust to the presumed longitudinal distance distribution of the scattering centers.

Based on the results of this paper, we also expect that the soft-gluon approximation can be reliably applied to the dynamical energy loss formalism, as implicitly suggested by the previous robust agreement~\citep{RunnC,CRHIC,HFLHC,NCLHC} of our theoretical predictions with a comprehensive set of experimental data. In particular, the effective cross section $v({\mathbf{q}})$ (which corresponds to interaction between the jet and exchanged gluon)~\citep{MagM} does not depend on $x$, so introduction of finite $x$ will not affect this term. We also expect that the rest of the energy loss expression (i.e. $f({\mathbf{k}}, {\mathbf{q}}, x)$, which corresponds to interaction between the jet and radiated gluon~\citep{MagM}) will be modified in the similar manner as in the static case, since when $x \rightarrow 0$, these two expressions coincide. However, relaxing the soft-gluon approximation in dynamical energy loss model is out of the scope of this paper, and this claim  still remains to be rigorously tested in the future.

{\em Acknowledgments:}
This work is supported by the European Research Council, grant ERC-2016-COG: 725741, and by the Ministry of Science and Technological
Development of the Republic of Serbia, under project numbers ON171004 and ON173052.
\appendix
\section{\label{sec:Notation}Notations and useful formulas}

In this paper we used the following notation for vectors, in consistency with both~\citep{DGLVstatic,GLV}:
\begin{itemize}
\item $\vec{\bf{p}}$ denotes momentum 3D vector
\item $\bf{p}$ denotes transverse momentum 2D vector
\item $p_z$ denotes component of momentum vector along the initial jet
\item $p=(p^0,p_z,{\bf{p}})=[p^+,p^-,\bf{p}]$ denotes momentum 4D vector in Minkowski and Light Cone coordinates,  respectively, where $p^+=p^0+p_z$ and $p^-=p^0-p^z$.
\end{itemize}

For simplicity, we here consider QCD medium consisting of static partons and model the interactions of the gluon jet with the  medium via static color-screened Yukawa potential, whose Fourier and color structure acquires the following form (\citep{GLV,GW}):
\begin{align}~\label{Yukawa}
V_n=  V(q_n)e^{i{q_n}{x_n}}= {} & 2 {\pi} {\delta(q^0_n)} v(\vec{\bf{q}}_n) e^{-i{\vec{\bf{q}}_n} \cdot {\vec{\bf{x}}_n}}  \nonumber \\
 \times {} & T_{a_n}(R)\otimes T_{a_n}(n),
\end{align}
\begin{eqnarray}
 v(\vec{\bf{q}}_n)=\frac{4\pi\alpha_s}{{\vec{\bf{q}}_n}^2+\mu^2},
\label{v}
\end{eqnarray}
where $x_n$ denotes space-time coordinate of the $n^{th}$ scatterer (target), $T_{a_n}(R)$ and $T_{a_n}(n)$ denote generators in $SU(N_c=3)$ color representation of jet and target, respectively, while $\mu$ is Debye screening mass and $\alpha_s=g^2_s/{4\pi}$ is strong coupling constant. In the following lines we will briefly display the identities and algebra that $SU(N_c=3)$ generators meet:
\begin{eqnarray}
\Tr(T^a(n))=0,
\label{tr1}
\end{eqnarray}
\begin{eqnarray}
\Tr(T^a(i)T^b(j))=\delta_{ij}\delta^{ab}\frac{C_2(i)d_i}{d_G},
\label{tr2}
\end{eqnarray}
where $d_G=8$ is the dimension of the adjoint representation $(G)$.
We assume that all target partons are in the same $d_T$ dimensional representation $(T)$ with Casimir operator $C_2(T)$, while the gluon jet is in the adjoint representation ($G$), with Casimir operator $C_2(G)$.

In $SU(N_c=3)$ color algebra, the following identities hold as well:
 \begin{eqnarray}
[T^a,T^b]=if^{abc}T^c,
\label{gen1}
\end{eqnarray}
while in the adjoint representation we have:
\begin{eqnarray}
(T^b)_{ac}=if^{abc},
\label{gen2}
\end{eqnarray}

\begin{eqnarray}
T^a(G)T^a(G)=C_2(G)I,
\label{gen3}
\end{eqnarray}
where {\it I} denotes identity matrix of dimension $d_G$ and the $SU(N_c=3)$ structure constants $f^{abc}$ are completely  antisymmetric to indices permutations, which we frequently apply. In the adjoint representation the following equalities also stand:
\begin{eqnarray}
C(G)=C_2(G)=N_c =3,
\label{gen4}
\end{eqnarray}
\begin{eqnarray}
\Tr(T^a(G)T^a(G))=d_GC_2(G).
\label{gen5}
\end{eqnarray}
And finally, in our computations we frequently make use of the fact that trace is invariant under cyclic permutations and that generators are Hermitian matrices.

Since our extensive calculations are done in pQCD at finite temperature and include only gluon interactions, below we list the necessary Feynman rules in covariant gauge that we employ:
\begin{itemize}
\item massless gluon propagator in Feynman gauge (note that all diagrams in this paper are plotted by using~\citep{tikzF}):
\begin{equation}
\begin{gathered}
\includegraphics[width=0.4\linewidth]{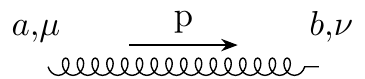}
\end{gathered}
= \frac{-i\delta_{ab} g_{\mu\nu}}{p^2 +i\epsilon},
\label{prop}
\end{equation}
\item 3-gluon vertex:
\begin{widetext}
\centering
\begin{equation}
\begin{gathered}
\includegraphics[width=0.18\linewidth]{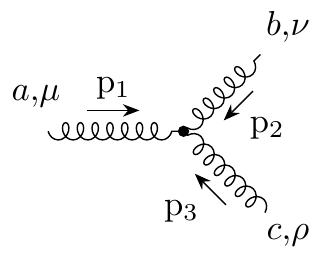}
\end{gathered}
=g_sf^{abc}\Big( g^{\mu\rho}(p_1-p_3)^{\nu}+g^{\mu\nu}(p_2-p_1)^{\rho}+g^{\nu\rho}(p_3-p_2)^{\mu}\Big).
\label{verteks}
\end{equation}
\end{widetext}
\end{itemize}

Since only physical transverse gluon states must be accounted, the transverse projector in the finite temperature case reduces to Eq.~(57) from~\citep{Altarelli}:
\begin{eqnarray}
P^{ij} = \sum_{pol} \epsilon^{i}(k) \epsilon^{j}(k)= \delta^{i j} - \frac{k^i k^j}{\vec{\bf{k}}^2},
\label{polariz}
\end{eqnarray}
where $i,j = 1, 2, 3$ correspond to spacial components of 4-vector.

\section{\label{sec:Assumptions}Assumptions}

 Throughout the paper we assume that initial gluon jet propagates along the $z$-axis, i.e. has transverse momentum equal to zero, while radiated gluon carries away a finite rate $x$ of initial gluon longitudinal momentum and energy, and final gluon emerges with momentum $p$. Therefore, instead of assuming soft-gluon approximation ($x \ll 1$), as it was done in~\citep{GLV, DGLVstatic}, 
we allow  $x$ to acquire finite non-zero values, thus relaxing the soft-gluon approximation.

Since we are calculating radiative energy loss within the (GLV) DGLV formalism apart from abandoning the soft-gluon approximation, the following assumptions remain:
\begin{itemize}
\item {\it{The soft-rescattering approximation.}} Consistently with~\citep{GLV,DGLVstatic} we assume that partons energies and longitudinal momenta are high compare to their transverse momenta, which disables the radiated and the final gluon to digress  much from the initial longitudinal direction (the eikonal approximation).
\begin{eqnarray}
E^+ \sim(1-x)E^+ \sim xE^+ \gg |{\mathbf{p}}|, |{\mathbf{k}}|, |{\mathbf{q}}_i|,
\label{app3}
\end{eqnarray}
\item{\it{The first order approximation.}} The gluon-jet radiative energy loss is calculated up to the first order in opacity expansion, as argued in~\citep{GLV,MGFprvi,prvi1}.
\item{\it{Scattering centers distribution and ensemble average.}} We consider that all scattering centers $x_i$ are distributed with the same transversely homogeneous density:
\begin{eqnarray}
\rho(\vec{\mathbf{x}}) = \frac{N}{A_{\perp}} \bar{\rho}(z),
\label{app7}
\end{eqnarray}
where $\int{dz \bar{\rho}(z)}=1$ and also that impact parameter (i.e. relative transverse coordinate) ${\mathbf{b}} = {\mathbf{x}}_i - {\mathbf{x}}_0 $ alters within a large transverse area $A_{\perp}$ compared to the interaction area $\frac{1}{\mu^2}$. Therefore, the ensemble average over the scattering center locations reduces to an impact parameter average:
\begin{eqnarray}
\mean{...} = \int{\frac{d^2{\mathbf{b}}}{A_{\perp}} ...},
\label{app8}
\end{eqnarray}
which in our case  is mainly used  in the following form:
\begin{eqnarray}
\mean {e^{-i({\mathbf{q}}_i + {\mathbf{q}}_j)\cdot {\mathbf{b}}}} = \frac{(2\pi)^2}{A_{\perp}} \delta^2({\mathbf{q}}_i + {\mathbf{q}}_j).
\label{app9}
\end{eqnarray}
\end{itemize}

We also assume that the energy of initial hard probe is large compared to the potential screening scale:
\begin{eqnarray}
E^+, (1-x)E^+, xE^+ \gg \mu . 
\label{app2}
\end{eqnarray}

Next, we assume that the distance between the source $J$ and the scattering centers is  large relative to the interaction  length:
  \begin{eqnarray}
z_i-z_0 \gg \frac{1}{\mu},
\label{app4}
\end{eqnarray}
then, that source current varies slowly with momentum:
  \begin{eqnarray}
J(p+k-q) \approx J(p+k),
\label{app5}
\end{eqnarray}
and that the source current can be written explicitly in terms of polarization vector:
\begin{align}
J^{\mu}_a(p+k-q) \equiv {} & J_a(p+k-q) \epsilon^{\mu}(p+k-q) \nonumber \\
 \approx {} & J_a(p+k) \epsilon^{\mu}(p+k-q).
\end{align}

In the following sections first we assume that gluons are massless (GLV) in order to make the comprehensive derivations more straightforward and easier to follow, but later we recalculate all the results with gluon mass~\citep{mg} included (DGLV) (Appendix~\ref{sec:Emg}).

It is worth noting that all diagrams are calculated by taking into account that each gluon can be in either of the two helicity states, and that final results are obtained by summing over helicities of final gluons $p$ and $k$ and averaging over helicity of the initial gluon. Note however that we obtained that thus calculated $\mean{|M_0|^2}$, $\mean{|M_1|^2}$ and $\mean{M_2 M^*_0}$ (for variables definition see the following Appendices) coincide with the corresponding quantities, when helicity (i.e. polarization) is considered unchanged during the process of gluon bremsstrahlung (which was the usual assumption in soft-gluon calculations~\citep{GLV,DGLVstatic}). We will explicitly demonstrate the equality of the results in these two approaches in the case of  $\mean{|M_0|^2}$ (see Appendix~\ref{sec:M0}), while in the consecutive sections (Appendices~\ref{sec:M1} to~\ref{sec:Emg}), for simplicity and easier comparison with previous studies, we assume that polarization does not change during the process, though we again note that the same results are obtained when helicity is explicitly accounted in the calculation. 

\section{\label{sec:M0}Gluon jet $M_0$}

First we calculate gluon-jet radiation amplitude to emit a gluon, carrying a finite fraction $x$ of initial jet energy, with momentum, polarization and color ({\it k, $\epsilon(k)$, c}) and without interactions with the medium $M_0$.

We assume that initial gluon ($p+k$) propagates along $z$-axis. By using $M_0$ amplitude as an example, we will implement the aforementioned assumptions in order to acquire momentum and polarization expressions. Thus, the initial gluon 4-momentum reads:
\begin{align}~\label{kin1}
 & p+k=(p^0+k^0,p_z+k_z,{\bf{0}}), \nonumber \\
 & p+k=[E^+,E^-,\bf{0}],
\end{align}
where $E^+=p^0+k^0+p_z+k_z$ and $E^-=p^0+k^0-p_z-k_z$.
\newline
\begin{figure}[h]
\includegraphics[scale=1]{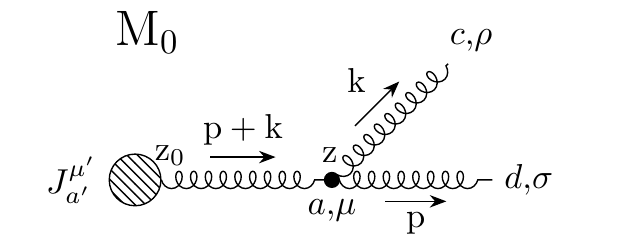}
\caption{\small Zeroth order diagram that includes no interaction with the QCD medium, and contributes to gluon radiation amplitude to the first order in opacity $L/{\lambda}$. The dashed circle represents the source $J$, which at longitudinal coordinate $z_0$ produces an off-shell gluon jet, propagating along $z$-axis. $z$ denotes longitudinal coordinate at which the gluon is radiated. Latin indices denote color charges, while Greek indices denote components of 4-vectors. $k$ denotes 4-momentum of the radiated gluon carrying the color $c$, and $p$ denotes 4-momentum of the final gluon jet carrying the color $d$.}
\label{M0f}
\end{figure}

Assuming massless (real) gluons for simplicity, the momentum vectors of the radiated ($k$) and the final ($p$) gluons acquire the following form:
 \begin{eqnarray}
k^2=0 \Rightarrow k =[xE^+,\frac{{\bf{k}}^2}{xE^+},{\bf{k}}],
\label{kin2}
\end{eqnarray}
 \begin{eqnarray}
p^2=0 \Rightarrow p = [(1-x)E^+,\frac{{\bf{p}}^2}{(1-x)E^+},{\bf{p}}].
\label{kin3}
\end{eqnarray}
We also assume that gluons are transversely polarized particles. Although we work in covariant gauge, we can choose any polarization vector for the external on-shell gluons, so in accordance with~\citep{DGLVstatic,GLV,Vitev_beyond_g} we choose   $n^{\mu}=[0, 2, {\bf{0}}]$, as stated above:
\begin{align}~\label{kin4}
& \epsilon(k)\cdot k=0, \qquad \epsilon(k)\cdot n=0, \qquad {\epsilon(k)}^2=-1, \nonumber \\
&  \epsilon(p)\cdot p=0, \qquad \epsilon(p)\cdot n=0, \qquad {\epsilon(p)}^2=-1,
\end{align}
while we assume that the source
has also the physical polarizations as the real gluons~\citep{Vitev_beyond_g}:
\begin{align}~\label{kin6}
& \epsilon(p+k)\cdot (p+k)=0, \qquad \epsilon(p+k)\cdot n=0, \nonumber \\ & {\epsilon(p+k)}^2=-1.
\end{align}

Using Eqs. (\ref{kin2}) to~(\ref{kin6}) we can now obtain the following expressions for the gluon polarizations:
 \begin{align}~\label{kin7}
& \epsilon_i^{\mu}(k)=[0, \frac{2{\bf{\boldsymbol{\epsilon}}}_i \cdot {\bf{\boldsymbol{k}}}}{xE^+},{\boldsymbol{\epsilon}}_i], \qquad \epsilon_i^{\mu}
(p)=[0, \frac{2{\bf{\boldsymbol{\epsilon}}}_i \cdot {\bf{\boldsymbol{p}}}}{(1-x)E^+},{\boldsymbol{\epsilon}}_i],\nonumber \\
& \epsilon_i^{\mu}(p+k)=[0,0,{\boldsymbol{\epsilon}}_i],
\end{align}
where $i=1, 2$ counts for polarization vectors. Note that ${\mathbf{\boldsymbol{\epsilon}}}_1$ and ${\mathbf{\boldsymbol{\epsilon}}}_2$ from Eq.~\eqref{kin7} are orthonormal~\citep{VitevEps}. Also, the 4-momentum is conserved, which leads to the relation:
\begin{align}~\label{zoiA}
  &{\bf{p}}+ {\bf{k}}=0,
  \end{align}
that we implement in Eqs.~(\ref{kin3}) and~(\ref{kin7}) in order to ensure that everything is expressed in terms of ${\bf{k}}$. Also, $E^+ \approx 2E$, $E^-= \frac{{\bf{k}}^2}{x(1-x)E^+}$, where $E = p^0 +k^0$ is the energy of initial jet.

Using the notation from Fig.~\ref{M0f} we obtain:
\begin{widetext}
\begin{align}~\label{nase}
M_{0}={}& \epsilon^*_{\sigma,i}(p)\epsilon^*_{\rho,j}(k) g_s f^{acd} \Big( g^{\mu \sigma} (2p +k)^{\rho} + g^{\mu \rho} (-p-2k)^{\sigma} + g^{\rho \sigma} (-p+k)^{\mu} \Big)
 \frac{-i \delta_{a a'} g_{\mu \mu'}}{(p+k)^2 +i\epsilon} i J_{a'}(p+k)  e^{i(p+k)x_0}  \nonumber \\
  \times {} & \epsilon^{\mu'}_l(p+k) = J_a(p+k)e^{i(p+k)x_0} g_s \frac{f^{acd}}{(p+k)^2 +i \epsilon} \Big(\big(\epsilon_i(p) \cdot \epsilon_l(p+k)\big) \big(\epsilon_j(k) \cdot (2p+k)\big)   \nonumber \\
 + &\big(\epsilon_j(k) \cdot \epsilon_l(p+k)\big) \big(\epsilon_i(p) \cdot (-p-2k)\big) + \big(\epsilon_i(p) \cdot \epsilon_j(k)\big) \big(\epsilon_l(p+k) \cdot (-p+k)\big) \Big)\nonumber \\
 = {} &  J_a(p+k)e^{i(p+k)x_0} (i g_s) \frac{(T^c)_{da}}{(p+k)^2 +i \epsilon} \Big( \big( -{\boldsymbol{\epsilon}}_i \cdot {\boldsymbol{\epsilon}}_l \big) \big( \epsilon_j(k) \cdot (2p) \big) + \big( -{\boldsymbol{\epsilon}}_j \cdot {\boldsymbol{\epsilon}}_l \big) \big( \epsilon_i(p) \cdot (-2k) \big) \nonumber \\
 + {} & \big( -{\boldsymbol{\epsilon}}_i \cdot {\boldsymbol{\epsilon}}_j \big) \big( \epsilon_l(p+k) \cdot (-p+k) \big) \Big),
\end{align}
\end{widetext}
where $i,j,l=1, 2$ now count for helicities, and where we used polarizations given by Eq.~\eqref{kin7}.  Then, the averaged value of $|M_0|^2$ reads:
\begin{widetext}
\begin{align}~\label{M0_aver}
\mean{|M_0|^2} ={}& \frac{1}{2} \sum_{i,j,l=1}^2  J_a(p+k) e^{i(p+k)x_0}  (i g_s) J^*_a(p+k) e^{-i(p+k)x_0}  (-i g_s) \frac{(T^c)_{da} (T^c)_{ad}}{{\bf{\boldsymbol{k}}}^4} x^2 (1-x)^2  \Big(  -\delta_{il} \big( 2 \frac{{\boldsymbol{\epsilon}}_j \cdot {\bf{\boldsymbol{k}}}}{x} \big) \nonumber \\
 - {} & \delta_{jl} \big( 2 \frac{{\boldsymbol{\epsilon}}_i \cdot {\bf{\boldsymbol{k}}}}{1-x} \big)  -\delta_{ij} \big(- 2 {\boldsymbol{\epsilon}}_l \cdot {\bf{\boldsymbol{k}}} \big) \Big)
\Big(  -\delta_{il} \big( 2 \frac{{\boldsymbol{\epsilon}}_j \cdot {\bf{\boldsymbol{k}}}}{x} \big) -  \delta_{jl} \big( 2 \frac{{\boldsymbol{\epsilon}}_i \cdot {\bf{\boldsymbol{k}}}}{1-x} \big)  -\delta_{ij} \big(- 2 {\boldsymbol{\epsilon}}_l \cdot {\bf{\boldsymbol{k}}} \big) \Big) \nonumber \\
  = {} & \sum \Big( J_a(p+k) e^{i(p+k)x_0} (-2ig_s)(1-x+x^2) \frac{{\mathbf{\boldsymbol{\epsilon}}}\cdot{\mathbf{k}}}{{\mathbf{k}}^2} (T^c)_{da} \Big)  \Big( J^*_a(p+k) e^{-i(p+k)x_0} (2ig_s)(1-x+x^2)  \frac{{\mathbf{\boldsymbol{\epsilon}}}\cdot{\mathbf{k}}}{{\mathbf{k}}^2} (T^c)_{ad} \Big) \nonumber \\
 = {} & |J(p+k)|^2  (4 g^2_s) \frac{C_2(G) d_G}{{\bf{\boldsymbol{k}}}^4} (1-x+x^2)^2 \sum \big( {\boldsymbol{\epsilon}} \cdot {\bf{\boldsymbol{k}}} \big)^2 \nonumber \\
 = {} & |J(p+k)|^2 (4 g^2_s) \frac{C_2(G)d_G}{{\mathbf{k}}^2} (1-x+x^2)^2,
\end{align}
\end{widetext}
where we used Eqs.~\eqref{kin1},~\eqref{kin2},~\eqref{kin3} and~\eqref{kin7}. Note that ${\boldsymbol{\epsilon}}$ in the third and the fourth line of Eq.~\eqref{M0_aver} (and in the further text) denotes either of the two vectors ${\boldsymbol{\epsilon}}_1$ and ${\boldsymbol{\epsilon}}_2$, and the summation is done over these two orthonormal polarizations (helicity states), where $\sum ({\mathbf{\boldsymbol{\epsilon}}}\cdot{\mathbf{k}})^2 = {\mathbf{k}}^2$. 
  Additionally, from the third line of this equation it is evident that the summation over helicities of final and radiated gluon, and averaging over helicity of initial gluon is equivalent to summation over two helicity states of $M_0 M^*_0$, when $M_0$ is expressed in the following simplified form:
\begin{align}~\label{M0_g}
 M_0 = {} &  J_a(p+k)e^{i(p+k)x_0} (-2g_s)(1-x+x^2)\frac{{\mathbf{\boldsymbol{\epsilon}}}\cdot{\mathbf{k}}}{{\mathbf{k}}^2} f^{acd} \nonumber \\
 = {} & J_a(p+k)e^{i(p+k)x_0} (-2ig_s)(1-x+x^2)\frac{{\mathbf{\boldsymbol{\epsilon}}}\cdot{\mathbf{k}}}{{\mathbf{k}}^2} (T^c)_{da},
 \end{align}
which is widely-accepted notation used in~\citep{GLV,DGLVstatic}, and which considered unchanged polarization in the process. After multiplying Eq.~\eqref{M0_g} by complex conjugate value, the summation over two helicity states gives: 
\begin{align}~\label{G2_3}
 \mean{|M_0|^2}= {} & |J(p+k)|^2 (4 g^2_s) \frac{C_2(G)d_G}{{\mathbf{k}}^2} (1-x+x^2)^2,
\end{align}
which is equivalent to Eq.~\eqref{M0_aver}.
Thus, as already explained in the last paragraph of Appendix~\ref{sec:Assumptions}, in order to make the calculations easier to follow, throughout this paper we adopt this condensed form (such as Eq.~\eqref{M0_g}) of expressing the amplitudes, while summation is done in the end (see Eqs.~\eqref{E_M1},~\eqref{E_M2},~\eqref{E_M1m} and~\eqref{E_M2m}). 



 Next we substitute Eq.~\eqref{G2_3} in:
 \begin{align}~\label{dn}
d^3N^{(0)}_g d^3N_J \approx & \Tr {\mean{|M_0|^2}} \frac{d^3\vec{\mathbf{p}}}{(2\pi)^3 2p^0} \frac{d^3\vec{\mathbf{k}}}{(2\pi)^3 2\omega}.
\end{align}
Note that, contrary to the soft-gluon approximation~\citep{DGLVstatic}, where:
\begin{align}~\label{dnj}
 d^3N_J \approx {} & d_G {|J(p+k)|^2} \frac{d^3\vec{\mathbf{p}}}{(2\pi)^3 2p^0},
\end{align}
  now $p$, denoting the momentum of the final gluon jet, is not approximately  equal to the momentum of initial gluon jet (i.e. the radiated gluon can carry away a substantial amount of the initial jet energy and longitudinal momentum). Thus instead of  Eq.~\eqref{dnj} throughout this paper we use the general one:
\begin{align}~\label{dnj2}
 d^3N_J = {} & d_G {|J(p+k)|^2} \frac{d^3\vec{\mathbf{p}}_J}{(2\pi)^3 2E_J},
\end{align}
where $E_J=E$ and ${\vec{\mathbf{p}}}_J$ denotes energy and 3D momentum of the initial gluon jet, respectively.
Knowing that the substitution of variables ($p_z, k_z \rightarrow p^J_z, xE$) gives:
\begin{align}~\label{cv}
 \frac{d^3\vec{\mathbf{p}}}{(2\pi)^3 2p^0} \frac{d^3\vec{\mathbf{k}}}{(2\pi)^3 2\omega} = {} & \frac{d^3\vec{\mathbf{p}}_J}{(2\pi)^3 2E_J} \frac{dx d^2{\mathbf{k}}}{(2\pi)^3 2x(1-x)},
\end{align}
and by substituting Eqs.~(\ref{G2_3}),~(\ref{dnj2}) and~(\ref{cv}) in Eq.~\eqref{dn}, for radiation spectrum we now obtain:
  \begin{align}~\label{G_2}
\frac{x d^3N^{(0)}_g}{dx d{\mathbf{k}}^2} = {} & \frac{\alpha_s}{\pi} \frac{C_2(G)}{{\mathbf{k}}^2} \frac{(1-x+x^2)^2}{1-x},
\end{align}
which recovers well-known Altarelli-Parisi result~\citep{Altarelli} and for $x \ll 1$ reduces to the massless soft-gluon limit of Eq.~(9) from~\citep{DGLVstatic}.
\newline
\section{\label{sec:M1}Diagrams $M_{1,1,0}$, $M_{1,0,0}$, $M_{1,0,1}$}

\begin{figure*}
\begin{subfigure}{0.365\textwidth}
\includegraphics[width=\linewidth]{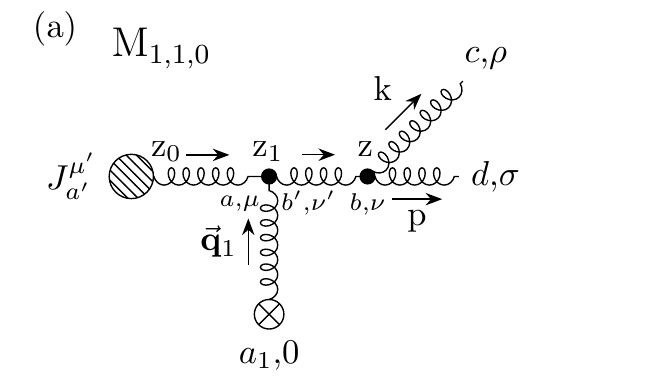}
\end{subfigure}
\begin{subfigure}{0.31\textwidth}
\includegraphics[width=\linewidth]{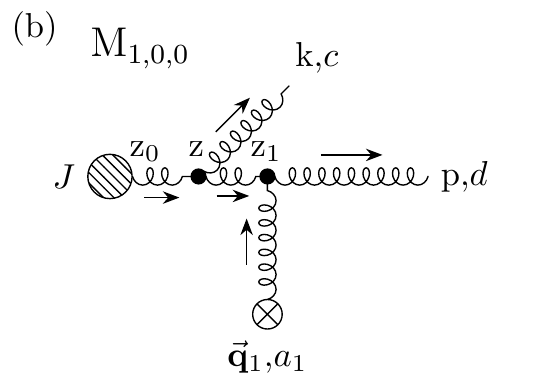}
\end{subfigure}
\begin{subfigure}{0.31\textwidth}
\includegraphics[width=\linewidth]{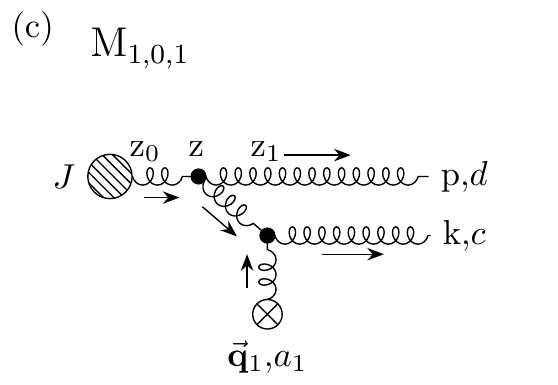}
\end{subfigure}
\caption{\small Three diagrams, corresponding to interaction with one static scattering center, that contribute to gluon-jet radiation amplitude to the first order in opacity $L/{\lambda}$. $z_1$ denotes longitudinal coordinate of the interactions with one scattering center. Crossed circle represents scatterer that exchanges 3D momentum ${\vec{\mathbf{q}}}_1$ with the jet. Note that, all three diagrams assume equivalently ordered Latin and Greek indices as indicated in (a). 
 Remaining labeling is the same as in Fig.~\ref{M0f}.} \label{fig:M1f}
\end{figure*}

In this section we provide a detailed calculations of Feynman amplitudes, corresponding to gluon-jet interaction with one scattering center, which are depicted in Fig.~\ref{fig:M1f}. Again for consistency, we assume that initial jet ($p+k-q$) propagates along z-axis. Throughout this section momentum and polarization vector for initial gluon read:
\begin{align}~\label{kin1M1}
 p+k-q_1=[E^+ - q_{1z},E^- +q_{1z},\bf{0}],
\end{align}
 \begin{align}~\label{kin4M1}
 \epsilon_i(p+k-q_1)=[0,0,{\boldsymbol{\epsilon}}_i],
\end{align}
where $q_1=[q_{1z}, -q_{1z}, {\bf{q}}_1]$, with $q^0_1=0$, denotes the momentum of exchanged gluon, while $p$, $k$ and corresponding polarization vectors retain the same expression as in Eqs.~\eqref{kin2},~\eqref{kin3} and~\eqref{kin7}, with the distinction that the following relation between gluon transverse momenta, due to 4-momentum conservation, holds:
\begin{align}~\label{timpulsiM1}
{\mathbf{q}}_1 = {\mathbf{p + k}}.
\end{align}

\subsection{Computation of $M_{1,1,0}$ diagram}

We chose to start with thorough derivation of the expression for $M_{1, 1, 0}$ amplitude, simply because it has no counterpart regarding the symmetry under ($p\leftrightarrow k, x\leftrightarrow(1-x), c\leftrightarrow d$) substitutions, and it provides all necessary steps for calculating the remaining two amplitudes from this chapter, apart from having one less singularity compared to the amplitudes $M_{1, 0, 0}$ and $M_{1,0,1}$. Thus, using the notation from 
Fig.~\ref{fig:M1f} (a), we write:
\begin{widetext}
\begin{align}~\label{m110_1}
M_{1, 1, 0}={}&\int{\frac{d^4q_1}{(2\pi)^4}\epsilon^*_{\sigma}(p)\epsilon^*_{\rho}(k)g_s f^{bcd} \Big(g^{\nu \sigma}(2p+k)^{\rho}+g^{\nu \rho}(-p-2k)^{\sigma}+ g^{\rho \sigma}(-p+k)^{\nu}\Big) {\frac{(-i)\delta_{bb'}g_{\nu \nu'}}{(p+k)^2  + i\epsilon}}}  \nonumber\\
\times {} & f^{ab'a_1}\Big(g^{\mu 0}(p+k-2q_1)^{\nu'}+g^{\mu \nu'}(-2p-2k+q_1)^0+g^{\nu' 0}(p+k+q_1)^{\mu}\Big) T_{a_1}V(q_1)e^{iq_1 x_1}  \nonumber \\
\times {} &  { \frac{(-i)\delta_{aa'}g_{\mu \mu'}}{(p+k-q_1)^2+i\epsilon}iJ_{a'}(p+k-q_1)\epsilon^{\mu'}(p+k-q_1)e^{i(p+k-q_1)x_0}} \nonumber \\
\approx{} &  J_a(p+k)e^{i(p+k)x_0}f^{bcd}f^{a_1ab}T_{a_1}(-i)(1-x+x^2) \int{\frac{d^2{\mathbf{q}}_1}{(2\pi)^2}e^{-i{\mathbf{q}}_1 \cdot({\mathbf{x}}_1-{\mathbf{x}}_0)} (2g_s)}\frac{(1-x)\,{\mathbf{\boldsymbol{\epsilon} \cdot k}}-x \, {\mathbf{\boldsymbol{\epsilon} \cdot p}}}{x(1-x)}  \nonumber \\
\times {} &  E^+\int{\frac{dq_{1z}}{2\pi}\frac{v(q_{1z},{\mathbf{q}}_1) e^{-iq_{1z}(z_1-z_0)}}{((p+k-q_1)^2+i\epsilon)((p+k)^2+i\epsilon)}},
\end{align}
\end{widetext}
where we used the equation:
\begin{align}~\label{momenti}
(p+k)^2=\frac{((1-x) {\mathbf{k}} - x{\mathbf{p}})^2}{x(1-x)},
\end{align}
and assumed that $J$ varies slowly with momentum $q_1$, i.e. Eq.~\eqref{app5}.
The longitudinal momentum transfer integral:
\begin{align}~\label{I3}
I_1(p, k, {\mathbf{q}}_1, z_1-z_0)\equiv {}  \int{\frac{dq_{1z}}{2\pi} \frac{ v(q_{1z}, {\mathbf{q}}_1) e^{-iq_{1z}(z_1-z_0)}}{(p+k-q_1)^2 +i\epsilon}}
\end{align}
has to be performed in the lower half-plane of the complex plain, since $z_1>z_0$. In order to determine the pole arising from potential, we rewrite Eq.~\eqref{v} in a more appropriate form:
\begin{eqnarray}
v(\vec{\bf{q}}_n)=\frac{4\pi\alpha_s}{(q_{nz} +i\mu_{n\perp})(q_{nz} -i\mu_{n\perp}) },
\label{v1}
\end{eqnarray}
where $\mu^2_{n\perp}= \mu^2 +{\bf{q}}^2_n $, with $n$ denoting the corresponding scattering center. 

Aside from the pole originating from Eq.~\eqref{v1} ($q_{1z}=-i\mu_{1\perp}$), there is also a singularity emerging from the gluon propagator:
\begin{align}~\label{m110_pol1}
\bar{q}_1= {} & -\frac{{\mathbf{k}}^2}{xE^+} -\frac{{\mathbf{p}}^2}{(1-x)E^+}-i\epsilon \nonumber \\
= & -\frac{{\mathbf{k}}^2}{2 \omega} -\frac{x}{1-x} \frac{({\mathbf{k}}-{\mathbf{q}}_1)^2}{2 \omega} -i\epsilon,
\end{align}
where $\omega = k_0 \approx \frac{xE^{+}}{2}$. 
The residue around the pole at $\bar{q}_1$ is computed as (the negative sign is due to the clock-wise orientation of the closed contour in the complex plain):
\begin{widetext}
\begin{align}~\label{m110_res1}
Res(\bar{q}_1)  \approx& -v(-\frac{{\mathbf{k}}^2}{xE^+} -\frac{{\mathbf{p}}^2}{(1-x)E^+}, {\mathbf{q}}_1)\frac{i}{ E^+} e^{i(\frac{{\mathbf{k}}^{\scalebox{.8}{$\scriptscriptstyle  2$}}}{xE^{\scalebox{.8}{$\scriptscriptstyle  +$}}} + \frac{{\mathbf{p}}^{\scalebox{.8}{$\scriptscriptstyle  2$}}}{(1-x)E^{\scalebox{.8}{$\scriptscriptstyle  +$}}})(z_1-z_0)}\nonumber \\
= & -v(-\frac{{\mathbf{k}}^2}{2 \omega} -\frac{x}{1-x} \frac{({\mathbf{k}}-{\mathbf{q}}_1)^2}{2 \omega}, {\mathbf{q}}_1) \frac{i}{E^+} e^{\frac{i}{2 \omega} ({\mathbf{k}}^2 + \frac{x}{1-x}({\mathbf{k}}-{\mathbf{q}}_1)^2)(z_1-z_0)}.
\end{align}
\end{widetext}
The pole originating from the potential ($q_{1z}=-i \mu_{1\perp}$) does not contribute to the longitudinal integral, since residue around that pole is exponentially suppressed due to Eq.~\eqref{app4}, i.e. $\mu (z_1-z_0)
 \gg 1$ (and $\mu \sim \mu_{1\perp}$):
\begin{align}~\label{m110_pol2}
Res(-i\mu_{1\perp}) \approx -i \frac{4 \pi \alpha_s}{(-2 i \mu_{1\perp})E^+ (-i \mu_{1\perp})} e^{-\mu_{1\perp}(z_1-z_0)} \rightarrow 0,
\end{align}
where we assumed that $E^+ \gg \mu$ and soft-rescattering approximation. 

This makes only $\bar{q}_1$ singularity relevant for calculating longitudinal integral. Therefore $I_1$ coincides with Eq.~\eqref{m110_res1}, i.e.:
\begin{widetext}
\begin{align}~\label{I3o}
I_1(p, k, {\mathbf{q}}_1, z_1-z_0)\approx& -v(-\frac{{\mathbf{k}}^2}{xE^+} -\frac{{\mathbf{p}}^2}{(1-x)E^+}, {\mathbf{q}}_1)\frac{i}{ E^+} e^{i(\frac{{\mathbf{k}}^{\scalebox{.8}{$\scriptscriptstyle  2$}}}{xE^{\scalebox{.8}{$\scriptscriptstyle  +$}}} + \frac{{\mathbf{p}}^{\scalebox{.8}{$\scriptscriptstyle  2$}}}{(1-x)E^{\scalebox{.8}{$\scriptscriptstyle  +$}}})(z_1-z_0)}
\approx  -v(0, {\mathbf{q}}_1) \frac{i}{E^+} e^{i(\frac{{\mathbf{k}}^{\scalebox{.8}{$\scriptscriptstyle  2$}}}{xE^{\scalebox{.8}{$\scriptscriptstyle  +$}}} + \frac{{\mathbf{p}}^{\scalebox{.8}{$\scriptscriptstyle  2$}}}{(1-x)E^{\scalebox{.8}{$\scriptscriptstyle  +$}}})(z_1-z_0)}\nonumber \\
= & -v(0, {\mathbf{q}}_1) \frac{i}{E^+} e^{\frac{i}{2 \omega} ( {\mathbf{k}}^2+ \frac{x}{1-x}({\mathbf{k}}-{\mathbf{q}}_1)^2)(z_1-z_0)},
\end{align}
\end{widetext}
where we used eikonal approximation (i.e. for a finite $x$:  $\frac{{\mathbf{k}}^2}{(xE^+)^2} \ll 1 $ and $\frac{{\mathbf{p}}^2}{((1-x)E^+)^2} \ll 1$). 
Finally, $M_{1, 1, 0}$ amplitude reads:
\begin{widetext}
\begin{align}~\label{m110_o}
M_{1, 1, 0}={}& J_a(p+k)e^{i(p+k)x_0}(-i)(1-x+x^2)f^{bcd} f^{a_1ab} T_{a_1} \int{\frac{d^2{\mathbf{q}}_1}{(2 \pi)^2} v(0, {\mathbf{q}}_1) e^{-i{\mathbf{q}}_1 \cdot {\mathbf{b}}_1}} (-2ig_s) \frac{{\mathbf{\boldsymbol{\epsilon}}}\cdot((1-x){\mathbf{k}}-x{\mathbf{p}})} {((1-x){\mathbf{k}}-x{\mathbf{p}})^2}  \nonumber \\
 \times {} & e^{i (\frac{{\mathbf{k}}^{\scalebox{.8}{$\scriptscriptstyle  2$}}}{xE^{\scalebox{.8}{$\scriptscriptstyle  +$}}}+\frac{{\mathbf{p}}^{\scalebox{.8}{$\scriptscriptstyle  2$}}}{(1-x)E^{\scalebox{.8}{$\scriptscriptstyle  +$}}})(z_1-z_0)}\nonumber \\
=& J_a(p+k)e^{i(p+k)x_0}(-i)(1-x+x^2)(T^c T^{a_1})_{da} T_{a_1} \int{\frac{d^2{\mathbf{q}}_1}{(2 \pi)^2} v(0, {\mathbf{q}}_1) e^{-i{\mathbf{q}}_1 \cdot {\mathbf{b}}_1}}(-2ig_s) \frac{{\mathbf{\boldsymbol{\epsilon}}}\cdot ({\mathbf{k}}-x{\mathbf{q}}_1)}{({\mathbf{k}}-x{\mathbf{q}}_1)^2}  \nonumber \\
 \times {} &  e^{\frac{i}{2\omega}({\mathbf{k}}^2 +\frac{x}{1-x}({\mathbf{k}}-{\mathbf{q}}_1)^2)(z_1-z_0)} ,
\end{align}
\end{widetext}
where we denoted ${\mathbf{b}}_1 \equiv {\mathbf{x}}_1 - {\mathbf{x}}_0$. In this subsection, we constantly make use of Eq.~\eqref{timpulsiM1} in the following form:
\begin{align}
{\mathbf{p}}^2=({\mathbf{k}} - {\mathbf{q}}_1)^2,
\end{align}
and also manipulate with $SU(N_c=3)$ structure constants  by using Eqs.~\eqref{gen1} and~\eqref{gen2}.
Note from Fig.~\ref{fig:M1f} (a) that, as expected, $M_{1, 1, 0}$ is symmetric under the substitutions: ($p\leftrightarrow k, x\leftrightarrow(1-x), c\leftrightarrow d$), where the symmetry can be straightforwardly verified by implementing these substitutions in the first two lines of Eq.~\eqref{m110_o}.

\subsection{Computation of $M_{1,0,0}$  and $M_{1,0,1}$ diagrams}

Applying the same procedure as in the previous subsection, we proceed with calculating $M_{1, 0, 0}$. Note that the order of the color and Dirac indices denoting vertices is the same for all three diagrams in Fig.~\ref{fig:M1f}, and are therefore omitted in Figs.~\ref{fig:M1f} (b) and (c). 
\begin{widetext}
\begin{align}~\label{m100_1}
M_{1,0,0}={}&\int{\frac{d^4q_1}{(2\pi)^4}\epsilon^*_{\sigma}(p) f^{bda_1} \Big(g^{\nu 0}(p-2q_1)^{\sigma}+g^{\nu \sigma}(-2p+q_1)^{0}+ g^{\sigma 0}(p+q_1)^{\nu}\Big) T_{a_1}V(q_1)e^{iq_1 x_1} {\frac{(-i)\delta_{bb'}g_{\nu \nu'}}{(p-q_1)^2  + i\epsilon}}} \nonumber\\
\times {} & g_s f^{acb'}\Big(g^{\mu \nu'}(2p+k-2q_1)^{\rho}+g^{\mu \rho}(-p-2k+q_1)^{\nu'}+g^{\rho \nu'}(-p+k+q_1)^{\mu}\Big) \epsilon^*_{\rho}(k) \frac{(-i)\delta_{aa'}g_{\mu \mu'}}{(p+k-q_1)^2+i\epsilon}  \nonumber \\
\times {} & {iJ_{a'}(p+k-q_1)\epsilon^{\mu'}(p+k-q_1)e^{i(p+k-q_1)x_0}} \nonumber \\
\approx{} &  J_a(p+k)e^{i(p+k)x_0} f^{bda_1}f^{acb}T_{a_1}  (-i)(1-x+x^2)E^+ \int{\frac{d^2{\mathbf{q}}_1}{(2\pi)^2}e^{-i{\mathbf{q}}_1 \cdot {\mathbf{b}}_1} (2 g_s) \frac{{\mathbf{\boldsymbol{\epsilon} \cdot k}}}{x}} I_2,
\end{align}
\end{widetext}
where:
\begin{align}~\label{I2}
I_2(p, k, {\mathbf{q}}_1, z_1-z_0)\equiv {} & \int{\frac{dq_{1z}}{2\pi}\frac{v(q_{1z},{\mathbf{q}}_1) e^{-iq_{1z}(z_1-z_0)}}{(p+k-q_1)^2+i\epsilon}}  \nonumber \\
 \times & \frac{1}{(p-q_1)^2+i\epsilon }.
\end{align}
In order to calculate the previous integral, due to $z_1 >  z_0 $ we again have to close the contour below the real axis. Similarly as in $M_{1,1,0}$ amplitude, again only the poles originating from the propagators contribute to the integral: $(-\frac{{\mathbf{k}}^2}{xE^+} -\frac{{\mathbf{p}}^2}{(1-x)E^+} -i\epsilon)$ and $(\frac{{\mathbf{k}}^2-{\mathbf{p}}^2}{(1-x)E^+} -i\epsilon)$, while ($-i \mu_{1\perp}$) is exponentially suppressed (due to $\mu (z_1-z_0) \gg 1$). Therefore we obtain:
\begin{widetext}
\begin{align}~\label{I2o}
I_2(p, k, {\mathbf{q}}_1, z_1-z_0) \approx {} & \frac{ix}{E^+ {\mathbf{k}}^2} v(0, {\mathbf{q}}_1) \Big(e^{i(\frac{{\mathbf{k}}^{\scalebox{.8}{$\scriptscriptstyle  2$}}}{xE^{\scalebox{.8}{$\scriptscriptstyle  +$}}} +\frac{{\mathbf{p}}^{\scalebox{.8}{$\scriptscriptstyle  2$}}}{(1-x)E^{\scalebox{.8}{$\scriptscriptstyle  +$}}})(z_1-z_0)} -  e^{i \frac{({\mathbf{p}}^{\scalebox{.8}{$\scriptscriptstyle  2$}} -{\mathbf{k}}^{\scalebox{.8}{$\scriptscriptstyle  2$}} ) }{(1-x)E^{\scalebox{.8}{$\scriptscriptstyle  +$}}} (z_1-z_0)}\Big) \nonumber \\
 \approx {} & \frac{ix}{E^+ {\mathbf{k}}^2} v(0, {\mathbf{q}}_1) \Big(e^{\frac{i}{2 \omega} ( {\mathbf{k}}^2+\frac{x}{1-x}({\mathbf{k}}-{\mathbf{q}}_1)^2 )(z_1-z_0)} -  e^{\frac{i}{2\omega} \frac{x}{1-x}(({\mathbf{k}}-{\mathbf{q}}_1)^2 -{\mathbf{k}}^2)(z_1-z_0)}\Big),
\end{align}
\end{widetext}
leading to:
\newpage
\begin{widetext}
\begin{align}~\label{m100_o}
M_{1,0,0}={}& J_a(p+k)e^{i(p+k)x_0}(-i)(1-x+x^2)f^{bda_1}f^{acb} T_{a_1}\int{\frac{d^2{\mathbf{q}}_1}{(2 \pi)^2} v(0, {\mathbf{q}}_1) e^{-i{\mathbf{q}}_1 \cdot {\mathbf{b}}_1}} (2ig_s) \frac{{\mathbf{\boldsymbol{\epsilon}\cdot k}}}{{\mathbf{k}}^2}  \nonumber \\
 \times {} & \Big( e^{i (\frac{{\mathbf{k}}^{\scalebox{.8}{$\scriptscriptstyle  2$}}}{xE^{\scalebox{.8}{$\scriptscriptstyle  +$}}}+\frac{{\mathbf{p}}^{\scalebox{.8}{$\scriptscriptstyle  2$}}}{(1-x)E^{\scalebox{.8}{$\scriptscriptstyle  +$}}})(z_1-z_0)} - e^{i \frac{({\mathbf{p}}^{\scalebox{.8}{$\scriptscriptstyle  2$}}-{\mathbf{k}}^{\scalebox{.8}{$\scriptscriptstyle  2$}})}{(1-x) E^{\scalebox{.8}{$\scriptscriptstyle  +$}}}(z_1-z_0)}\Big)\nonumber \\
={}& J_a(p+k)e^{i(p+k)x_0}(-i)(1-x+x^2)(T^{a_1}T^c)_{da}T_{a_1} \int{\frac{d^2{\mathbf{q}}_1}{(2 \pi)^2} v(0, {\mathbf{q}}_1) e^{-i{\mathbf{q}}_1 \cdot {\mathbf{b}}_1}} (2ig_s) \frac{{\mathbf{\boldsymbol{\epsilon}\cdot k}}}{{\mathbf{k}}^2}  \nonumber \\
 \times  {} & \Big( e^{\frac{i}{2\omega} ({\mathbf{k}}^2 +\frac{x}{1-x}({\mathbf{k}}-{\mathbf{q}}_1)^2)(z_1-z_0)} - e^{-\frac{i}{2\omega} \frac{x}{1-x} ({\mathbf{k}}^2 -({\mathbf{k}}-{\mathbf{q}}_1)^2)(z_1-z_0)} \Big).
\end{align}
\end{widetext}

By applying similar procedure for $M_{1,0,1}$ we obtain:
\begin{widetext}
\begin{align}~\label{m101_o}
M_{1,0,1}={}& J_a(p+k)e^{i(p+k)x_0}(-i)(1-x+x^2)f^{bca_1}f^{adb} T_{a_1}\int{\frac{d^2{\mathbf{q}}_1}{(2 \pi)^2} v(0, {\mathbf{q}}_1) e^{-i{\mathbf{q}}_1 \cdot {\mathbf{b}}_1}} (2ig_s) \frac{{\mathbf{\boldsymbol{\epsilon}\cdot p}}}{{\mathbf{p}}^2}  \nonumber \\
 \times {} & \Big( e^{i (\frac{{\mathbf{k}}^{\scalebox{.8}{$\scriptscriptstyle  2$}}}{xE^{\scalebox{.8}{$\scriptscriptstyle  +$}}}+\frac{{\mathbf{p}}^{\scalebox{.8}{$\scriptscriptstyle  2$}}}{(1-x)E^{\scalebox{.8}{$\scriptscriptstyle  +$}}})(z_1-z_0)} - e^{i \frac{({\mathbf{k}}^{\scalebox{.8}{$\scriptscriptstyle  2$}}-{\mathbf{p}}^{\scalebox{.8}{$\scriptscriptstyle  2$}})}{xE^{\scalebox{.8}{$\scriptscriptstyle  +$}}}(z_1-z_0)}\Big)\nonumber \\
={}& J_a(p+k)e^{i(p+k)x_0}(-i)(1-x+x^2)[T^c,T^{a_1}]_{da}T_{a_1} \int{\frac{d^2{\mathbf{q}}_1}{(2 \pi)^2} v(0, {\mathbf{q}}_1) e^{-i{\mathbf{q}}_1 \cdot {\mathbf{b}}_1}} (2ig_s) \frac{{\mathbf{\boldsymbol{\epsilon}}}\cdot ({\mathbf{k}}-{\mathbf{q}}_1)}{({\mathbf{k}}-{\mathbf{q}}_1)^2}  \nonumber \\
 \times {} &  \Big( e^{\frac{i}{2\omega} ({\mathbf{k}}^2 +\frac{x}{1-x}({\mathbf{k}}-{\mathbf{q}}_1)^2)(z_1-z_0)} - e^{\frac{i}{2\omega} ({\mathbf{k}}^2 -({\mathbf{k}}-{\mathbf{q}}_1)^2)(z_1-z_0)} \Big).
\end{align}
\end{widetext}

Notice from Figs.~\ref{fig:M1f} (b) and (c) that $M_{1,0,1}$ and $M_{1,0,0}$ are symmetric under the following substitutions: ($p\leftrightarrow k, x\leftrightarrow(1-x), c\leftrightarrow d$); it can be straightforwardly verified that Eqs.~\eqref{m100_o} and~\eqref{m101_o} are symmetric under these substitutions.

\section{\label{sec:M220}Diagram $M_{2,2,0}$}

Next we concentrate on the diagrams containing two interactions with the static scattering centers, since they also contribute to the gluon radiative energy loss to the first order in opacity, when multiplied by $M^*_0$. There are seven such diagrams, that we gather into four groups, each of which contains two (or one) diagrams symmetric under ($p\leftrightarrow k, x\leftrightarrow(1-x), c\leftrightarrow d$) substitutions.

For consistency the initial gluon jet (with momentum $p+k-q_1-q_2$) propagates along z-axis, i.e.:
\begin{align}~\label{kin1M2}
 p+k-q_1 - q_2=[E^+ - q_{1z} - q_{2z},E^- +q_{1z}+q_{2z},\bf{0}],
\end{align}
\begin{align}~\label{kin4M2}
\epsilon_i(p+k-q_1-q_2)=[0,0,{\boldsymbol{\epsilon}}_i],
\end{align}
where $q_i=[q_{iz}, -q_{iz}, {\bf{q}}_i]$, $i=1,2$ with $q^0_i=0$ denote momenta of exchanged gluons, while $p$, $k$ and corresponding polarizations retain the same expressions as in Eqs.~\eqref{kin2},~\eqref{kin3} and~\eqref{kin7}, with distinction that, due to 4-momentum conservation, the following relation between gluon transverse momenta holds:
\begin{align}~\label{timpulsiM2}
{\mathbf{p + k}} = {\mathbf{q}}_1 + {\mathbf{q}}_2.
\end{align}

\begin{figure*}
\begin{subfigure}{0.48\textwidth}
\includegraphics[width=\linewidth]{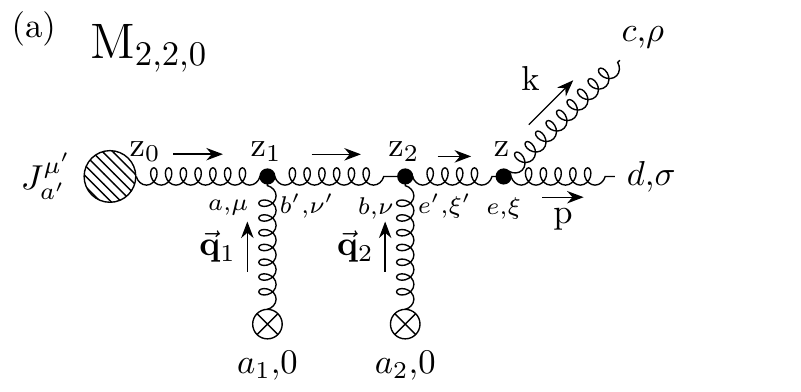}
\end{subfigure}
\begin{subfigure}{0.48\textwidth}
\includegraphics[width=\linewidth]{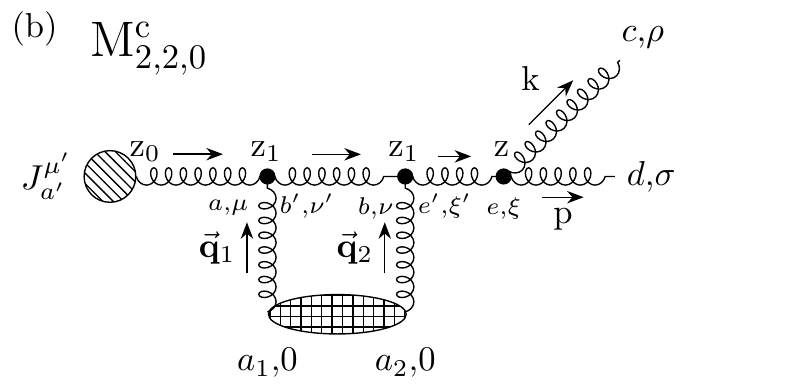}
\end{subfigure}
\caption{\small (a) Feynman diagram $M_{2, 2, 0}$ and its contribution to the first order in opacity gluon-jet radiative energy loss: (b) contact-limit $M^c_{2, 2, 0}$.  $z_i$, where $i=1,2$, denotes longitudinal coordinate of the interactions with  the consecutive scattering centers (or in the contact limit $z_1=z_2$). Crossed circles represent scatterers that exchange 3D momentum ${\vec{\mathbf{q}}}_i$ with the jet, which in contact-limit case merge into one gridded ellipse. Note that, all the following figures assume equivalently ordered Latin and Greek indices as in this figure. Remaining labeling is the same as in~\cref{M0f,fig:M1f}.} \label{M2_1}
\end{figure*}

Again, from seven diagrams we chose one model diagram $M_{2, 2, 0}$, based on the same reason as in Appendix~\ref{sec:M1}, for thorough derivation of the final amplitude expression. From Fig.~\ref{M2_1}, where gluon jet after two consecutive interactions with scattering centers radiates a gluon with momentum $k$, we observe that there are two limiting cases that we consider.

Using the notation from Fig.~\ref{M2_1} we write:
\begin{widetext}
\begin{align}~\label{m220_1}
M_{2, 2, 0}={}& \int{\frac{d^4q_1}{(2\pi)^4}\frac{d^4q_2}{(2\pi)^4}\epsilon^*_{\sigma}(p) \epsilon^*_{\rho}(k) g_s f^{ecd}\Big(g^{\xi \sigma} (2p+k)^{\rho} + g^{\xi \rho} (-p-2k)^{\sigma} + g^{\rho \sigma} (-p+k)^{\xi} \Big) }  \frac{-i \delta_{e e'} g_{\xi \xi'}}{(p+k)^2 +i\epsilon} \nonumber \\
 \times {} & f^{be'a_2} \Big(g^{\nu 0} (p+k-2q_2)^{\xi'} + g^{\nu \xi'}(-2p-2k+q_2)^0 +g^{\xi' 0} (p+k+q_2)^{\nu} \Big)  T_{a_2}V(q_2)e^{iq_2x_2} \frac{-i\delta_{bb'}g_{\nu \nu'}}{(p+k-q_2)^2 +i\epsilon}   \nonumber \\
 \times{} & f^{ab'a_1}   \Big(g^{\mu 0} (p+k-2q_1 -q_2)^{\nu'} + g^{\mu \nu'} (-2p-2k+q_1+2q_2)^0 + g^{\nu' 0} (p+k+q_1-q_2)^{\mu} \Big) T_{a_1}V(q_1) e^{iq_1 x_1}  \nonumber \\
 \times {} & \frac{-i\delta_{aa'}g_{\mu \mu'}}{(p+k-q_1-q_2)^2 +i\epsilon} i J_{a'}(p+k-q_1-q_2) \epsilon^{\mu'}(p+k-q_1-q_2) e^{i(p+k-q_1-q_2)x_0} \nonumber \\
\approx {} & i J_a(p+k) e^{i(p+k)x_0}f^{ecd}f^{bea_2}f^{aba_1}T_{a_2}T_{a_1}(1-x+x^2) (-i) \int{\frac{d^2{\mathbf{q}}_1}{(2\pi)^2}}  (-i) \int{\frac{d^2{\mathbf{q}}_2}{(2\pi)^2}} (2 ig_s) \frac{{\mathbf{\boldsymbol{\epsilon}}} \cdot((1-x){\mathbf{k}}-x{\mathbf{p}})}{((1-x){\mathbf{k}}-x{\mathbf{p}})^2} \nonumber \\
 \times{} &  e^{-i{\mathbf{q}}_1 \cdot {\mathbf{b}}_1}e^{-i{\mathbf{q}}_2 \cdot {\mathbf{b}}_2}(E^+)^2
  \int{\frac{dq_{1z}}{2\pi} \frac{dq_{2z}}{2\pi}}\frac{v(q_{1z},{\mathbf{q}}_1)v(q_{2z},{\mathbf{q}}_2)e^{-iq_{1z}(z_1-z_0)}e^{-iq_{2z}(z_2-z_0)}}{((p+k-q_1-q_2)^2 +i\epsilon)((p+k-q_2)^2 +i\epsilon)},
\end{align}
\end{widetext}
where ${\mathbf{b}}_i \equiv {\mathbf{x}}_i- {\mathbf{x}}_0$, $i=1,2$ denote transverse impact parameters. We used Eq.~\eqref{momenti} and assumed that $J$ varies slowly with momentum $q_i$, i.e. $J(p+k-q_1-q_2)\approx J(p+k)$.

Regarding the longitudinal $q_{1z}$ integral, we introduce a new variable: $q_z = q_{1z} + q_{2z}$ throughout this, and the following sections involving Feynman amplitudes which include interactions with two scattering centers. Therefore, we rewrite the exponent in the following manner:  $e^{-iq_{1z}(z_1-z_0)}e^{-iq_{2z}(z_2-z_0)} = e^{-iq_{z}(z_1-z_0)}e^{-iq_{2z}(z_2-z_1)}$. Rewriting $q_{1z}$ longitudinal integral in terms of $q_z$, i.e. changing the variables, we obtain:
\begin{align}~\label{I2_1}
I_2(p, k, {\mathbf{q}}_1, \vec{{\mathbf{q}}}_2, z_1-z_0) = \int{\frac{dq_{z}}{2\pi}}\frac{v(q_z-q_{2z},{\mathbf{q}}_1)e^{-iq_{z}(z_1-z_0)}}{(p+k-q_1-q_2)^2 +i\epsilon}.
\end{align}
Again, due to $z_1 > z_0$, the contour must be closed in the lower half-plane of complex $q_z$ plain, so additional minus sign arises from the negative orientation of the contour and also we neglect the pole at $q_z=-i\mu_{1\perp} + q_{2z}$, since it is exponentially suppressed due to Eq.~\eqref{app4}. Thus, only one pole, originating from the gluon propagator, contributes to the first longitudinal integral:
\begin{align}~\label{m220_pol}
\bar{q}= {} & -\frac{{\mathbf{k}}^2}{xE^+} -\frac{{\mathbf{p}}^2}{(1-x)E^+} -i\epsilon \nonumber \\
= & -\frac{{\mathbf{k}}^2}{2\omega} -\frac{x}{1-x}\frac{({\mathbf{k}}-{\mathbf{q}}_1-{\mathbf{q}}_2)^2}{2\omega} -i\epsilon,
\end{align}
where we used, as well as throughout the Appendices~\ref{sec:M203}-\ref{sec:M210} the relation between transverse momenta Eq.~\eqref{timpulsiM2}.
The residue at Eq.~\eqref{m220_pol} then gives:
\begin{widetext}
\begin{align}~\label{I2_1o}
I_2(p, k, {\mathbf{q}}_1, \vec{{\mathbf{q}}}_2, z_1-z_0) \approx -v(-q_{2z}-\frac{{\mathbf{k}}^2}{2\omega} -\frac{x}{1-x}\frac{({\mathbf{k}}-{\mathbf{q}}_1-{\mathbf{q}}_2)^2}{2\omega}, {\mathbf{q}}_1) \frac{i}{E^+} e^{\frac{i}{2\omega} ({\mathbf{k}}^2 + \frac{x}{1-x}({\mathbf{k}}-{\mathbf{q}}_1-{\mathbf{q}}_2)^2)(z_1-z_0)}.
\end{align}
\end{widetext}
Next we need to solve the remaining $q_{2z}$ longitudinal momentum transfer integral:
\begin{align}~\label{I3_1}
I_3(p, k, {\mathbf{q}}_1, {\mathbf{q}}_2 & ,  z_2-z_1) = {}  \int{\frac{dq_{2z}}{2\pi}}\frac{v(q_{2z},{\mathbf{q}}_2)e^{-iq_{2z}(z_2-z_1)}}{(p+k-q_2)^2 +i\epsilon}  \nonumber \\
& \times {}  v(-q_{2z}-\frac{{\mathbf{k}}^2}{2\omega}-\frac{x}{1-x}\frac{({\mathbf{k}}-{\mathbf{q}}_1-{\mathbf{q}}_2)^2}{2\omega},{\mathbf{q}}_1).
\end{align}
Luckily, we are interested only in two limiting cases:
\begin{itemize}
\item The limit of well-separated scattering centers $z_2 -z_1\gg 1/\mu$, where poles originating from Yukawa potentials are exponentially suppressed,
\item The contact limit $z_1=z_2$, where these poles contribute to the final results.
\end{itemize}
In the case of two distinct scatterers ($z_1 \neq z_2$) and in the limit of well-separated scattering centers there is only one pole that contributes to the residue (the singularities originating  from Yukawa potential once again are exponentially suppressed):
\begin{align}~\label{m220_pol2}
\bar{q}_{2z}={} & -\frac{{\mathbf{k}}^2}{xE^+} -\frac{{\mathbf{p}}^2}{(1-x)E^+} +\frac{{\mathbf{q}}_1^2}{E^+}-i\epsilon \nonumber \\
 = & -\frac{{\mathbf{k}}^2}{2\omega} -\frac{x}{1-x}\frac{({\mathbf{k}}-{\mathbf{q}}_1-{\mathbf{q}}_2)^2}{2\omega} +\frac{{\mathbf{q}}^2_1}{E^+} -i\epsilon.
\end{align}
Since $z_2 > z_1$ again we close the contour below the real $q_{2z}$ axis and thus obtain:
\begin{align}~\label{I3_1n}
I_3(p, k, {\mathbf{q}}_1, {\mathbf{q}}_2&, z_2-z_1) \approx {}  -v(0,{\mathbf{q}}_1) v(0, {\mathbf{q}}_2)\frac{i}{E^+}  \nonumber \\
& \times {}  e^{\frac{i}{2\omega} ({\mathbf{k}}^2 +\frac{x}{1-x}({\mathbf{k}}-{\mathbf{q}}_1-{\mathbf{q}}_2)^2 -x{\mathbf{q}}_1^2)(z_2-z_1)}.
\end{align}
In the special case of  contact limit, i.e. when $z_1=z_2$, instead of Eq.~\eqref{I3_1} we need to calculate the following $q_{2z}$ integral:
\begin{align}~\label{I3_co}
I^c_3(p, & k, {\mathbf{q}}_1, {\mathbf{q}}_2,0) = {}  \int{\frac{dq_{2z}}{2\pi}}\frac{v(q_{2z},{\mathbf{q}}_2)}{(p+k-q_2)^2 +i\epsilon}  \nonumber \\
& \times v(-q_{2z}-\frac{{\mathbf{k}}^2}{2\omega}-\frac{x}{1-x}\frac{({\mathbf{k}}-{\mathbf{q}}_1-{\mathbf{q}}_2)^2}{2\omega},{\mathbf{q}}_1).
\end{align}
Now, the contributions from Yukawa singularities ($q_{2z}=-i\mu_{1\perp}$, $q_{2z}=-i\mu_{2\perp}$) are not negligible and need to be included together with Eq.~\eqref{m220_pol2}. By choosing the same integration contour we obtain:
\begin{widetext}
\begin{align}~\label{I3_co1}
I^c_3(p, k, {\mathbf{q}}_1, {\mathbf{q}}_2,0) \approx{} & \frac{-i}{E^+}\Big( v(0, {\mathbf{q}}_1)v(0, {\mathbf{q}}_2) -\frac{(4\pi \alpha_s)^2}{2}\frac{1}{{\mu^2_{2\perp}}-{\mu^2_{1\perp}}}(\frac{1}{{\mu^2_{1\perp}}} -\frac{1}{{\mu^2_{2\perp}}}) \Big) = -v(0,{\mathbf{q}}_1)v(0,{\mathbf{q}}_2)\frac{i}{2E^+},
\end{align}
\end{widetext}
which is exactly $\frac{1}{2}$ of the strength of Eq.~\eqref{I3_1n}.
Note that, in previous calculations we applied soft-rescattering approximation and also assumed $E^+ \gg \mu_{i\perp}$, $i=1,2$.

Finally, contact limit of this amplitude reads:
\begin{widetext}
\begin{align}~\label{m220_o}
M^c_{2, 2, 0}={}&
-i J_a(p+k) e^{i(p+k)x_0}f^{ecd}f^{bea_2}f^{aba_1}T_{a_2}T_{a_1}(1-x+x^2)  (-i)\int{\frac{d^2{\mathbf{q}}_1}{(2\pi)^2}} (-i) \int{\frac{d^2{\mathbf{q}}_2}{(2\pi)^2}} v(0,{\mathbf{q}}_1) v(0, {\mathbf{q}}_2)e^{-i({\mathbf{q}}_1+{\mathbf{q}}_2) \cdot {\mathbf{b}}_1} \nonumber \\
 \times {} & \frac{1}{2} (2 ig_s) \frac{{\mathbf{\boldsymbol{\epsilon}}}\cdot((1-x){\mathbf{k}}-x{\mathbf{p}})}{((1-x){\mathbf{k}}-x{\mathbf{p}})^2} e^{i(\frac{{\mathbf{k}}^{\scalebox{.8}{$\scriptscriptstyle  2$}}}{xE^{\scalebox{.8}{$\scriptscriptstyle  +$}}} +\frac{{\mathbf{p}}^{\scalebox{.8}{$\scriptscriptstyle  2$}}}{(1-x)E^{\scalebox{.8}{$\scriptscriptstyle  +$}}})(z_1-z_0)}\nonumber \\
={} & -J_a(p+k)e^{i(p+k)x_0} (T^cT^{a_2}T^{a_1})_{da} T_{a_2}T_{a_1}(1-x+x^2)  (-i)\int{\frac{d^2{\mathbf{q}}_1}{(2\pi)^2}} (-i) \int{\frac{d^2{\mathbf{q}}_2}{(2\pi)^2}} v(0,{\mathbf{q}}_1) v(0, {\mathbf{q}}_2)e^{-i({\mathbf{q}}_1+{\mathbf{q}}_2) \cdot {\mathbf{b}}_1} \nonumber \\
 \times {} & \frac{1}{2} (2 ig_s) \frac{{\mathbf{\boldsymbol{\epsilon}}}\cdot({\mathbf{k}}-x({\mathbf{q}}_1 + {\mathbf{q}}_2))}{({\mathbf{k}}-x({\mathbf{q}}_1+ {\mathbf{q}}_2))^2} e^{\frac{i}{2\omega}({\mathbf{k}}^2 + \frac{x}{1-x}({\mathbf{k}}-{\mathbf{q}}_1-{\mathbf{q}}_2)^2)(z_1-z_0)},
\end{align}
\end{widetext}
where we applied Eq.~\eqref{timpulsiM2} and manipulated with $SU(N_c=3)$ structure constants  by using Eqs.~\eqref{gen1} and~\eqref{gen2}. Also we assumed that ${\mathbf{x}}_1={\mathbf{x}}_2$, since diagrams with two different centers will not contribute to the final result due to Eqs.~\eqref{tr1} and~\eqref{tr2}. 

Note from Fig.~\ref{M2_1} that $M_{2, 2, 0}$ is symmetric under the substitutions: ($p\leftrightarrow k, x\leftrightarrow(1-x), c\leftrightarrow d$), which can be straightforwardly verified by implementing these substitutions in the first two lines of Eq.~\eqref{m220_o}.
\newline
\section{\label{sec:M203}Diagrams $M_{2, 0, 3}$ and $M_{2, 0, 0}$}

\begin{figure*}
\begin{subfigure}{0.42\textwidth}
\includegraphics[width=\linewidth]{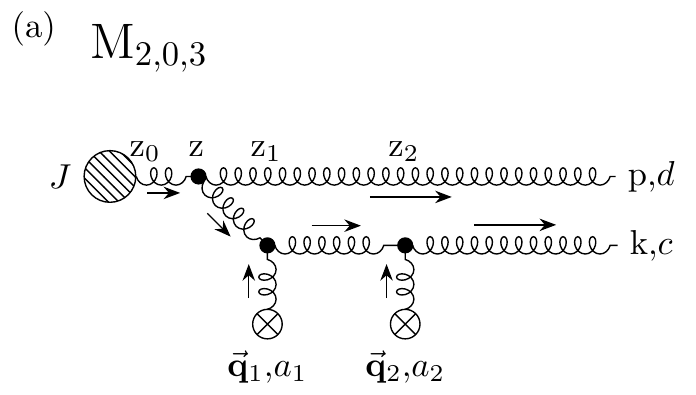}
\end{subfigure}
\begin{subfigure}{0.42\textwidth}
\includegraphics[width=\linewidth]{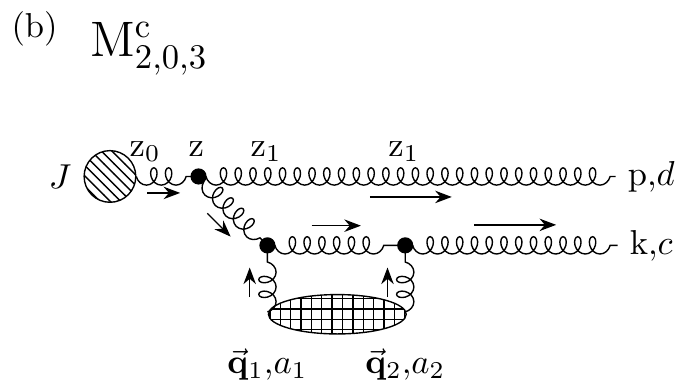}
\end{subfigure}  
\begin{subfigure}{0.42\textwidth}
\includegraphics[width=\linewidth]{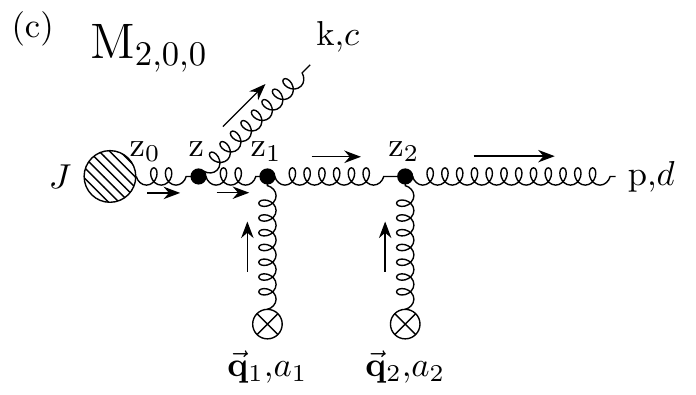}
\end{subfigure}
\begin{subfigure}{0.42\textwidth}
\includegraphics[width=\linewidth]{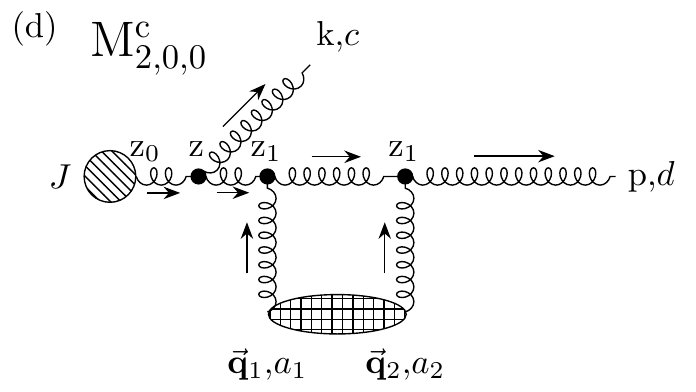}
\end{subfigure}
\caption{\small Feynman diagrams $M_{2, 0, 3}$ and $M_{2, 0, 0}$ in well-separated (
(a) and (c)) and in contact-limit case ($z_1=z_2$), which contributes to the first order in opacity of gluon-jet radiative energy loss: $M^c_{2, 0, 3}$ and $M^c_{2, 0, 0}$ (
(b) and (d)). Remaining labeling is the same as in Fig.~\ref{M2_1}.}\label{M2_2}
\end{figure*}

Next we consider $M_{2, 0, 3}$ diagram, where the radiated gluon suffers two consecutive interactions with the QCD medium (
Figs.~\ref{M2_2} (a) and (b)).

Note that the order of the color and Dirac indices denoting vertices is the same for all the remaining diagrams containing two interactions with the scatterers as in Fig.~\ref{M2_1}, and therefore omitted onward.
\begin{widetext}
\begin{align}~\label{m203_1}
M_{2, 0, 3}={}& \int{\frac{d^4q_1}{(2\pi)^4}\frac{d^4q_2}{(2\pi)^4}\epsilon^*_{\rho}(k) f^{eca_2}\Big(g^{\xi 0} (k-2q_2)^{\rho} + g^{\xi \rho} (-2k + q_2)^{0} + g^{\rho 0} (k+q_2)^{\xi} \Big) T_{a_2} V(q_2) e^{iq_2x_2}} \frac{-i \delta_{e e'} g_{\xi \xi'}}{(k-q_2)^2 +i\epsilon}  \nonumber \\
 \times {} & f^{be'a_1} \Big(g^{\nu 0} (k-2q_1-q_2)^{\xi'} + g^{\nu \xi'}(-2k+q_1+2q_2)^0 +g^{\xi' 0} (k+q_1-q_2)^{\nu} \Big) T_{a_1}V(q_1)e^{iq_1x_1} \frac{-i\delta_{bb'}g_{\nu \nu'}}{(k-q_1-q_2)^2 +i\epsilon}    \nonumber \\
 \times {} & \epsilon^*_{\sigma}(p) g_s f^{adb'}\Big(g^{\mu \nu'} (p+2k-2q_1 -2q_2)^{\sigma} +      g^{\mu \sigma} (-2p-k+q_1+q_2)^{\nu'} + g^{\sigma \nu'} (p-k+q_1+q_2)^{\mu} \Big)  \nonumber \\
 \times {} & \frac{-i\delta_{aa'}g_{\mu \mu'}}{(p+k-q_1-q_2)^2 +i\epsilon} i J_{a'}(p+k-q_1-q_2) \epsilon^{\mu'}(p+k-q_1-q_2) e^{i(p+k-q_1-q_2)x_0} \nonumber \\
\approx {} & i J_a(p+k) e^{i(p+k)x_0}f^{eca_2}f^{bea_1}f^{adb}T_{a_2}T_{a_1}\frac{(1-x+x^2)}{1-x}   (-i) \int{\frac{d^2{\mathbf{q}}_1}{(2\pi)^2}}(-i) \int{\frac{d^2{\mathbf{q}}_2}{(2\pi)^2}} (2 ig_s) \, {\mathbf{\boldsymbol{\epsilon}}}\cdot{\mathbf{p}} \;  e^{-i{\mathbf{q}}_1 \cdot {\mathbf{b}}_1}e^{-i{\mathbf{q}}_2 \cdot {\mathbf{b}}_2}  \nonumber \\
 \times {} & \int{\frac{dq_{1z}}{2\pi} \frac{dq_{2z}}{2\pi}}\frac{E^+ k^+ v(q_{1z},{\mathbf{q}}_1)v(q_{2z},{\mathbf{q}}_2)e^{-iq_{1z}(z_1-z_0)}e^{-iq_{2z}(z_2-z_0)}}{((p+k-q_1-q_2)^2 +i\epsilon)((k-q_1-q_2)^2 +i\epsilon)((k-q_2)^2 +i\epsilon)}.
\end{align}
\end{widetext}
Next, again by changing the variables $q_{1z}\rightarrow q_z=q_{1z}+q_{2z}$, we define the following  integral:
\begin{widetext}
\begin{align}~\label{I2_2}
I_2(p, k, {\mathbf{q}}_1, \vec{{\mathbf{q}}}_2, z_1-z_0) = {} & \int{\frac{dq_{z}}{2\pi}}\frac{v(q_z-q_{2z},{\mathbf{q}}_1)e^{-iq_{z}(z_1-z_0)}}{((p+k-q_1-q_2)^2 +i\epsilon)((k-q_1-q_2)^2 +i\epsilon)}.
\end{align}
\end{widetext}
Again, as explained in the previous section, we close the contour in lower half-plane, and since $\mu (z_1 -z_0) \gg 1$   the pole at $q_z= -i\mu_{1\perp} + q_{2z}$ is again exponentially suppressed. Therefore the remaining $q_z$ singularities originating from gluon propagators are:
\begin{widetext}
\begin{align}~\label{m203_pol1}
\bar{q}_1= {} & -\frac{{\mathbf{k}}^2}{xE^+} -\frac{{\mathbf{p}}^2}{(1-x)E^+} -i\epsilon 
 = -\frac{{\mathbf{k}}^2}{2\omega} -\frac{x}{1-x}\frac{({\mathbf{k}}-{\mathbf{q}}_1-{\mathbf{q}}_2)^2}{2\omega} -i\epsilon, \nonumber \\
 \bar{q}_2= {} & -\frac{{\mathbf{k}}^2}{xE^+} +\frac{{\mathbf{p}}^2}{xE^+} -i\epsilon  = -\frac{{\mathbf{k}}^2}{2\omega} +\frac{({\mathbf{k}}-{\mathbf{q}}_1-{\mathbf{q}}_2)^2}{2\omega}  -i\epsilon.
\end{align}
\end{widetext}
After performing the integration, i.e. summing the residues at these two poles, $I_2$ now reads:
\begin{widetext}
\begin{align}~\label{I2_2o}
I_2(p, k, {\mathbf{q}}_1, \vec{{\mathbf{q}}}_2, z_1-z_0) \approx  v(-q_{2z}, {\mathbf{q}}_1) \frac{i(1-x)}{E^+({\mathbf{k}}-{\mathbf{q}}_1-{\mathbf{q}}_2)^2}\Big( & e^{\frac{i}{2\omega} ({\mathbf{k}}^2 + \frac{x}{1-x}({\mathbf{k}}-{\mathbf{q}}_1-{\mathbf{q}}_2)^2)(z_1-z_0)} -  e^{\frac{i}{2\omega} ({\mathbf{k}}^2 -({\mathbf{k}}-{\mathbf{q}}_1-{\mathbf{q}}_2)^2)(z_1-z_0)} \Big).
\end{align}
\end{widetext}
The remaining integral over $q_{2z}$ is:
\begin{align}~\label{I3_2}
I_3(p, k, {\mathbf{q}}_1, {\mathbf{q}}_2, z_2-z_1) = {} & \int{\frac{dq_{2z}}{2\pi}}\frac{v(q_{2z},{\mathbf{q}}_2)e^{-iq_{2z}(z_2-z_1)}}{(k-q_2)^2 +i\epsilon}  \nonumber \\
 \times {} & v(-q_{2z},{\mathbf{q}}_1),
\end{align}
and since we are interested only in the contact-limit case  (i.e. $z_1=z_2$), we need to calculate:
\begin{align}~\label{I3_2c}
I^c_3(p, k, {\mathbf{q}}_1, {\mathbf{q}}_2, 0) = \int{\frac{dq_{2z}}{2\pi}}\frac{v(q_{2z},{\mathbf{q}}_2)v(-q_{2z},{\mathbf{q}}_1) }{(k-q_2)^2 +i\epsilon},
\end{align}
which gives:
\begin{align}~\label{I3_co2}
I^c_3(p, k, {\mathbf{q}}_1, {\mathbf{q}}_2,0) \approx -v(0,{\mathbf{q}}_1)v(0,{\mathbf{q}}_2)\frac{i}{2xE^+},
\end{align}
which can readily be shown to represent exactly $\frac{1}{2}$ of the strength of the well-separated limit Eq.~\eqref{I3_2}, as for $M_{2,2,0}$ amplitude.
The contact limit of this amplitude reduces to:
\begin{widetext}
\begin{align}~\label{m203_co}
M^c_{2, 0, 3}={}&
i J_a(p+k) e^{i(p+k)x_0}f^{eca_2}f^{bea_1}f^{adb}T_{a_2}T_{a_1}(1-x+x^2)   (-i)\int{\frac{d^2{\mathbf{q}}_1}{(2\pi)^2}} (-i) \int{\frac{d^2{\mathbf{q}}_2}{(2\pi)^2}} v(0,{\mathbf{q}}_1) v(0, {\mathbf{q}}_2)e^{-i({\mathbf{q}}_1+{\mathbf{q}}_2)\cdot {\mathbf{b}}_1} \nonumber \\
 \times {} & \frac{1}{2} (2 ig_s) \frac{{\mathbf{\boldsymbol{\epsilon}}}\cdot{\mathbf{p}}}{{\mathbf{p}}^2}\Big( e^{i(\frac{{\mathbf{k}}^{\scalebox{.8}{$\scriptscriptstyle  2$}}}{xE^{\scalebox{.8}{$\scriptscriptstyle  +$}}} +\frac{{\mathbf{p}}^{\scalebox{.8}{$\scriptscriptstyle  2$}}}{(1-x)E^{\scalebox{.8}{$\scriptscriptstyle  +$}}})(z_1-z_0)} -e^{i\frac{({\mathbf{k}}^{\scalebox{.8}{$\scriptscriptstyle  2$}}-{\mathbf{p}}^{\scalebox{.8}{$\scriptscriptstyle  2$}})}{xE^{\scalebox{.8}{$\scriptscriptstyle  +$}}} (z_1-z_0)} \Big) \nonumber \\
={} & J_a(p+k)e^{i(p+k)x_0} [[T^c, T^{a_2}],T^{a_1}]_{da} T_{a_2}T_{a_1}(1-x+x^2)  (-i)\int{\frac{d^2{\mathbf{q}}_1}{(2\pi)^2}} (-i) \int{\frac{d^2{\mathbf{q}}_2}{(2\pi)^2}} v(0,{\mathbf{q}}_1) v(0, {\mathbf{q}}_2)e^{-i({\mathbf{q}}_1+{\mathbf{q}}_2)\cdot {\mathbf{b}}_1} \nonumber \\
 \times {} & \frac{1}{2} (2 ig_s) \frac{{\mathbf{\boldsymbol{\epsilon}}}\cdot({\mathbf{k}}-{\mathbf{q}}_1 - {\mathbf{q}}_2)}{({\mathbf{k}}-{\mathbf{q}}_1- {\mathbf{q}}_2)^2} \Big( e^{\frac{i}{2\omega}({\mathbf{k}}^2 + \frac{x}{1-x}({\mathbf{k}}-{\mathbf{q}}_1-{\mathbf{q}}_2)^2)(z_1-z_0)} - e^{\frac{i}{2\omega}({\mathbf{k}}^2 - ({\mathbf{k}}-{\mathbf{q}}_1-{\mathbf{q}}_2)^2)(z_1-z_0)} \Big).
\end{align}
\end{widetext}

Proceeding in the same manner, for $M^c_{2, 0, 0}$ amplitude (
Figs.~\ref{M2_2} (c) and (d)) we obtain:
\begin{widetext}
\begin{align}~\label{m200_co}
M^c_{2, 0, 0}={}&
i J_a(p+k) e^{i(p+k)x_0}f^{eda_2}f^{bea_1}f^{acb}T_{a_2}T_{a_1}(1-x+x^2)    (-i)\int{\frac{d^2{\mathbf{q}}_1}{(2\pi)^2}} (-i) \int{\frac{d^2{\mathbf{q}}_2}{(2\pi)^2}} v(0,{\mathbf{q}}_1) v(0, {\mathbf{q}}_2)e^{-i({\mathbf{q}}_1+{\mathbf{q}}_2)\cdot {\mathbf{b}}_1} \nonumber \\
 \times {} & \frac{1}{2} (2 ig_s) \frac{{\mathbf{\boldsymbol{\epsilon}}}\cdot{\mathbf{k}}}{{\mathbf{k}}^2}\Big( e^{i(\frac{{\mathbf{k}}^{\scalebox{.8}{$\scriptscriptstyle  2$}}}{xE^{\scalebox{.8}{$\scriptscriptstyle  +$}}} +\frac{{\mathbf{p}}^{\scalebox{.8}{$\scriptscriptstyle  2$}}}{(1-x)E^{\scalebox{.8}{$\scriptscriptstyle  +$}}})(z_1-z_0)} -e^{i\frac{({\mathbf{p}}^{\scalebox{.8}{$\scriptscriptstyle  2$}}-{\mathbf{k}}^{\scalebox{.8}{$\scriptscriptstyle  2$}})}{(1-x)E^{\scalebox{.8}{$\scriptscriptstyle  +$}}} (z_1-z_0)} \Big) \nonumber \\
={} & J_a(p+k)e^{i(p+k)x_0} (T^{a_2} T^{a_1}T^{c})_{da} T_{a_2}T_{a_1}(1-x+x^2)  (-i)\int{\frac{d^2{\mathbf{q}}_1}{(2\pi)^2}} (-i) \int{\frac{d^2{\mathbf{q}}_2}{(2\pi)^2}} v(0,{\mathbf{q}}_1) v(0, {\mathbf{q}}_2)e^{-i({\mathbf{q}}_1+{\mathbf{q}}_2) \cdot {\mathbf{b}}_1} \nonumber \\
 \times {} & \frac{1}{2} (2 ig_s) \frac{{\mathbf{\boldsymbol{\epsilon}}}\cdot{\mathbf{k}}}{{\mathbf{k}}^2} \Big( e^{\frac{i}{2\omega}({\mathbf{k}}^2 + \frac{x}{1-x}({\mathbf{k}}-{\mathbf{q}}_1-{\mathbf{q}}_2)^2)(z_1-z_0)} - e^{\frac{i}{2\omega} \frac{x}{1-x}(({\mathbf{k}}-{\mathbf{q}}_1-{\mathbf{q}}_2)^2-{\mathbf{k}}^2)(z_1-z_0)} \Big).
\end{align}
\end{widetext}
From Fig.~\ref{M2_2} we infer that $M_{2,0,3}$ and $M_{2, 0, 0}$ are symmetric under the following substitutions: ($p\leftrightarrow k, x\leftrightarrow(1-x), c\leftrightarrow d$), which can be straightforwardly verified by implementing these substitutions in Eqs.~\eqref{m203_co} and~\eqref{m200_co}.

\section{\label{sec:M201}Diagrams $M_{2, 0, 1}$ and $M_{2, 0, 2}$}

\begin{figure*}
\begin{subfigure}{0.42\textwidth}
\includegraphics[width=\linewidth]{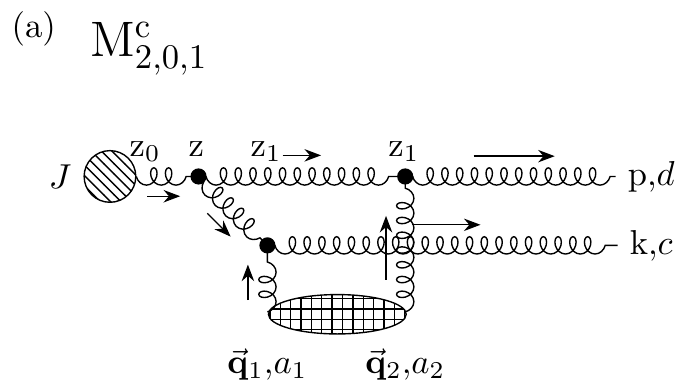}
\end{subfigure}
\begin{subfigure}{0.42\textwidth}
\includegraphics[width=\linewidth]{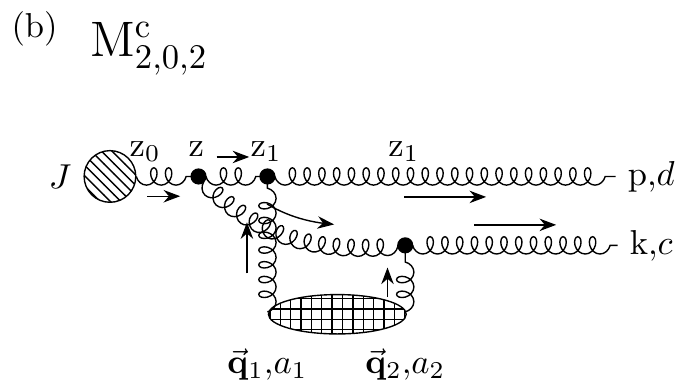}
\end{subfigure}
\caption{\small Topologically indistinct Feynman diagrams: (a) $M^c_{2, 0, 1}$ and (b) $M^c_{2, 0, 2}$ in contact  limit ($z_1=z_2$), which contribute to the first order in opacity of gluon-jet radiative energy loss. Remaining labeling is the same as in Fig.~\ref{M2_1}.}\label{M2_3} 
\end{figure*}
 Here we consider the case when both initial gluon jet and radiated gluon interact with one scattering center. We provide only the contact-limit case diagrams $M^c_{2, 0, 1}$ and $M^c_{2,0,2}$ (Fig.~\ref{M2_3}), since, in the end, only they are used in calculating radiative energy loss to the first order in opacity.
 \newpage
 \begin{widetext}
\begin{align}~\label{m201_1}
M_{2, 0, 1}={}& \int{\frac{d^4q_1}{(2\pi)^4}\frac{d^4q_2}{(2\pi)^4}\epsilon^*_{\sigma}(p) f^{eda_2}\Big(g^{\xi 0} (p-2q_2)^{\sigma} + g^{\xi \sigma} (-2p + q_2)^{0} + g^{\sigma 0} (p+q_2)^{\xi} \Big) T_{a_2} V(q_2) e^{iq_2x_2}}\nonumber \\
 \times {} & \frac{-i \delta_{e e'} g_{\xi \xi'}}{(p-q_2)^2 +i\epsilon} g_s f^{ae'b'} \Big(g^{\mu \nu'} (p+2k-2q_1-q_2)^{\xi'} + g^{\mu \xi'}(-2p-k+q_1+2q_2)^{\nu'} + g^{\xi' \nu'} (p-k+q_1-q_2)^{\mu} \Big)    \nonumber \\
 \times {} & \epsilon^*_{\rho}(k) f^{bca_1}\Big(g^{\nu 0} (k-2q_1)^{\rho} + g^{\nu \rho} (-2k+q_1)^{0} + g^{\rho 0} (k+q_1)^{\nu} \Big)T_{a_1}V(q_1)e^{iq_1x_1}\frac{-i\delta_{bb'}g_{\nu \nu'}}{(k-q_1)^2 +i\epsilon}  \nonumber \\
 \times  {} &  \frac{-i\delta_{aa'}g_{\mu \mu'}}{(p+k-q_1-q_2)^2 +i\epsilon}   i J_{a'}(p+k-q_1-q_2) \epsilon^{\mu'}(p+k-q_1-q_2) e^{i(p+k-q_1-q_2)x_0} \nonumber \\
\approx & -i J_a(p+k) e^{i(p+k)x_0}f^{eda_2}f^{aeb}f^{bca_1}T_{a_2}T_{a_1}(1-x+x^2)   (-i) \int{\frac{d^2{\mathbf{q}}_1}{(2\pi)^2}}(-i) \int{\frac{d^2{\mathbf{q}}_2}{(2\pi)^2}}  (2 ig_s)\, {\mathbf{\boldsymbol{\epsilon}}}\cdot ({\mathbf{k}}-{\mathbf{q}}_1) \; e^{-i{\mathbf{q}}_1 \cdot {\mathbf{b}}_1}  \nonumber \\
 \times {} & e^{-i{\mathbf{q}}_2 \cdot {\mathbf{b}}_2} (E^+)^2 \int{\frac{dq_{1z}}{2\pi} \frac{dq_{2z}}{2\pi}}\frac{v(q_{1z},{\mathbf{q}}_1)v(q_{2z},{\mathbf{q}}_2)e^{-iq_{1z}(z_1-z_0)}e^{-iq_{2z}(z_2-z_0)}}{((p+k-q_1-q_2)^2 +i\epsilon)((k-q_1)^2 +i\epsilon)((p-q_2)^2 +i\epsilon)}.
\end{align}
\end{widetext}
Again, by changing the variables $q_{1z}\rightarrow q_z=q_{1z}+q_{2z}$, we define the following  integral:
\begin{align}~\label{I2_3}
I_2(p, k, {\mathbf{q}}_1, \vec{{\mathbf{q}}}_2, z_1-z_0) = {} & \int{\frac{dq_{z}}{2\pi}}\frac{v(q_z-q_{2z},{\mathbf{q}}_1)e^{-iq_{z}(z_1-z_0)}}{(p+k-q_1-q_2)^2 +i\epsilon} \nonumber \\
 \times {} & \frac{1}{(k-q_1)^2 +i\epsilon}.
\end{align}
Since $z_1>z_0$ we must close the contour in lower half-plane, and since $\mu (z_1 -z_0) \gg 1$ again we neglect the pole at $q_z= -i\mu_{1\perp} + q_{2z}$. Therefore the remaining $q_z$ singularities originating from gluon propagators are:
\begin{equation}~\label{m201_pol1}
\begin{aligned}
& \bar{q}_1=  -\frac{{\mathbf{k}}^2}{2\omega} -\frac{x}{1-x}\frac{({\mathbf{k}}-{\mathbf{q}}_1-{\mathbf{q}}_2)^2}{2\omega} -i\epsilon, \\
& \bar{q}_2=  -\frac{{\mathbf{k}}^2}{2\omega} + \frac{({\mathbf{k}}-{\mathbf{q}}_1)^2}{2\omega}+q_{2z} -i\epsilon.
\end{aligned}
\end{equation}
Summing the residues gives:
\begin{widetext}
\begin{align}~\label{I2_3o}
 I_2(p,& k,  {\mathbf{q}}_1, \vec{{\mathbf{q}}}_2, z_1-z_0) \approx    \frac{i e^{i\frac{{\mathbf{k}}^2}{2\omega} (z_1-z_0)}}{E^+k^+(q_{2z}+\frac{({\mathbf{k}}-{\mathbf{q}}_1)^2}{2\omega}+ \frac{x}{1-x}\frac{({\mathbf{k}}-{\mathbf{q}}_1-{\mathbf{q}}_2)^2}{2\omega})}  \nonumber \\
& \times \Big(v(- q_{2z}-\frac{{\mathbf{k}}^2}{2\omega} -\frac{x}{1-x}\frac{({\mathbf{k}}-{\mathbf{q}}_1-{\mathbf{q}}_2)^2}{2\omega}  , {\mathbf{q}}_1) e^{i \frac{x}{1-x} \frac{({\mathbf{k}}-{\mathbf{q}}_{\scalebox{.8}{$\scriptscriptstyle  1$}}-{\mathbf{q}}_{\scalebox{.8}{$\scriptscriptstyle  2$}})^2}{2\omega} (z_1-z_0)}  - v(\frac{({\mathbf{k}}-{\mathbf{q}}_1)^2}{2\omega}-\frac{{\mathbf{k}}^2}{2\omega} , {\mathbf{q}}_1) e^{-i ( q_{2z} + \frac{({\mathbf{k}}-{\mathbf{q}}_{\scalebox{.8}{$\scriptscriptstyle  1$}})^2}{2\omega}) (z_1-z_0)} \Big).
\end{align}
\end{widetext}
The remaining $q_{2z}$ integral is:
\begin{widetext}
\begin{align}~\label{I3_3}
 I_3(p,  k, {\mathbf{q}}_1, {\mathbf{q}}_2, & z_2-z_0, z_2-z_1) = \int{\frac{dq_{2z}}{2\pi}}\frac{1}{q_{2z} +  \frac{({\mathbf{k}}-{\mathbf{q}}_1)^2}{2\omega}+ \frac{x}{1-x}\frac{({\mathbf{k}}-{\mathbf{q}}_1-{\mathbf{q}}_2)^2}{2\omega} } \, \frac{v(q_{2z},{\mathbf{q}}_2)}{(p-q_2)^2 +i\epsilon} \nonumber \\
& \times  \Big( e^{-iq_{2z}(z_2-z_1)}{}e^{\frac{i}{2\omega} ({\mathbf{k}}^2 +\frac{x}{1-x}({\mathbf{k}}-{\mathbf{q}}_1-{\mathbf{q}}_2)^2 )(z_1-z_0)}
v(-q_{2z} -\frac{{\mathbf{k}}^2}{2\omega} -\frac{x}{1-x}\frac{({\mathbf{k}}-{\mathbf{q}}_1-{\mathbf{q}}_2)^2}{2\omega} ,{\mathbf{q}}_1)   \nonumber \\
& - e^{-iq_{2z}(z_2-z_0)} e^{-\frac{i}{2\omega}(({\mathbf{k}}-{\mathbf{q}}_1)^2 - {\mathbf{k}}^2)(z_1-z_0)} v(\frac{({\mathbf{k}}-{\mathbf{q}}_1)^2}{2\omega} -\frac{{\mathbf{k}}^2}{2\omega} ,{\mathbf{q}}_1) \Big),
\end{align}
\end{widetext}
where the singularity on $q_{2z}$ real axis: $q_{2z}= -\frac{({\mathbf{k}}-{\mathbf{q}}_1)^2}{2\omega} -\frac{x}{1-x}\frac{({\mathbf{k}}-{\mathbf{q}}_1-{\mathbf{q}}_2)^2}{2\omega}  \equiv -a$, ($a>0$) has to be avoided by taking Cauchy principal value of $I_3$ according to the Fig.~\ref{PVf}, i.e.:
\begin{figure*}
\centering
\begin{equation}
\begin{subfigure}{0.2\textwidth}
\includegraphics[width=\linewidth]{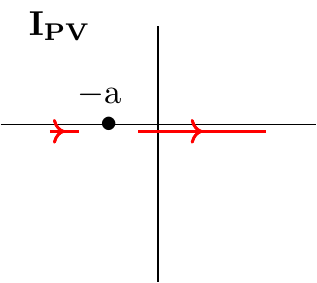}
\end{subfigure} = 
\begin{subfigure}{0.2\textwidth}
\includegraphics[width=\linewidth]{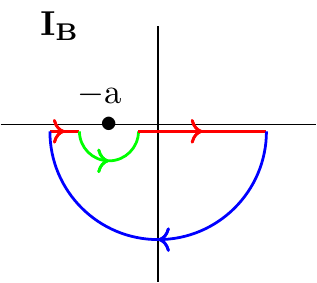}
\end{subfigure}
 - 
\begin{subfigure}{0.2\textwidth} \includegraphics[width=\linewidth]{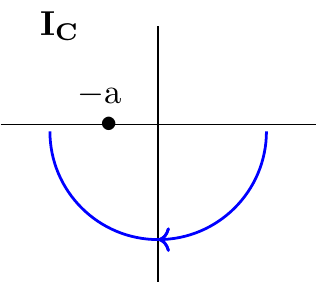} 
\end{subfigure} - 
\begin{subfigure}{0.2\textwidth}
\includegraphics[width=\linewidth]{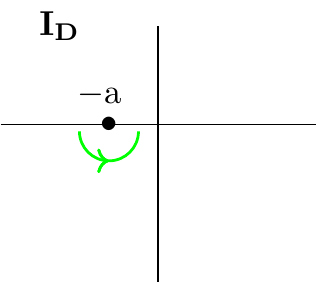}
\end{subfigure} \nonumber
\end{equation}
\caption{Illustration of calculating Cauchy principal value ($I_{PV}$) in the case when singularity (-a) on the real (horizontal) axis arises.}
\label{PVf}
\end{figure*}
\begin{eqnarray}
I_3 \equiv I_{PV}= I_B -I_C-I_D,
\end{eqnarray}
where $I_B=-2 \pi i \sum_i Res(I_3(\bar{q}_i))$, with $i$ counting the poles in the lower-half plane. Additionally $I_C=0$, and it's straightforward to show, that after the following substitution $q_{2z}= -a + r e^{i \varphi} $, where $r \rightarrow 0$, also $I_D =0$.
Therefore, principal value of $I_3$ reduces to $I_B$, i.e. $-2 \pi i \sum_i Res(I_3(\bar{q}_i))$.

In the well-separated case Eq.~\eqref{I3_3} poles originating from Yukawa potentials ($q_{2z}= -\frac{{\mathbf{k}}^2}{2\omega} -\frac{x}{1-x}\frac{({\mathbf{k}}-{\mathbf{q}}_1-{\mathbf{q}}_2)^2}{2\omega} -i\mu_{1\perp}$ and $q_{2z}= -i\mu_{2\perp}$) are again exponentially suppressed ($e^{-\mu_{i\perp}(z_2-z_{0,1})}\rightarrow0$, $i=1, 2$) and therefore can be neglected, so  only the pole from the propagator survives $q_{2z}= \frac{x}{1-x} ( \frac{({\mathbf{k}}-{\mathbf{q}}_1)^2}{2\omega}-\frac{({\mathbf{k}}-{\mathbf{q}}_1-{\mathbf{q}}_2)^2}{2\omega} ) -i\epsilon$. However, since we are interested only in the contact-limit case  (i.e. $z_1=z_2$), instead of Eq.~\eqref{I3_3} we need to calculate the principal value of the following integral:
\begin{widetext}
\begin{align}~\label{I3_3c}
 I^c_3(p,  k, {\mathbf{q}}_1, {\mathbf{q}}_2, & z_1-z_0) = \int{\frac{dq_{2z}}{2\pi}}\frac{1}{q_{2z} +\frac{({\mathbf{k}}-{\mathbf{q}}_1)^2}{2\omega} + \frac{x}{1-x}\frac{({\mathbf{k}}-{\mathbf{q}}_1-{\mathbf{q}}_2)^2}{2\omega}  }  \nonumber \\
& \times  \Big( \frac{e^{\frac{i}{2\omega} ( {\mathbf{k}}^2+\frac{x}{1-x}({\mathbf{k}}-{\mathbf{q}}_1-{\mathbf{q}}_2)^2 )(z_1-z_0)}}{(p-q_2)^2 +i\epsilon}  v(-q_{2z} - \frac{{\mathbf{k}}^2}{2\omega}-\frac{x}{1-x}\frac{({\mathbf{k}}-{\mathbf{q}}_1-{\mathbf{q}}_2)^2}{2\omega} ,{\mathbf{q}}_1) v(q_{2z},{\mathbf{q}}_2)  \nonumber \\
& -\frac{e^{-iq_{2z}(z_1-z_0)}{} e^{-\frac{i}{2\omega}(({\mathbf{k}}-{\mathbf{q}}_1)^2 - {\mathbf{k}}^2)(z_1-z_0)}}{(p-q_2)^2 +i\epsilon} v( \frac{({\mathbf{k}}-{\mathbf{q}}_1)^2}{2\omega}- \frac{{\mathbf{k}}^2}{2\omega},{\mathbf{q}}_1) v(q_{2z},{\mathbf{q}}_2) \Big),
\end{align}
\end{widetext}
which again reduces to the sum of residua, with $-a$ effectively not being a pole (Fig.~\ref{PVf}). Particularly, for the second term in the bracket of Eq.~\eqref{I3_3c}, only the propagator pole survives, while for the first term in the bracket all three poles have to be accounted, although residues at poles from potentials sum to the order of ${\mathcal{O}}(\frac{({\mathbf{k-q_1}})^{\scalebox{.8}{$\scriptscriptstyle  2$}}}{x(1-x)E^{\scalebox{.8}{$\scriptscriptstyle  +$}}(\mu_{1\perp} +\mu_{2\perp})})$, and thus can be neglected compared to the remaining residue.

Finally, in the contact-limit case we obtain:
\begin{widetext}
\begin{align}~\label{m201_co}
M^c_{2, 0, 1} \approx {}&
-i J_a(p+k) e^{i(p+k)x_0}f^{eda_2}f^{aeb}f^{bca_1}T_{a_2}T_{a_1}(1-x+x^2)    (-i)\int{\frac{d^2{\mathbf{q}}_1}{(2\pi)^2}} (-i) \int{\frac{d^2{\mathbf{q}}_2}{(2\pi)^2}} v(0,{\mathbf{q}}_1) v(0, {\mathbf{q}}_2)e^{-i({\mathbf{q}}_1+ {\mathbf{q}}_2) \cdot {\mathbf{b}}_1} \nonumber \\
 \times {} & (2 ig_s) \frac{{\mathbf{\boldsymbol{\epsilon}}}\cdot({\mathbf{k}}- {\mathbf{q}}_1)}{({\mathbf{k}}-{\mathbf{q}}_1)^2}\Big( e^{i(\frac{{\mathbf{k}}^{\scalebox{.8}{$\scriptscriptstyle  2$}}}{xE^{\scalebox{.8}{$\scriptscriptstyle  +$}}} +\frac{{\mathbf{p}}^{\scalebox{.8}{$\scriptscriptstyle  2$}}}{(1-x)E^{\scalebox{.8}{$\scriptscriptstyle  +$}}})(z_1-z_0)} -e^{i(\frac{{\mathbf{k}}^{\scalebox{.8}{$\scriptscriptstyle  2$}}}{xE^{\scalebox{.8}{$\scriptscriptstyle  +$}}} + \frac{{\mathbf{p}}^{\scalebox{.8}{$\scriptscriptstyle  2$}}}{(1-x)E^{\scalebox{.8}{$\scriptscriptstyle  +$}}} - \frac{({\mathbf{k}}- {\mathbf{q}}_{\scalebox{.8}{$\scriptscriptstyle  1$}})^{\scalebox{.8}{$\scriptscriptstyle  2$}}}{x(1-x)E^{\scalebox{.8}{$\scriptscriptstyle  +$}}} )(z_1-z_0)} \Big) \nonumber \\
={} & J_a(p+k)e^{i(p+k)x_0} (T^{a_2} [T^{c},T^{a_1}])_{da} T_{a_2}T_{a_1}(1-x+x^2)  (-i)\int{\frac{d^2{\mathbf{q}}_1}{(2\pi)^2}} (-i) \int{\frac{d^2{\mathbf{q}}_2}{(2\pi)^2}} v(0,{\mathbf{q}}_1) v(0, {\mathbf{q}}_2)e^{-i({\mathbf{q}}_1+ {\mathbf{q}}_2) \cdot {\mathbf{b}}_1} \nonumber \\
 \times {} & (2 ig_s) \frac{{\mathbf{\boldsymbol{\epsilon}}}\cdot({\mathbf{k}}- {\mathbf{q}}_1)}{({\mathbf{k}}-{\mathbf{q}}_1)^2} \Big( e^{\frac{i}{2\omega}({\mathbf{k}}^2 +\frac{x}{1-x}({\mathbf{k}}- {\mathbf{q}}_1- {\mathbf{q}}_2)^2)(z_1-z_0)} -e^{\frac{i}{2\omega}({\mathbf{k}}^2-\frac{({\mathbf{k}}-{\mathbf{q}}_{\scalebox{.8}{$\scriptscriptstyle  1$}})^{\scalebox{.8}{$\scriptscriptstyle  2$}}}{1-x} +\frac{x}{1-x}({\mathbf{k}}-{\mathbf{q}}_1- {\mathbf{q}}_2)^2)(z_1-z_0) } \Big).
\end{align}
\end{widetext}
Notice that, contrary to the previous three amplitudes that also included two scattering centers, in Eq.~\eqref{m201_co} no factor $\frac{1}{2}$ when comparing to well-separated limit appears. 

Proceeding in the same manner, for $M^c_{2, 0, 2}$  we obtain:
\begin{widetext}
 \begin{align}~\label{m202_co}
M^c_{2, 0, 2} \approx {}&
i J_a(p+k) e^{i(p+k)x_0}f^{eca_2}f^{abe}f^{bda_1}T_{a_2}T_{a_1}(1-x+x^2)    (-i)\int{\frac{d^2{\mathbf{q}}_1}{(2\pi)^2}} (-i) \int{\frac{d^2{\mathbf{q}}_2}{(2\pi)^2}} v(0,{\mathbf{q}}_1) v(0, {\mathbf{q}}_2)e^{-i({\mathbf{q}}_1+{\mathbf{q}}_2)\cdot {\mathbf{b}}_1} \nonumber \\
 \times {} & (2 ig_s) \frac{{\mathbf{\boldsymbol{\epsilon}}}\cdot({\mathbf{p}}-{\mathbf{q}}_1)}{({\mathbf{p}}- {\mathbf{q}}_1)^2}\Big( e^{i(\frac{{\mathbf{k}}^{\scalebox{.8}{$\scriptscriptstyle  2$}}}{xE^{\scalebox{.8}{$\scriptscriptstyle  +$}}} +\frac{{\mathbf{p}}^{\scalebox{.8}{$\scriptscriptstyle  2$}}}{(1-x)E^{\scalebox{.8}{$\scriptscriptstyle  +$}}})(z_1-z_0)} -e^{i(\frac{{\mathbf{k}}^{\scalebox{.8}{$\scriptscriptstyle  2$}}}{xE^{\scalebox{.8}{$\scriptscriptstyle  +$}}} + \frac{{\mathbf{p}}^{\scalebox{.8}{$\scriptscriptstyle  2$}}}{(1-x)E^{\scalebox{.8}{$\scriptscriptstyle  +$}}} - \frac{({\mathbf{p}}- {\mathbf{q}}_{\scalebox{.8}{$\scriptscriptstyle  1$}})^{\scalebox{.8}{$\scriptscriptstyle  2$}}}{x(1-x)E^{\scalebox{.8}{$\scriptscriptstyle  +$}}} )(z_1-z_0)} \Big) \nonumber \\
={} & J_a(p+k)e^{i(p+k)x_0} (T^{a_1} [T^{c},T^{a_2}])_{da} T_{a_2}T_{a_1}(1-x+x^2) (-i)\int{\frac{d^2{\mathbf{q}}_1}{(2\pi)^2}} (-i) \int{\frac{d^2{\mathbf{q}}_2}{(2\pi)^2}} v(0,{\mathbf{q}}_1) v(0, {\mathbf{q}}_2)e^{-i({\mathbf{q}}_1+ {\mathbf{q}}_2) \cdot {\mathbf{b}}_1} \nonumber \\
 \times {} & (2 ig_s) \frac{{\mathbf{\boldsymbol{\epsilon}}}\cdot({\mathbf{k}}- {\mathbf{q}}_2)}{({\mathbf{k}}- {\mathbf{q}}_2)^2} \Big( e^{\frac{i}{2\omega}({\mathbf{k}}^2 +\frac{x}{1-x}({\mathbf{k}}- {\mathbf{q}}_1- {\mathbf{q}}_2)^2)(z_1-z_0)} -e^{\frac{i}{2\omega}({\mathbf{k}}^2-\frac{({\mathbf{k}}- {\mathbf{q}}_{\scalebox{.8}{$\scriptscriptstyle  2$}})^{\scalebox{.8}{$\scriptscriptstyle  2$}}}{1-x} +\frac{x}{1-x}({\mathbf{k}}-{\mathbf{q}}_1- {\mathbf{q}}_2)^2)(z_1-z_0) } \Big).
\end{align}
\end{widetext}
As for $M^c_{2,0,1}$ amplitude, no factor of $\frac{1}{2}$ appears.
From well-separated analogon of Fig.~\ref{M2_3} we could infer that $M_{2,0,1}$ and $M_{2, 0, 2}$ are symmetric under the following substitutions: ($p\leftrightarrow k, x\leftrightarrow(1-x), c\leftrightarrow d$), which can readily be verified by implementing these substitutions in the first two lines of either of the two Eqs.~\eqref{m201_co} and~\eqref{m202_co} and by using structure constant asymmetry. Note that, in Eq.~\eqref{m202_co} we applied Eq.~\eqref{timpulsiM2}. Also, since in contact-limit case these two diagrams are topologically indistinct, we need to either omit one of them in order to avoid over counting,  or to include both, but multiply each by a factor $\frac{1}{2}$ (we will do the latter).

\section{\label{sec:M210}Diagrams $M_{2, 1, 0}$ and $M_{2, 1, 1}$}

\begin{figure*}
\begin{subfigure}{0.42\textwidth}
\includegraphics[width=\linewidth]{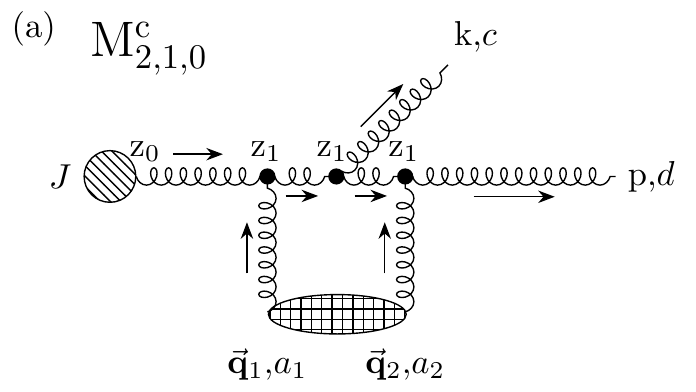}
\end{subfigure}
\begin{subfigure}{0.42\textwidth}
\includegraphics[width=\linewidth]{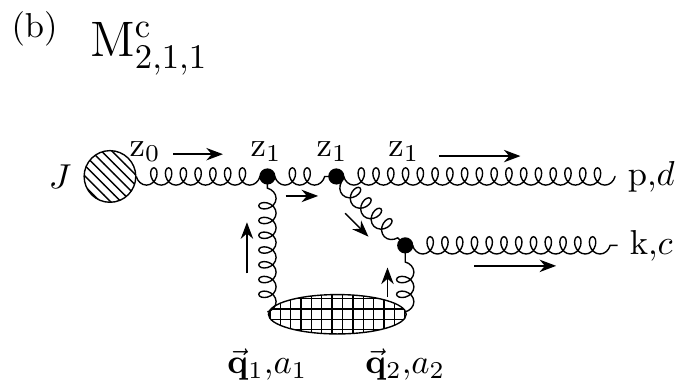}
\end{subfigure}
\caption{\small Feynman diagrams: (a) $M^c_{2, 1, 0}$ and (b) $M^c_{2, 1, 1}$ in contact limit ($z_1=z_2$), which have negligible contribution to the first order in opacity gluon-jet radiative energy loss. Remaining labeling is the same as in Fig.~\ref{M2_1}.}\label{M2_4}
\end{figure*}
The contact-limit case of the remaining two diagrams is presented in Fig.~\ref{M2_4}. These diagrams correspond to the case when one interaction with the scattering center located at $\vec{{\mathbf{x}}}_1$ occurs before and the other interaction at the same place occurs after the gluon has been radiated.

In order to avoid redundant derivations (i.e. repetition of the above calculations) we briefly outline our derivation of Feynman amplitudes for only contact-limit case.

In the light of time-ordered perturbation theory from~\citep{GLV1,timeordered1} these two diagrams are identically equal to zero, since $\int_{t_1}^{t_1}dt...=0$, but for the consistency we will provide a brief verification of this argument.
\begin{widetext}
\begin{align}~\label{m210_1}
M_{2, 1, 0}={}& \int{\frac{d^4q_1}{(2\pi)^4}\frac{d^4q_2}{(2\pi)^4}\epsilon^*_{\sigma}(p) f^{eda_2}\Big(g^{\xi 0} (p-2q_2)^{\sigma} + g^{\xi \sigma} (-2p + q_2)^{0} + g^{\sigma 0} (p+q_2)^{\xi} \Big) }  T_{a_2} V(q_2) e^{iq_2x_2} \frac{-i \delta_{e e'} g_{\xi \xi'}}{(p-q_2)^2 +i\epsilon} \nonumber \\
 \times {} & \epsilon^*_{\rho}(k) g_s f^{bce'} \Big(g^{\nu \xi'} (2p+k-2q_2)^{\rho} + g^{\nu \rho}(-p-2k+q_2)^{\xi'} +g^{\rho \xi'} (-p+k+q_2)^{\nu} \Big) \frac{-i\delta_{bb'}g_{\nu \nu'}}{(p+k-q_2)^2 +i\epsilon}  \nonumber \\
 \times {} & f^{ab'a_1}\Big(g^{\mu 0} (p+k-2q_1-q_2)^{\nu'} + g^{\mu \nu'} (-2p-2k+q_1+2q_2)^{0} + g^{\nu' 0} (p+k+q_1-q_2)^{\mu} \Big)  T_{a_1}V(q_1)e^{iq_1x_1}   \nonumber \\
 \times {} & \frac{-i\delta_{aa'}g_{\mu \mu'}}{(p+k-q_1-q_2)^2 +i\epsilon}  i J_{a'}(p+k-q_1-q_2) \epsilon^{\mu'}(p+k-q_1-q_2) e^{i(p+k-q_1-q_2)x_0} \nonumber \\
\approx {} & i J_a(p+k) e^{i(p+k)x_0}f^{eda_2}f^{bce}f^{aba_1}T_{a_2}T_{a_1}\frac{(1-x+x^2)}{x}   (-i) \int{\frac{d^2{\mathbf{q}}_1}{(2\pi)^2}}(-i) \int{\frac{d^2{\mathbf{q}}_2}{(2\pi)^2}} (2 ig_s) \, {\mathbf{\boldsymbol{\epsilon}}} \cdot ({\mathbf{k}}-x {\mathbf{q}}_1) \; e^{-i{\mathbf{q}}_1 \cdot {\mathbf{b}}_1}  \nonumber \\
 \times {} & e^{-i{\mathbf{q}}_2 \cdot {\mathbf{b}}_2}  (E^+)^2 \int{\frac{dq_{2z}}{2\pi}}\frac{v(q_{2z},{\mathbf{q}}_2)e^{-iq_{2z}(z_2-z_1)}}{((p+k-q_2)^2 +i\epsilon)((p-q_2)^2 +i\epsilon)}\int{\frac{dq_{z}}{2\pi}\frac{v(q_{z}-q_{2z},{\mathbf{q}}_1)e^{-iq_{z}(z_1-z_0)}}{(p+k-q_1-q_2)^2 +i\epsilon}} \nonumber \\
\approx {} & J_a(p+k) e^{i(p+k)x_0}f^{eda_2}f^{bce}f^{aba_1}T_{a_2}T_{a_1}(1-x+x^2)  (-i) \int{\frac{d^2{\mathbf{q}}_1}{(2\pi)^2}}(-i) \int{\frac{d^2{\mathbf{q}}_2}{(2\pi)^2}} (2 ig_s) \, {\mathbf{\boldsymbol{\epsilon}}} \cdot({\mathbf{k}}-x {\mathbf{q}}_1) \; e^{-i{\mathbf{q}}_1 \cdot {\mathbf{b}}_1}   \nonumber \\
 \times {} & e^{-i{\mathbf{q}}_2 \cdot {\mathbf{b}}_2} (E^+)^2  \int{\frac{dq_{2z}}{2\pi}}\frac{v(q_{2z},{\mathbf{q}}_2)e^{-iq_{2z}(z_2-z_1)}}{((p+k-q_2)^2 +i\epsilon)((p-q_2)^2 +i\epsilon)}\frac{1}{k^+}  v(-\frac{{\mathbf{k}}^2}{2\omega}-\frac{x}{1-x} \frac{({\mathbf{k}}- {\mathbf{q}}_1- {\mathbf{q}}_2)^2}{2\omega}-q_{2z},{\mathbf{q}}_1) \nonumber \\
\times {} & e^{\frac{i}{2\omega}({\mathbf{k}}^2+ \frac{x}{1-x}{(\mathbf{k}}- {\mathbf{q}}_1- {\mathbf{q}}_2)^2)(z_1-z_0)}.
\end{align}
\end{widetext}
In the contact-limit case there are four $q_{2z}$ poles of the above integral in the lower half-plane: $- \frac{{\mathbf{k}}^2}{2\omega}-\frac{x}{1-x}\frac{({\mathbf{k}}-{\mathbf{q}}_1-{\mathbf{q}}_2)^2}{2\omega} + \frac{x{\mathbf{q}}_1^2}{2 \omega} -i\epsilon$, $\frac{x}{1-x}(\frac{({\mathbf{k}}-{\mathbf{q}}_1)^2}{2\omega} - \frac{({\mathbf{k}}-{\mathbf{q}}_1-{\mathbf{q}}_2)^2}{2\omega}) -i\epsilon$, $-i\mu_{1\perp}$ and $-i\mu_{2\perp}$, which give:
\begin{widetext}
\begin{align}~\label{m210_co}
M^c_{2, 1, 0} ={} & i J_a(p+k)e^{i(p+k)x_0} (T^{a_2} T^{c} T^{a_1})_{da} T_{a_2}T_{a_1}(1-x+x^2) (-i)\int{\frac{d^2{\mathbf{q}}_1}{(2\pi)^2}} (-i) \int{\frac{d^2{\mathbf{q}}_2}{(2\pi)^2}} v(0,{\mathbf{q}}_1) v(0, {\mathbf{q}}_2)e^{-i({\mathbf{q}}_1+ {\mathbf{q}}_2) \cdot {\mathbf{b}}_1} \nonumber \\
 \times {} &  (ig_s) \frac{{\mathbf{\boldsymbol{\epsilon}}}\cdot({\mathbf{k}}-x{\mathbf{q}}_1)}{({\mathbf{k}}-x{\mathbf{q}}_1)^2} e^{\frac{i}{2\omega}({\mathbf{k}}^2 +\frac{x}{1-x}({\mathbf{k}}- {\mathbf{q}}_1- {\mathbf{q}}_2)^2)(z_1-z_0)} \frac{\mu^2_{1\perp} + \mu_{1\perp} \mu_{2\perp} +\mu^2_{2\perp}}{\mu_{1\perp} \mu_{2\perp}}  \frac{({\mathbf{k}}-x{\mathbf{q}}_1)^2}{x(1-x)E^+(\mu_{1\perp} + \mu_{2\perp})} ,
\end{align}
\end{widetext}
where the residues at first two poles (i.e. originating from the gluon propagators) cancel each  other exactly, leading to the result Eq.~\eqref{m210_co} that is suppressed by a factor of $\mathcal{O}(\frac{({\mathbf{k}}-x{\mathbf{q}}_1)^{\scalebox{.8}{$\scriptscriptstyle  2$}}}{x(1-x)E^{\scalebox{.8}{$\scriptscriptstyle  +$}}(\mu_{1\perp} + \mu_{2\perp})})$ compared to the all previous amplitudes (note that $x$ is finite), as in the case of soft-gluon approximation~\citep{GLV,DGLVstatic}. 

The same conclusion applies to $M^c_{2, 1, 1}$ amplitude, which can be straightforwardly verified by repeating the analogous procedure as for $M^c_{2, 1, 0}$, and by the fact that these two amplitudes are symmetric (see Fig.~\ref{M2_4}) to the exchange  ($p\leftrightarrow k, x\leftrightarrow(1-x), c\leftrightarrow d$).

\section{\label{sec:E}Calculation of radiative energy loss}

In this section we provide concise outline of calculating the first order in opacity radiative energy loss. We start with the equation:
\def\mean#1{\left< #1 \right>}
\begin{align}~\label{E1}
d^3N^{(1)}_g d^3N_J ={} & \Big( \frac{1}{d_T} \Tr\mean{|M_1|^2}  + \frac{2}{d_T} Re \Tr\mean{M_2 M^*_0} \Big)  \nonumber \\  \times {} & \frac{d^3\vec{\mathbf{p}}}{(2\pi)^3 2p^0} \frac{d^3\vec{\mathbf{k}}}{(2\pi)^3 2\omega}, 
\end{align}
where $M_1$ is sum of all diagrams with one scattering center from Appendix~\ref{sec:M1}, $M_2$ is sum of all diagrams with two scattering centers in the contact limit from Appendices~\ref{sec:M220},~\ref{sec:M203},~\ref{sec:M201} and $M^*_0$ is obtained from  Appendix~\ref{sec:M0}.

The final results from Appendix~\ref{sec:M1} yield:
\begin{widetext}
\begin{align}~\label{E2}
M_1 = {} & M_{1, 1, 0} + M_{1, 0, 0} + M_{1, 0, 1}= J_a(p+k) e^{i(p+k)x_0} (1-x+x^2)T_{a_1}(-i) \int{\frac{d^2{\mathbf{q}}_1}{(2\pi)^2}} v(0, {\mathbf{q}}_1) e^{-i{\mathbf{q}}_1 \cdot {\mathbf{b}}_1} (2ig_s)  \nonumber \\
 \times {} &  \Big\{ \Big( \frac{{\mathbf{\boldsymbol{\epsilon}}} \cdot({\mathbf{k}}-{\mathbf{q}}_1)}{({\mathbf{k}}- {\mathbf{q}}_1)^2} [T^c, T^{a_1}]_{da} - \frac{{\mathbf{\boldsymbol{\epsilon}}} \cdot({\mathbf{k}}-x {\mathbf{q}}_1)}{({\mathbf{k}}-x {\mathbf{q}}_1)^2} (T^cT^{a_1})_{da} +  \frac{{\mathbf{\boldsymbol{\epsilon}}}\cdot {\mathbf{k}}}{{\mathbf{k}}^2} (T^{a_1}T^c)_{da} \Big) e^{\frac{i}{2\omega} ({\mathbf{k}}^2 + \frac{x}{1-x}({\mathbf{k}}- {\mathbf{q}}_1)^2 )(z_1-z_0)}  \nonumber \\
 - {} & \frac{{\mathbf{\boldsymbol{\epsilon}}} \cdot({\mathbf{k}}-{\mathbf{q}}_1)}{({\mathbf{k}}-{\mathbf{q}}_1)^2} [T^c, T^{a_1}]_{da} e^{\frac{i}{2\omega} ({\mathbf{k}}^2- ({\mathbf{k}}- {\mathbf{q}}_1)^2 )(z_1-z_0)} - \frac{{\mathbf{\boldsymbol{\epsilon}}} \cdot {\mathbf{k}}}{{\mathbf{k}}^2} (T^{a_1}T^c)_{da} e^{\frac{i}{2\omega} \frac{x}{1-x}(({\mathbf{k}}- {\mathbf{q}}_1)^2-{\mathbf{k}}^2 )(z_1 - z_0)} \Big\},
\end{align}
\end{widetext}
leading to:
\newpage
\begin{widetext}
\begin{align}~\label{E_M1}
\frac{1}{d_T}\Tr & \mean{|M_1|^2} = {}  \sum N |J(p+k)|^2(4g^2_s) \frac{1}{A_{\perp}} (1-x+x^2)^2 \int{\frac{d^2{\mathbf{q}}_1}{(2\pi)^2}} |v(0, {\mathbf{q}}_1)|^2 \frac{C_2(T)}{d_G}  \Big\{ (\frac{{\mathbf{\boldsymbol{\epsilon}}} \cdot ({\mathbf{k}}-x {\mathbf{q}}_1)}{({\mathbf{k}}-x {\mathbf{q}}_1)^2})^2 \Tr((T^c)^2(T^{a_1})^2)  \nonumber \\
&  + 2\alpha \Big(2\frac{{\mathbf{\boldsymbol{\epsilon}}} \cdot ({\mathbf{k}}- {\mathbf{q}}_1)}{({\mathbf{k}}- {\mathbf{q}}_1)^2} - \frac{{\mathbf{\boldsymbol{\epsilon}}} \cdot {\mathbf{k}}}{{\mathbf{k}}^2} -\frac{{\mathbf{\boldsymbol{\epsilon}}}\cdot ({\mathbf{k}}-x {\mathbf{q}}_1)}{({\mathbf{k}}-x {\mathbf{q}}_1)^2} \Big)\frac{{\mathbf{\boldsymbol{\epsilon}}} \cdot ({\mathbf{k}}- {\mathbf{q}}_1)}{({\mathbf{k}}- {\mathbf{q}}_1)^2}  - \alpha \frac{{\mathbf{\boldsymbol{\epsilon}}}\cdot {\mathbf{k}}}{{\mathbf{k}}^2} \, \frac{{\mathbf{\boldsymbol{\epsilon}}} \cdot ({\mathbf{k}}- {\mathbf{q}}_1)}{({\mathbf{k}}- {\mathbf{q}}_1)^2} 2\cos{(\frac{ {\mathbf{k}}^2- ({\mathbf{k}}- {\mathbf{q}}_1)^2}{x(1-x)E^+}(z_1-z_0))}   \nonumber \\
& +2 \Big(\frac{{\mathbf{\boldsymbol{\epsilon}}}\cdot {\mathbf{k}}}{{\mathbf{k}}^2} \Tr((T^c)^2(T^{a_1})^2) -\frac{{\mathbf{\boldsymbol{\epsilon}}} \cdot ({\mathbf{k}}-x {\mathbf{q}}_1)}{({\mathbf{k}}-x {\mathbf{q}}_1)^2} \Tr(T^c T^{a_1} T^c T^{a_1}) \Big)\frac{{\mathbf{\boldsymbol{\epsilon}}} \cdot {\mathbf{k}}}{{\mathbf{k}}^2}  \nonumber \\
& - 2\alpha \Big(\frac{{\mathbf{\boldsymbol{\epsilon}}}\cdot ({\mathbf{k}}- {\mathbf{q}}_1)}{({\mathbf{k}}-{\mathbf{q}}_1)^2} -\frac{1}{2}\frac{{\mathbf{\boldsymbol{\epsilon}}} \cdot {\mathbf{k}}}{{\mathbf{k}}^2} -\frac{1}{2}\frac{{\mathbf{\boldsymbol{\epsilon}}} \cdot ({\mathbf{k}}-x {\mathbf{q}}_1)}{({\mathbf{k}}-x {\mathbf{q}}_1)^2} \Big)\frac{{\mathbf{\boldsymbol{\epsilon}}} \cdot ({\mathbf{k}}- {\mathbf{q}}_1)}{({\mathbf{k}}- {\mathbf{q}}_1)^2}  2\cos{(\frac{({\mathbf{k}}- {\mathbf{q}}_1)^2}{x (1-x) E^+}(z_1 - z_0))}  \nonumber \\
& + \Big( \alpha \frac{{\mathbf{\boldsymbol{\epsilon}}} \cdot ({\mathbf{k}}- {\mathbf{q}}_1)}{({\mathbf{k}}-{\mathbf{q}}_1)^2} -  \frac{{\mathbf{\boldsymbol{\epsilon}}} \cdot {{\mathbf{k}}}}{{\mathbf{k}}^2}\Tr((T^c)^2(T^{a_1})^2) + \frac{{\mathbf{\boldsymbol{\epsilon}}} \cdot ({\mathbf{k}}-x {\mathbf{q}}_1)}{({\mathbf{k}}-x {\mathbf{q}}_1)^2}\Tr(T^c T^{a_1} T^c T^{a_1}) \Big)  \frac{{\mathbf{\boldsymbol{\epsilon}}} \cdot {\mathbf{k}}}{{\mathbf{k}}^2} 2 \cos{(\frac{{\mathbf{k}}^2}{x(1-x)E^+} (z_1-z_0))}
\Big\},
\end{align}
\end{widetext}
where the number of scattering centers $N$ comes from summation over scattering centers Eqs.~\eqref{app7} and~\eqref{app8}, then $\alpha \equiv \Tr((T^c)^2(T^{a_1})^2 -T^cT^{a_1}T^cT^{a_1})$, and we also used the definition of commutator, the fact that trace is invariant under the cyclic permutations, Eq.~\eqref{tr2} (with $i=j$  and $d_i=d_T$) and the relation $E^+ \approx 2E$. We verified that this result is also symmetric  under the substitutions: ($p\leftrightarrow k, x\leftrightarrow(1-x), c\leftrightarrow d$) when written in terms of structure constants.

Next, we summarize contact limits of all diagrams that contain two scattering centers from Appendices~\ref{sec:M220}-\ref{sec:M201} and then take their ensemble average according to Eqs.~\eqref{app7} to~\eqref{app9} in order to obtain $M_2$:
\begin{widetext}
\begin{align}~\label{E3}
M_2 = {} & M^c_{2, 2, 0} + M^c_{2, 0, 3} + M^c_{2, 0, 0}+ \frac{1}{2} (M^c_{2, 0, 1}+ M^c_{2, 0, 2})=  \frac{1}{2} N J_a(p+k) e^{i(p+k)x_0} (-2ig_s) \frac{1}{A_{\perp}}  (1-x+x^2) T_{a_2} T_{a_1}  \nonumber \\
& \times\int{\frac{d^2{\mathbf{q}}_1}{(2\pi)^2}} |v(0, {\mathbf{q}}_1)|^2  \Big\{\frac{{\mathbf{\boldsymbol{\epsilon}}} \cdot {\mathbf{k}}}{{\mathbf{k}}^2} \Big(e^{\frac{i}{2\omega} \frac{{\mathbf{k}}^{\scalebox{.8}{$\scriptscriptstyle  2$}}}{1-x} (z_1-z_0)} \big( [[T^c,T^{a_2}],T^{a_1}]_{da} +  [T^{a_2}T^{a_1}, T^c]_{da} \big) -  [[T^c,T^{a_2}],T^{a_1}]_{da}  \nonumber \\
& -(T^{a_2}T^{a_1}T^c)_{da} \Big) +  \frac{{\mathbf{\boldsymbol{\epsilon}}} \cdot ({\mathbf{k}}- {\mathbf{q}}_1)}{({\mathbf{k}}- {\mathbf{q}}_1)^2} \Big(e^{\frac{i}{2\omega}\frac{{\mathbf{k}}^{\scalebox{.8}{$\scriptscriptstyle  2$}}}{1-x}(z_1-z_0)} - e^{\frac{i}{2\omega} \frac{{\mathbf{k}}^{\scalebox{.8}{$\scriptscriptstyle  2$}} - ({\mathbf{k}}- {\mathbf{q}}_{\scalebox{.8}{$\scriptscriptstyle  1$}})^{\scalebox{.8}{$\scriptscriptstyle  2$}}}{1-x}(z_1-z_0)}\Big)  \big( (T^{a_2}[T^c,T^{a_1}])_{da} + (T^{a_1} [T^c,T^{a_2}])_{da} \big) \Big\}.
\end{align}
\end{widetext}
Then, by multiplying the previous expression by $M^*_0$, we obtain:
\begin{widetext}
\begin{align}~\label{E_M2}
\frac{2}{d_T} Re \Tr  \mean{M_2 M^*_0} = {} & \sum N |J(p+k)|^2 (4g^2_s) \frac{1}{A_{\perp}} (1-x+x^2)^2  \int{\frac{d^2{\mathbf{q}}_1}{(2\pi)^2}} |v(0, {\mathbf{q}}_1)|^2 \frac{C_2(T)}{d_G}   \nonumber \\
 \times {} & \Big\{(\frac{{\mathbf{\boldsymbol{\epsilon}}} \cdot {\mathbf{k}}}{{\mathbf{k}}^2})^2 \Big(2 \alpha \cos{(\frac{{\mathbf{k}}^2}{x(1-x)E^+}(z_1-z_0))} - 2\alpha -\Tr((T^c)^2(T^{a_1})^2) \Big)  \nonumber \\
  - {} & 2\alpha \frac{{\mathbf{\boldsymbol{\epsilon}}}\cdot {\mathbf{k}}}{{\mathbf{k}}^2} \, \frac{{\mathbf{\boldsymbol{\epsilon}}} \cdot ({\mathbf{k}}- {\mathbf{q}}_1)}{({\mathbf{k}}- {\mathbf{q}}_1)^2}\Big(\cos{(\frac{{\mathbf{k}}^2}{x(1-x)E^+}(z_1 -z_0))} - \cos{(\frac{{\mathbf{k}}^2-({\mathbf{k}}- {\mathbf{q}}_1)^2}{x(1-x)E^+}(z_1-z_0))} \Big) \Big\},
\end{align}
\end{widetext}
which can easily be verified to be symmetric to the exchange ($p\leftrightarrow k, x\leftrightarrow(1-x), c\leftrightarrow d$), when written in terms of structure constants.
By summing the expressions Eqs.~\eqref{E_M1} and~\eqref{E_M2} we obtain:
\begin{widetext}
\begin{align}~\label{E_M1_2}
\frac{1}{d_T}\Tr & \mean{|M_1|^2} + \frac{2}{d_T} Re \Tr\mean{M_2 M^*_0} =   N d_G |J(p+k)|^2 (4g^2_s)\frac{C_2(T)}{d_G} C^2_2(G) \frac{1}{A_{\perp}} (1-x+x^2)^2 \sum \int{\frac{d^2{\mathbf{q}}_1}{(2\pi)^2}} |v(0, {\mathbf{q}}_1)|^2  \nonumber \\
& \times   \Big\{\Big( 1-\cos{(\frac{{\mathbf{k}}^2}{x(1-x)E^+}(z_1-z_0))} \Big)
 \Big(\frac{{\mathbf{\boldsymbol{\epsilon}}}\cdot {\mathbf{k}}}{{\mathbf{k}}^2} -\frac{{\mathbf{\boldsymbol{\epsilon}}} \cdot ({\mathbf{k}}-x{\mathbf{q}}_1)}{({\mathbf{k}}-x{\mathbf{q}}_1)^2} \Big)\frac{{\mathbf{\boldsymbol{\epsilon}}} \cdot {\mathbf{k}}}{{\mathbf{k}}^2}   + \Big( (\frac{{\mathbf{\boldsymbol{\epsilon}}} \cdot ({\mathbf{k}} -x{\mathbf{q}}_1)}{({\mathbf{k}} -x{\mathbf{q}}_1)^2})^2 -(\frac{{\mathbf{\boldsymbol{\epsilon}}} \cdot {\mathbf{k}}}{{\mathbf{k}}^2})^2 \Big)  \nonumber \\
&  + \Big( 1-\cos{(\frac{({\mathbf{k}}-{\mathbf{q}}_1)^2}{x(1-x)E^+}(z_1-z_0))} \Big) \Big(2\frac{{\mathbf{\boldsymbol{\epsilon}}}\cdot ({\mathbf{k}}- {\mathbf{q}}_1)}{({\mathbf{k}}- {\mathbf{q}}_1)^2} - \frac{{\mathbf{\boldsymbol{\epsilon}}} \cdot {\mathbf{k}}}{{\mathbf{k}}^2} - \frac{{\mathbf{\boldsymbol{\epsilon}}}\cdot ({\mathbf{k}} -x{\mathbf{q}}_1)}{({\mathbf{k}} -x{\mathbf{q}}_1)^2} \Big)\frac{{\mathbf{\boldsymbol{\epsilon}}}\cdot ({\mathbf{k}}-{\mathbf{q}}_1)}{({\mathbf{k}}- {\mathbf{q}}_1)^2} \Big\},
\end{align}
\end{widetext}
which in the soft-gluon approximation coincides with massless limit of Eq.~(82) from~\citep{DGLVstatic} and
where we used the following equalities that are valid in adjoint representation: $\Tr(T^cT^{a_1}T^cT^{a_1})=\frac{1}{2}C^2_2(G)d_G=\alpha= \frac{1}{2}\Tr((T^c)^2(T^{a_1})^2)$, which follow from Eqs.~\eqref{tr2} to \eqref{gen5} and the commutator definition.

Since we are considering optically "thin" QCD plasma, it would be convenient to expand energy loss in powers of opacity, which is defined by the mean number of collisions in QCD medium~\citep{GLV}:
\begin{eqnarray}
\bar{n}=\frac{L}{\lambda} = \frac{N \sigma_{el}}{A_{\perp}},
\label{opacity}
\end{eqnarray}
where the small transverse momentum transfer elastic cross section between the jet and the target partons is taken from GW model (Eq.~(6) from~\citep{GLV}), which in our case reads:
\begin{eqnarray}
\frac{d\sigma_{el}}{d^2{\mathbf{q}}_1} = \frac{C_2(G)C_2(T)}{d_G} \frac{| v(0,{\mathbf{q}}_1)|^2}{(2\pi)^2}.
\label{elcs}
\end{eqnarray}
Combining Eqs.~\eqref{opacity} and~\eqref{elcs} we obtain:
\begin{eqnarray}
\frac{L}{\lambda} = \frac{N}{A_{\perp}} \frac{C_2(G)C_2(T)}{4 \pi d_G} \frac{(4\pi \alpha_s)^2}{\mu^2}.
\label{opacity1}
\end{eqnarray}
Next we incorporate Eq.~\eqref{opacity1} in  Eq.~\eqref{E_M1_2}, substitute  obtained expression in Eq.~\eqref{E1}, keeping in mind that $\vec{\mathbf{p}}$ is 3D momentum of a final jet, and that we need to apply Eqs.~\eqref{dnj2} and~\eqref{cv}. 

Thus, in the case of simple exponential distribution $\frac{2}{L} e^{-2\frac{z_1-z_0}{L}}$ of the scattering centers (as in~\citep{DGLVstatic}) the single gluon radiation spectrum in the first order in opacity  becomes:
\begin{widetext}
\begin{align}~\label{dN_dx1}
\frac{dN^{(1)}_g}{dx}={} & \frac{C_2(G) \alpha_s}{\pi} \frac{L}{\lambda} \frac{(1-x+x^2)^2}{x(1-x)} \sum \int{\frac{d^2{\mathbf{q}}_1}{\pi} \frac{\mu^2}{({\mathbf{q}}_1^2 + \mu^2)^2}}
\int{\frac{ d^2{\mathbf{k}}}{\pi} }   \Big\{ - \frac{{\mathbf{\boldsymbol{\epsilon}}}\cdot ({\mathbf{k}}- {\mathbf{q}}_1)}{({\mathbf{k}}-{\mathbf{q}}_1)^2} \Big(\frac{{\mathbf{\boldsymbol{\epsilon}}} \cdot {\mathbf{k}}}{{\mathbf{k}}^2} +\frac{{\mathbf{\boldsymbol{\epsilon}}} \cdot ({\mathbf{k}}-x{\mathbf{q}}_1)}{({\mathbf{k}}-x{\mathbf{q}}_1)^2} -2\frac{{\mathbf{\boldsymbol{\epsilon}}} \cdot ({\mathbf{k}}- {\mathbf{q}}_1)}{({\mathbf{k}}- {\mathbf{q}}_1)^2} \Big)  \nonumber \\
 \times {} & \int{dz_1}(1-\cos({\frac{({\mathbf{k}}- {\mathbf{q}}_1)^2}{x(1-x)E^+}(z_1-z_0)})) \frac{2}{L}e^{-\frac{2(z_1-z_0)}{L}} + \frac{{\mathbf{\boldsymbol{\epsilon}}} \cdot {\mathbf{k}}}{{\mathbf{k}}^2} \Big(\frac{{\mathbf{\boldsymbol{\epsilon}}} \cdot {\mathbf{k}}}{{\mathbf{k}}^2} - \frac{{\mathbf{\boldsymbol{\epsilon}}} \cdot ({\mathbf{k}}-x{\mathbf{q}}_1)}{({\mathbf{k}}-x{\mathbf{q}}_1)^2}\Big)  \nonumber \\
 \times {} & \int{dz_1(1-\cos(\frac{{\mathbf{k}}^2}{x(1-x)E^+}(z_1-z_0)))\frac{2}{L}e^{-\frac{2(z_1-z_0)}{L}}} +  \Big((\frac{{\mathbf{\boldsymbol{\epsilon}}}\cdot ({\mathbf{k}}-x{\mathbf{q}}_1)}{({\mathbf{k}}-x{\mathbf{q}}_1)^2})^2 -(\frac{{\mathbf{\boldsymbol{\epsilon}}} \cdot {\mathbf{k}}}{{\mathbf{k}}^2})^2 \Big)\int{dz_1 \frac{2}{L}e^{-\frac{2(z_1-z_0)}{L}}}  \Big\},
\end{align}
\end{widetext}
and the differential radiative energy loss $ \frac{dE^{(1)}}{dx} \equiv \omega \frac{d^3N^{(1)}_g}{dx} \approx xE \frac{d^3N^{(1)}_g}{dx}$ acquires the form:
\begin{widetext}
\begin{align}~\label{dE_dx1}
\frac{dE^{(1)}}{dx}={} & \frac{C_2(G) \alpha_s}{\pi} \frac{L}{\lambda}E \frac{(1-x+x^2)^2}{1-x} \sum \int{\frac{d^2{\mathbf{q}}_1}{\pi} \frac{\mu^2}{({\mathbf{q}}_1^2 + \mu^2)^2}}
\int{\frac{ d^2{\mathbf{k}}}{\pi} }   \Big\{ - \frac{{\mathbf{\boldsymbol{\epsilon}}}\cdot ({\mathbf{k}}- {\mathbf{q}}_1)}{({\mathbf{k}}-{\mathbf{q}}_1)^2} \Big(\frac{{\mathbf{\boldsymbol{\epsilon}}} \cdot {\mathbf{k}}}{{\mathbf{k}}^2} +\frac{{\mathbf{\boldsymbol{\epsilon}}} \cdot ({\mathbf{k}}-x{\mathbf{q}}_1)}{({\mathbf{k}}-x{\mathbf{q}}_1)^2} -2\frac{{\mathbf{\boldsymbol{\epsilon}}} \cdot ({\mathbf{k}}- {\mathbf{q}}_1)}{({\mathbf{k}}- {\mathbf{q}}_1)^2} \Big) \nonumber \\
 \times {} & \int{dz_1}(1-\cos({\frac{({\mathbf{k}}- {\mathbf{q}}_1)^2}{x(1-x)E^+}(z_1-z_0)})) \frac{2}{L}e^{-\frac{2(z_1-z_0)}{L}} + \frac{{\mathbf{\boldsymbol{\epsilon}}} \cdot {\mathbf{k}}}{{\mathbf{k}}^2} \Big(\frac{{\mathbf{\boldsymbol{\epsilon}}} \cdot {\mathbf{k}}}{{\mathbf{k}}^2} - \frac{{\mathbf{\boldsymbol{\epsilon}}} \cdot ({\mathbf{k}}-x{\mathbf{q}}_1)}{({\mathbf{k}}-x{\mathbf{q}}_1)^2}\Big)  \nonumber \\
 \times {} & \int{dz_1(1-\cos(\frac{{\mathbf{k}}^2}{x(1-x)E^+}(z_1-z_0)))\frac{2}{L}e^{-\frac{2(z_1-z_0)}{L}}} +  \Big((\frac{{\mathbf{\boldsymbol{\epsilon}}}\cdot ({\mathbf{k}}-x{\mathbf{q}}_1)}{({\mathbf{k}}-x{\mathbf{q}}_1)^2})^2 -(\frac{{\mathbf{\boldsymbol{\epsilon}}} \cdot {\mathbf{k}}}{{\mathbf{k}}^2})^2 \Big)\int{dz_1 \frac{2}{L}e^{-\frac{2(z_1-z_0)}{L}}}  \Big\}.
\end{align}
\end{widetext}
So we finally obtain:
\begin{widetext}
\begin{align}~\label{dN_dx_massless}
\frac{dN^{(1)}_g}{dx^{}}={} & \frac{C_2(G) \alpha_s}{\pi} \frac{L}{\lambda} \frac{(1-x+x^2)^2}{x(1-x)}\int{\frac{d^2{\mathbf{q}}_1}{\pi} \frac{\mu^2}{({\mathbf{q}}_1^2 + \mu^2)^2}}
\int{d{\mathbf{k}}^2}  \nonumber \\
 \times {} & \Big\{  \frac{({\mathbf{k}}- {\mathbf{q}}_1)^2}{(\frac{4x(1-x)E}{L})^2 +({\mathbf{k}}- {\mathbf{q}}_1)^4} \Big( 2 -\frac{{\mathbf{k}} \cdot ({\mathbf{k}}- {\mathbf{q}}_1)}{{\mathbf{k}}^2} - \frac{({\mathbf{k}}- {\mathbf{q}}_1) \cdot ({\mathbf{k}}- x{\mathbf{q}}_1)}{({\mathbf{k}}- x{\mathbf{q}}_1)^2} \Big)  \nonumber \\
 + {} & \frac{{\mathbf{k}}^2}{(\frac{4x(1-x)E}{L})^2 + {\mathbf{k}}^4} \Big( 1 - \frac{{\mathbf{k}} \cdot ({\mathbf{k}}- x{\mathbf{q}}_1)}{({\mathbf{k}}- x{\mathbf{q}}_1)^2} \Big) + \Big( \frac{1}{({\mathbf{k}}- x{\mathbf{q}}_1)^2} - \frac{1}{{\mathbf{k}}^2} \Big) \Big\},
\end{align}
\end{widetext}
which is symmetric to the exchange of $p$ and $k$ gluons, and:
\begin{widetext}
\begin{align}~\label{dE_dx2}
\frac{dE^{(1)}}{dx^{}}={} & \frac{C_2(G) \alpha_s}{\pi} \frac{L}{\lambda} E \frac{(1-x+x^2)^2}{1-x}\int{\frac{d^2{\mathbf{q}}_1}{\pi} \frac{\mu^2}{({\mathbf{q}}_1^2 + \mu^2)^2}}
\int{d{\mathbf{k}}^2}  \nonumber \\
 \times {} & \Big\{  \frac{({\mathbf{k}}- {\mathbf{q}}_1)^2}{(\frac{4x(1-x)E}{L})^2 +({\mathbf{k}}- {\mathbf{q}}_1)^4} \Big( 2 -\frac{{\mathbf{k}} \cdot ({\mathbf{k}}- {\mathbf{q}}_1)}{{\mathbf{k}}^2} - \frac{({\mathbf{k}}- {\mathbf{q}}_1) \cdot ({\mathbf{k}}- x{\mathbf{q}}_1)}{({\mathbf{k}}- x{\mathbf{q}}_1)^2} \Big)  \nonumber \\
 + {} & \frac{{\mathbf{k}}^2}{(\frac{4x(1-x)E}{L})^2 + {\mathbf{k}}^4} \Big( 1 - \frac{{\mathbf{k}} \cdot ({\mathbf{k}}- x{\mathbf{q}}_1)}{({\mathbf{k}}- x{\mathbf{q}}_1)^2} \Big) + \Big( \frac{1}{({\mathbf{k}}- x{\mathbf{q}}_1)^2} - \frac{1}{{\mathbf{k}}^2} \Big) \Big\},
\end{align}
\end{widetext}
which in soft-gluon approximation reduces to massless limit of Eq.~(84) from~\citep{DGLVstatic}.

\section{\label{sec:Emg}Diagrams and radiative energy loss in finite T QCD medium}

Next we recalculate the results from Appendices~\ref{sec:M0}-\ref{sec:M210} when the gluon mass $m_g = \frac{\mu}{\sqrt{2}}$ is included, i.e. gluon propagator has the following form~\citep{mg}:
\begin{itemize}
\item gluon propagator with mass $m_g$ in Feynman gauge:
\begin{equation}
\begin{gathered}
\includegraphics[width=0.4\linewidth]{gluon_propagator.pdf}
\end{gathered} 
= \frac{i\delta_{ab} P_{\mu\nu}}{p^2 -m^2_g +i\epsilon},
\end{equation}
\end{itemize}
where $P_{\mu \nu}$, given by Eq.~(12) from~\citep{mg} (specifically $P_{\mu \nu}= - \Big( g_{\mu \nu} - \frac{ p_{\mu} p_{\nu} n^2 +n_{\mu} n_{\nu} p^2 -n_{\mu}p_{\nu} (np) -n_{\nu} p_{\mu}(np)}{n^2 p^2-(np)^2} \Big)$, which reduces to Eq.~\eqref{polariz}), represents the transverse projector. Note that, since the transverse projectors act directly or indirectly on transverse polarization vectors one may immediately replace $P_{\mu \nu}$ with $-g_{\mu \nu} $  in gluon propagators, in order to facilitate the calculations. This observation is obvious for off-shell gluon propagator, whereas the derivation for the remaining internal gluon lines is  straightforward. 

Consistently throughout this section, initial jet propagates long z-axis, 4-momentum is conserved and minus Light cone coordinate of $p$ and $k$ momenta acquire an additional term $+m^2_g$ in the numerator compared to massless case (Appendices~\ref{sec:M0}-\ref{sec:M210}), due to relations $k^2=p^2=m^2_g$, while the polarizations remain the same.

We provide only the final expressions for all 11 Feynman diagrams beyond soft-gluon approximation, when the gluon mass is included, since its derivation  is similar to the case of massless gluons and in order to avoid unnecessary repetition (Appendices~\ref{sec:M0}-\ref{sec:M210}).

Thus, for $M_0$ we obtain: 
\begin{align}~\label{M0_gnbsg_m}
M_{0}= {} & J_a(p+k)e^{i(p+k)x_0} (-2ig_s)(1-x+x^2)  \nonumber \\
 \times {} & \frac{{\mathbf{\boldsymbol{\epsilon}}}\cdot{\mathbf{k}}}{{\mathbf{k}}^2 + m^2_g (1-x+x^2)} (T^c)_{da}.
\end{align}

The expression for  $M_{1, 1, 0}$ now reads:
\begin{widetext}
\begin{align}~\label{m110_m}
M_{1, 1, 0}={}& J_a(p+k)e^{i(p+k)x_0}(-i)(1-x+x^2)(T^c T^{a_1})_{da} T_{a_1} \int{\frac{d^2{\mathbf{q}}_1}{(2 \pi)^2} v(0, {\mathbf{q}}_1) e^{-i{\mathbf{q}}_1 \cdot {\mathbf{b}}_1}} \nonumber \\
 \times {} & (-2ig_s) \frac{{\mathbf{\boldsymbol{\epsilon}}}\cdot ({\mathbf{k}}-x{\mathbf{q}}_1)}{({\mathbf{k}}-x{\mathbf{q}}_1)^2 + m^2_g (1-x+x^2)} e^{\frac{i}{2\omega}({\mathbf{k}}^2 +\frac{x}{1-x}({\mathbf{k}}-{\mathbf{q}}_1)^2 + \frac{m^{\scalebox{.8}{$\scriptscriptstyle  2$}}_{\scalebox{.8}{$\scriptscriptstyle  g$}} (1-x+x^{\scalebox{.8}{$\scriptscriptstyle  2$}})}{1-x})(z_1-z_0)},
\end{align}
\end{widetext}
 which differs from Eq.~\eqref{m110_o} in the term $\chi \equiv m^2_g (1-x+x^2)$, which now appears in the denominator and in exponent, accompanying the squared transverse momentum. Further on, we will use the shorthand notation $\chi$.

Similarly, for $M_{1, 0, 0}$ and $M_{1, 0, 1}$ we obtain, respectively:
\begin{widetext}
\begin{align}~\label{m100_m}
M_{1,0,0}={}& J_a(p+k)e^{i(p+k)x_0}(-i)(1-x+x^2)(T^{a_1}T^c)_{da}T_{a_1} \int{\frac{d^2{\mathbf{q}}_1}{(2 \pi)^2} v(0, {\mathbf{q}}_1) e^{-i{\mathbf{q}}_1 \cdot {\mathbf{b}}_1}} \nonumber \\
 \times {} & (2ig_s) \frac{{\mathbf{\boldsymbol{\epsilon}\cdot k}}}{{\mathbf{k}}^2 + \chi} \Big( e^{\frac{i}{2\omega} ({\mathbf{k}}^2 +\frac{x}{1-x}({\mathbf{k}}-{\mathbf{q}}_1)^2 + \frac{\chi}{1-x})(z_1-z_0)} - e^{-\frac{i}{2\omega} \frac{x}{1-x} ({\mathbf{k}}^2 -({\mathbf{k}}-{\mathbf{q}}_1)^2)(z_1-z_0)} \Big),
\end{align}
\end{widetext}
\begin{widetext}
\begin{align}~\label{m101_m}
M_{1,0,1}={}& J_a(p+k)e^{i(p+k)x_0}(-i)(1-x+x^2)[T^c,T^{a_1}]_{da}T_{a_1} \int{\frac{d^2{\mathbf{q}}_1}{(2 \pi)^2} v(0, {\mathbf{q}}_1) e^{-i{\mathbf{q}}_1 \cdot {\mathbf{b}}_1}} \nonumber \\
 \times {} & (2ig_s) \frac{{\mathbf{\boldsymbol{\epsilon}}} \cdot({\mathbf{k}}- {\mathbf{q}}_1)}{({\mathbf{k}}- {\mathbf{q}}_1)^2 + \chi} \Big( e^{\frac{i}{2\omega} ({\mathbf{k}}^2 +\frac{x}{1-x}({\mathbf{k}}-{\mathbf{q}}_1)^2 + \frac{\chi}{1-x})(z_1-z_0)} - e^{\frac{i}{2\omega} ({\mathbf{k}}^2 -({\mathbf{k}}-{\mathbf{q}}_1)^2)(z_1-z_0)} \Big).
\end{align}
\end{widetext}

Proceeding in the similar manner, we obtain the following expressions for contact-limit diagrams which include interactions with two scattering centers:
\begin{widetext}
\begin{align}~\label{m220_m}
M^c_{2, 2, 0}={} & -J_a(p+k)e^{i(p+k)x_0} (T^cT^{a_2}T^{a_1})_{da} T_{a_2}T_{a_1}(1-x+x^2)  (-i)\int{\frac{d^2{\mathbf{q}}_1}{(2\pi)^2}} (-i) \int{\frac{d^2{\mathbf{q}}_2}{(2\pi)^2}} v(0,{\mathbf{q}}_1) v(0, {\mathbf{q}}_2)e^{-i({\mathbf{q}}_1+ {\mathbf{q}}_2) \cdot {\mathbf{b}}_1}  \nonumber \\
 \times {} & \frac{1}{2} (2 ig_s) \frac{{\mathbf{\boldsymbol{\epsilon}}}\cdot({\mathbf{k}}-x({\mathbf{q}}_1 + {\mathbf{q}}_2))}{({\mathbf{k}}-x({\mathbf{q}}_1+ {\mathbf{q}}_2))^2 +\chi} e^{\frac{i}{2\omega}({\mathbf{k}}^2 + \frac{x}{1-x}({\mathbf{k}}- {\mathbf{q}}_1- {\mathbf{q}}_2)^2 + \frac{\chi}{1-x})(z_1-z_0)},
\end{align}
\end{widetext}
\begin{widetext}
\begin{align}~\label{m203_clm}
M^c_{2, 0, 3}={} & J_a(p+k)e^{i(p+k)x_0} [[T^c, T^{a_2}],T^{a_1}]_{da} T_{a_2}T_{a_1}(1-x+x^2)(-i)\int{\frac{d^2{\mathbf{q}}_1}{(2\pi)^2}}  (-i) \int{\frac{d^2{\mathbf{q}}_2}{(2\pi)^2}} v(0,{\mathbf{q}}_1) v(0, {\mathbf{q}}_2)e^{-i({\mathbf{q}}_1+ {\mathbf{q}}_2) \cdot {\mathbf{b}}_1}  \nonumber \\
 \times {} & \frac{1}{2} (2 ig_s) \frac{{\mathbf{\boldsymbol{\epsilon}}}\cdot({\mathbf{k}}-{\mathbf{q}}_1 - {\mathbf{q}}_2)}{({\mathbf{k}}-{\mathbf{q}}_1- {\mathbf{q}}_2)^2 +\chi} \Big( e^{\frac{i}{2\omega}({\mathbf{k}}^2 + \frac{x}{1-x}({\mathbf{k}}- {\mathbf{q}}_1- {\mathbf{q}}_2)^2 + \frac{\chi}{1-x})(z_1-z_0)} - e^{\frac{i}{2\omega}({\mathbf{k}}^2 - ({\mathbf{k}}- {\mathbf{q}}_1- {\mathbf{q}}_2)^2)(z_1-z_0)} \Big),
\end{align}
\end{widetext}
\begin{widetext}
\begin{align}~\label{m200_clm}
M^c_{2, 0, 0}={} & J_a(p+k)e^{i(p+k)x_0} (T^{a_2} T^{a_1}T^{c})_{da} T_{a_2}T_{a_1}(1-x+x^2)(-i)\int{\frac{d^2{\mathbf{q}}_1}{(2\pi)^2}}  (-i) \int{\frac{d^2{\mathbf{q}}_2}{(2\pi)^2}} v(0,{\mathbf{q}}_1) v(0, {\mathbf{q}}_2)e^{-i({\mathbf{q}}_1+ {\mathbf{q}}_2) \cdot {\mathbf{b}}_1}  \nonumber \\
 \times {} & \frac{1}{2} (2 ig_s) \frac{{\mathbf{\boldsymbol{\epsilon}}}\cdot{\mathbf{k}}}{{\mathbf{k}}^2 + \chi} \Big( e^{\frac{i}{2\omega}({\mathbf{k}}^2 + \frac{x}{1-x}({\mathbf{k}}- {\mathbf{q}}_1- {\mathbf{q}}_2)^2 + \frac{\chi}{1-x})(z_1-z_0)} - e^{\frac{i}{2\omega} \frac{x}{1-x}(({\mathbf{k}}- {\mathbf{q}}_1- {\mathbf{q}}_2)^2-{\mathbf{k}}^2)(z_1-z_0)} \Big),
\end{align}
\end{widetext}
\begin{widetext}
\begin{align}~\label{m201_clm}
M^c_{2, 0, 1} = {} & J_a(p+k)e^{i(p+k)x_0} (T^{a_2} [T^{c},T^{a_1}])_{da} T_{a_2}T_{a_1}(1-x+x^2)(-i)\int{\frac{d^2{\mathbf{q}}_1}{(2\pi)^2}}  (-i) \int{\frac{d^2{\mathbf{q}}_2}{(2\pi)^2}} v(0,{\mathbf{q}}_1) v(0, {\mathbf{q}}_2)e^{-i({\mathbf{q}}_1+ {\mathbf{q}}_2) \cdot {\mathbf{b}}_1} \nonumber \\
\times & (2 ig_s)\frac{{\mathbf{\boldsymbol{\epsilon}}}\cdot({\mathbf{k}}-{\mathbf{q}}_1)}{({\mathbf{k}}- {\mathbf{q}}_1)^2 + \chi} \Big( e^{\frac{i}{2\omega}({\mathbf{k}}^2 +\frac{x}{1-x}({\mathbf{k}}- {\mathbf{q}}_1- {\mathbf{q}}_2)^2 + \frac{\chi}{1-x})(z_1-z_0)} -e^{\frac{i}{2\omega}({\mathbf{k}}^2-\frac{({\mathbf{k}}- {\mathbf{q}}_{\scalebox{.8}{$\scriptscriptstyle  1$}})^{\scalebox{.8}{$\scriptscriptstyle  2$}}}{1-x} +\frac{x}{1-x}({\mathbf{k}}- {\mathbf{q}}_1-{\mathbf{q}}_2)^2)(z_1-z_0) } \Big),
\end{align}
\end{widetext}
\begin{widetext}
\begin{align}~\label{m202_clm}
M^c_{2, 0, 2}= {} & J_a(p+k)e^{i(p+k)x_0} (T^{a_1} [T^{c},T^{a_2}])_{da} T_{a_2}T_{a_1}(1-x+x^2)(-i)\int{\frac{d^2{\mathbf{q}}_1}{(2\pi)^2}} (-i) \int{\frac{d^2{\mathbf{q}}_2}{(2\pi)^2}} v(0,{\mathbf{q}}_1) v(0, {\mathbf{q}}_2)e^{-i({\mathbf{q}}_1+ {\mathbf{q}}_2) \cdot {\mathbf{b}}_1}  \nonumber \\
\times  & (2 ig_s)\frac{{\mathbf{\boldsymbol{\epsilon}}}\cdot({\mathbf{k}}-{\mathbf{q}}_2)}{({\mathbf{k}}- {\mathbf{q}}_2)^2 + \chi}\Big( e^{\frac{i}{2\omega}({\mathbf{k}}^2 +\frac{x}{1-x}({\mathbf{k}}- {\mathbf{q}}_1- {\mathbf{q}}_2)^2 + \frac{\chi}{1-x})(z_1-z_0)} -e^{\frac{i}{2\omega}({\mathbf{k}}^2-\frac{({\mathbf{k}}- {\mathbf{q}}_{\scalebox{.8}{$\scriptscriptstyle  2$}})^{\scalebox{.8}{$\scriptscriptstyle  2$}}}{1-x} +\frac{x}{1-x}({\mathbf{k}}- {\mathbf{q}}_1- {\mathbf{q}}_2)^2)(z_1-z_0) } \Big).
\end{align}
\end{widetext}
The amplitudes $M^c_{2, 1, 0}$ and $M^c_{2, 1, 1}$ are omitted as they are suppressed compared to the remaining amplitudes.

After adding Eqs.~\eqref{m110_m} to~\eqref{m101_m}, we obtain:
\begin{widetext}
\begin{align}~\label{E_M1m}
\frac{1}{d_T} & {} \Tr  \mean{|M_1|^2} = {} \sum N |J(p+k)|^2(4g^2_s) \frac{1}{A_{\perp}} (1-x+x^2)^2 \int{\frac{d^2{\mathbf{q}}_1}{(2\pi)^2}} |v(0, {\mathbf{q}}_1)|^2 \frac{C_2(T)}{d_G}  \Big\{  (\frac{{\mathbf{\boldsymbol{\epsilon}}} \cdot ({\mathbf{k}}-x {\mathbf{q}}_1)}{({\mathbf{k}}-x {\mathbf{q}}_1)^2 +\chi})^2 \Tr((T^c)^2(T^{a_1})^2) \nonumber \\
 + & 2\alpha \Big(2\frac{{\mathbf{\boldsymbol{\epsilon}}} \cdot ({\mathbf{k}}- {\mathbf{q}}_1)}{({\mathbf{k}}- {\mathbf{q}}_1)^2 + \chi} - \frac{{\mathbf{\boldsymbol{\epsilon}}} \cdot {\mathbf{k}}}{{\mathbf{k}}^2 + \chi} - \frac{{\mathbf{\boldsymbol{\epsilon}}}\cdot ({\mathbf{k}}-x {\mathbf{q}}_1)}{({\mathbf{k}}-x {\mathbf{q}}_1)^2 + \chi} \Big)\frac{{\mathbf{\boldsymbol{\epsilon}}} \cdot ({\mathbf{k}}-{\mathbf{q}}_1)}{({\mathbf{k}}- {\mathbf{q}}_1)^2 +\chi}  - \alpha \frac{{\mathbf{\boldsymbol{\epsilon}}}\cdot {\mathbf{k}}}{{\mathbf{k}}^2 + \chi} \frac{{\mathbf{\boldsymbol{\epsilon}}} \cdot ({\mathbf{k}}-{\mathbf{q}}_1)}{({\mathbf{k}}- {\mathbf{q}}_1)^2 + \chi} 2\cos{(\frac{{\mathbf{k}}^2 -({\mathbf{k}}- {\mathbf{q}}_1)^2 }{x(1-x)E^+}(z_1-z_0))}   \nonumber \\
 + & 2 \Big(\frac{{\mathbf{\boldsymbol{\epsilon}}}\cdot {\mathbf{k}}}{{\mathbf{k}}^2 + \chi} \Tr((T^c)^2(T^{a_1})^2) -\frac{{\mathbf{\boldsymbol{\epsilon}}} \cdot ({\mathbf{k}}-x {\mathbf{q}}_1)}{({\mathbf{k}}-x {\mathbf{q}}_1)^2 + \chi} \Tr(T^c T^{a_1} T^c T^{a_1}) \Big) \frac{{\mathbf{\boldsymbol{\epsilon}}} \cdot {\mathbf{k}}}{{\mathbf{k}}^2 + \chi}  \nonumber \\
 - {} & 2\alpha \Big(\frac{{\mathbf{\boldsymbol{\epsilon}}}\cdot ({\mathbf{k}}- {\mathbf{q}}_1)}{({\mathbf{k}}- {\mathbf{q}}_1)^2 + \chi} -
 \frac{1}{2}\frac{{\mathbf{\boldsymbol{\epsilon}}} \cdot {\mathbf{k}}}{{\mathbf{k}}^2 + \chi} -\frac{1}{2}\frac{{\mathbf{\boldsymbol{\epsilon}}} \cdot ({\mathbf{k}}-x {\mathbf{q}}_1)}{({\mathbf{k}}-x {\mathbf{q}}_1)^2 + \chi} \Big) \frac{{\mathbf{\boldsymbol{\epsilon}}} \cdot ({\mathbf{k}}-{\mathbf{q}}_1)}{({\mathbf{k}}- {\mathbf{q}}_1)^2 + \chi}
 2\cos{(\frac{({\mathbf{k}} -{\mathbf{q}}_1)^2 +\chi}{x(1-x)E^+}(z_1 - z_0))}  \nonumber \\
+ &  \Big( \alpha \frac{{\mathbf{\boldsymbol{\epsilon}}} \cdot ({\mathbf{k}}-  {\mathbf{q}}_1)}{({\mathbf{k}}-{\mathbf{q}}_1)^2 + \chi} -  \frac{{\mathbf{\boldsymbol{\epsilon}}} \cdot {{\mathbf{k}}}}{{\mathbf{k}}^2 + \chi}\Tr((T^c)^2(T^{a_1})^2) + \frac{{\mathbf{\boldsymbol{\epsilon}}} \cdot ({\mathbf{k}}-x {\mathbf{q}}_1)}{({\mathbf{k}}-x {\mathbf{q}}_1)^2 + \chi}\Tr(T^c T^{a_1} T^c T^{a_1}) \Big)\frac{{\mathbf{\boldsymbol{\epsilon}}} \cdot {\mathbf{k}}}{{\mathbf{k}}^2 +\chi} 2 \cos{(\frac{{\mathbf{k}}^2 + \chi}{x(1-x)E^+} (z_1-z_0))}   \Big\}.
\end{align}
\end{widetext}
\newpage

Likewise, after adding Eqs.~\eqref{m220_m} to~\eqref{m202_clm}, we obtain:
\newpage
\begin{widetext}
\begin{align}~\label{E_M2m}
 \frac{2}{d_T} Re \Tr  \mean{M_2 M^*_0} = {} & \sum N |J(p+k)|^2 (4g^2_s) \frac{1}{A_{\perp}} (1-x+x^2)^2  \int{\frac{d^2{\mathbf{q}}_1}{(2\pi)^2}} |v(0, {\mathbf{q}}_1)|^2 \frac{C_2(T)}{d_G}  \nonumber \\
 \times {} & \Big\{(\frac{{\mathbf{\boldsymbol{\epsilon}}} \cdot {\mathbf{k}}}{{\mathbf{k}}^2 + \chi})^2
 \Big(2 \alpha \cos{(\frac{{\mathbf{k}}^2 + \chi}{x(1-x)E^+}(z_1-z_0))} - 2\alpha -\Tr((T^c)^2(T^{a_1})^2) \Big)  \nonumber \\
  - {} &
 2\alpha \frac{{\mathbf{\boldsymbol{\epsilon}}}\cdot {\mathbf{k}}}{{\mathbf{k}}^2 + \chi}\frac{{\mathbf{\boldsymbol{\epsilon}}} \cdot ({\mathbf{k}}-{\mathbf{q}}_1)}{({\mathbf{k}}- {\mathbf{q}}_1)^2 + \chi} \Big(\cos{(\frac{{\mathbf{k}}^2 + \chi}{x(1-x)E^+}(z_1 -z_0))}
 - \cos{(\frac{{\mathbf{k}}^2 - ({\mathbf{k}}- {\mathbf{q}}_1)^2}{x(1-x)E^+}(z_1-z_0))} \Big) \Big\},
\end{align}
\end{widetext}
leading to:
\newpage
\begin{widetext}
\begin{align}~\label{E_M1_2m}
\frac{1}{d_T}\Tr & \mean{|M_1|^2} + \frac{2}{d_T} Re \Tr\mean{M_2 M^*_0} =    N d_G |J(p+k)|^2 (4g^2_s)\frac{C_2(T)}{d_G} C^2_2(G) \frac{1}{A_{\perp}} (1-x+x^2)^2 \sum \int{\frac{d^2{\mathbf{q}}_1}{(2\pi)^2}} |v(0, {\mathbf{q}}_1)|^2  \nonumber \\
 \times &  \Big\{\Big( 1-\cos{(\frac{{\mathbf{k}}^2+ \chi}{x(1-x)E^+}(z_1-z_0))} \Big)
 \Big(\frac{{\mathbf{\boldsymbol{\epsilon}}}\cdot {\mathbf{k}}}{{\mathbf{k}}^2+ \chi} -\frac{{\mathbf{\boldsymbol{\epsilon}}} \cdot ({\mathbf{k}}-x{\mathbf{q}}_1)}{({\mathbf{k}}-x{\mathbf{q}}_1)^2+ \chi} \Big)\frac{{\mathbf{\boldsymbol{\epsilon}}} \cdot {\mathbf{k}}}{{\mathbf{k}}^2+ \chi} + \Big( (\frac{{\mathbf{\boldsymbol{\epsilon}}} \cdot ({\mathbf{k}} -x{\mathbf{q}}_1)}{({\mathbf{k}} -x{\mathbf{q}}_1)^2+ \chi})^2 -(\frac{{\mathbf{\boldsymbol{\epsilon}}} \cdot {\mathbf{k}}}{{\mathbf{k}}^2+ \chi})^2 \Big)  \nonumber \\
 + &  \Big( 1-\cos{(\frac{({\mathbf{k}}- {\mathbf{q}}_1)^2+ \chi}{x(1-x)E^+}(z_1-z_0))} \Big)
  \Big( 2\frac{{\mathbf{\boldsymbol{\epsilon}}}\cdot ({\mathbf{k}}- {\mathbf{q}}_1)}{({\mathbf{k}}-{\mathbf{q}}_1)^2+ \chi} - \frac{{\mathbf{\boldsymbol{\epsilon}}} \cdot {\mathbf{k}}}{{\mathbf{k}}^2+ \chi}  -  \frac{{\mathbf{\boldsymbol{\epsilon}}}\cdot ({\mathbf{k}} -x{\mathbf{q}}_1)}{({\mathbf{k}} -x{\mathbf{q}}_1)^2+ \chi} \Big) \frac{{\mathbf{\boldsymbol{\epsilon}}}\cdot ({\mathbf{k}}-{\mathbf{q}}_1)}{({\mathbf{k}}- {\mathbf{q}}_1)^2+ \chi}
  \Big\}.
\end{align}
\end{widetext}

In the soft-gluon approximation the previous expression coincides with  Eq.~(82) from~\citep{DGLVstatic} (note that contrary to the cited paper, we here consider gluon jet, so that $M$ no longer denotes heavy quark mass, but instead $M \equiv m_g$ and therefore the term $M^2x^2$ is also negligible).

If we further apply the same procedure as in Appendix~\ref{sec:E}, and again assume the simple exponential distribution $\frac{2}{L} e^{-2\frac{z_1-z_0}{L}}$ of the scattering centers, we obtain:
\begin{widetext}
\begin{align}~\label{dN_dxmassiveA}
\frac{dN^{(1)}_{g}}{dx^{}}={} & \frac{C_2(G) \alpha_s}{\pi} \frac{L}{\lambda} \frac{(1-x+x^2)^2}{x(1-x)}\int{\frac{d^2{\mathbf{q}}_1}{\pi} \frac{\mu^2}{({\mathbf{q}}^2_1 + \mu^2)^2}}
\int{d{\mathbf{k}}^2} \nonumber \\
 \times {} & \Big\{  \frac{({\mathbf{k}}-{\mathbf{q}}_1)^2 + \chi}{(\frac{4x(1-x)E}{L})^2 +(({\mathbf{k}}-{\mathbf{q}}_1)^2+ \chi)^2} \Big(2\frac{({\mathbf{k}}-{\mathbf{q}}_1)^2}{({\mathbf{k}}-{\mathbf{q}}_1)^2 + \chi} -\frac{{\mathbf{k}} \cdot ({\mathbf{k}}-{\mathbf{q}}_1)}{{\mathbf{k}}^2 + \chi} - \frac{({\mathbf{k}}-{\mathbf{q}}_1) \cdot ({\mathbf{k}}-x{\mathbf{q}}_1)}{({\mathbf{k}}-x{\mathbf{q}}_1)^2 + \chi} \Big)  \nonumber \\
   + {} & \frac{{\mathbf{k}}^2 + \chi}{(\frac{4x(1-x)E}{L})^2 +({\mathbf{k}}^2+ \chi)^2}
 \Big(\frac{{\mathbf{k}}^2}{{\mathbf{k}}^2+ \chi} - \frac{{\mathbf{k}} \cdot ({\mathbf{k}}-x{\mathbf{q}}_1) }{({\mathbf{k}}-x{\mathbf{q}}_1)^2 + \chi} \Big)  + \Big( \frac{({\mathbf{k}}-x{\mathbf{q}}_1)^2}{(({\mathbf{k}}-x{\mathbf{q}}_1)^2 + \chi)^2}
  - \frac{{\mathbf{k}}^2}{({\mathbf{k}}^2+ \chi)^2} \Big) \Big\},
\end{align}
\end{widetext}
which is symmetric to the exchange of $p$ and $k$ gluons, and which for $m_g \rightarrow 0$ coincides with  Eq.~\eqref{dN_dx_massless}. Also:
\begin{widetext}
\begin{align}~\label{dE_dx2m}
\frac{dE^{(1)}}{dx^{}}={} & \frac{C_2(G) \alpha_s}{\pi} \frac{L}{\lambda}E \frac{(1-x+x^2)^2}{1-x}\int{\frac{d^2{\mathbf{q}}_1}{\pi} \frac{\mu^2}{({\mathbf{q}}^2_1 + \mu^2)^2}}
\int{d{\mathbf{k}}^2}  \nonumber \\
 \times {} & \Big\{  \frac{({\mathbf{k}}-{\mathbf{q}}_1)^2 + \chi}{(\frac{4x(1-x)E}{L})^2 +(({\mathbf{k}}-{\mathbf{q}}_1)^2+ \chi)^2} \Big(2\frac{({\mathbf{k}}-{\mathbf{q}}_1)^2}{({\mathbf{k}}-{\mathbf{q}}_1)^2 + \chi} -\frac{{\mathbf{k}} \cdot ({\mathbf{k}}-{\mathbf{q}}_1)}{{\mathbf{k}}^2 + \chi} - \frac{({\mathbf{k}}-{\mathbf{q}}_1) \cdot ({\mathbf{k}}-x{\mathbf{q}}_1)}{({\mathbf{k}}-x{\mathbf{q}}_1)^2 + \chi} \Big)  \nonumber \\
   + {} & \frac{{\mathbf{k}}^2 + \chi}{(\frac{4x(1-x)E}{L})^2 +({\mathbf{k}}^2+ \chi)^2}
 \Big(\frac{{\mathbf{k}}^2}{{\mathbf{k}}^2+ \chi} - \frac{{\mathbf{k}} \cdot ({\mathbf{k}}-x{\mathbf{q}}_1) }{({\mathbf{k}}-x{\mathbf{q}}_1)^2 + \chi} \Big)  + \Big( \frac{({\mathbf{k}}-x{\mathbf{q}}_1)^2}{(({\mathbf{k}}-x{\mathbf{q}}_1)^2 + \chi)^2}
  - \frac{{\mathbf{k}}^2}{({\mathbf{k}}^2+ \chi)^2} \Big) \Big\},
\end{align}
\end{widetext}
which, in soft-gluon approximation, reduces to Eq.~(84) from~\citep{DGLVstatic}, and which for $m_g \rightarrow 0$ coincides with our massless beyond soft-gluon approximation expression Eq.~\eqref{dE_dx2}.

Further, we display the beyond soft-gluon approximation expressions  needed for numerical evaluation of the corresponding variables.
So, the number of radiated gluons to the first order in opacity for gluons with effective mass $m_g$ and for finite $x$ reads:
\begin{widetext}
\begin{align}~\label{N_dxmassiveA}
N^{(1)}_{g}={} & \frac{C_2(G) \alpha_s}{\pi} \frac{L}{\lambda} \int_{0}^{\frac{1}{2}} dx{\frac{(1-x+x^2)^2}{x(1-x)}}\int{\frac{d^2{\mathbf{q}}_1}{\pi} \frac{\mu^2}{({\mathbf{q}}^2_1 + \mu^2)^2}}
\int{d{\mathbf{k}}^2}  \nonumber \\
 \times {} & \Big\{  \frac{({\mathbf{k}}-{\mathbf{q}}_1)^2 + \chi}{(\frac{4x(1-x)E}{L})^2 +(({\mathbf{k}}-{\mathbf{q}}_1)^2+ \chi)^2} \Big(2\frac{({\mathbf{k}}-{\mathbf{q}}_1)^2}{({\mathbf{k}}-{\mathbf{q}}_1)^2 + \chi} -\frac{{\mathbf{k}} \cdot ({\mathbf{k}}-{\mathbf{q}}_1)}{{\mathbf{k}}^2 + \chi} - \frac{({\mathbf{k}}-{\mathbf{q}}_1) \cdot ({\mathbf{k}}-x{\mathbf{q}}_1)}{({\mathbf{k}}-x{\mathbf{q}}_1)^2 + \chi} \Big)  \nonumber \\
   + {} & \frac{{\mathbf{k}}^2 + \chi}{(\frac{4x(1-x)E}{L})^2 +({\mathbf{k}}^2+ \chi)^2}
 \Big(\frac{{\mathbf{k}}^2}{{\mathbf{k}}^2+ \chi} - \frac{{\mathbf{k}} \cdot ({\mathbf{k}}-x{\mathbf{q}}_1) }{({\mathbf{k}}-x{\mathbf{q}}_1)^2 + \chi} \Big)  + \Big( \frac{({\mathbf{k}}-x{\mathbf{q}}_1)^2}{(({\mathbf{k}}-x{\mathbf{q}}_1)^2 + \chi)^2}
  - \frac{{\mathbf{k}}^2}{({\mathbf{k}}^2+ \chi)^2} \Big) \Big\}.
\end{align}
\end{widetext}
Similarly, the fractional radiative energy loss is given by:  
\begin{widetext}
\begin{align}~\label{dE2massiveA}
\frac{\Delta E^{(1)}}{E}={} & \frac{C_2(G) \alpha_s}{\pi} \frac{L}{\lambda} \int_{0}^{\frac{1}{2}}{dx\frac{(1-x+x^2)^2}{1-x}}\int{\frac{d^2{\mathbf{q}}_1}{\pi} \frac{\mu^2}{({\mathbf{q}}^2_1 + \mu^2)^2}}
\int{d{\mathbf{k}}^2}  \nonumber \\
 \times {} & \Big\{  \frac{({\mathbf{k}}-{\mathbf{q}}_1)^2 + \chi}{(\frac{4x(1-x)E}{L})^2 +(({\mathbf{k}}-{\mathbf{q}}_1)^2+ \chi)^2} \Big(2\frac{({\mathbf{k}}-{\mathbf{q}}_1)^2}{({\mathbf{k}}-{\mathbf{q}}_1)^2 + \chi} -\frac{{\mathbf{k}} \cdot ({\mathbf{k}}-{\mathbf{q}}_1)}{{\mathbf{k}}^2 + \chi} - \frac{({\mathbf{k}}-{\mathbf{q}}_1) \cdot ({\mathbf{k}}-x{\mathbf{q}}_1)}{({\mathbf{k}}-x{\mathbf{q}}_1)^2 + \chi} \Big)  \nonumber \\
   + {} & \frac{{\mathbf{k}}^2 + \chi}{(\frac{4x(1-x)E}{L})^2 +({\mathbf{k}}^2+ \chi)^2}
 \Big(\frac{{\mathbf{k}}^2}{{\mathbf{k}}^2+ \chi} - \frac{{\mathbf{k}} \cdot ({\mathbf{k}}-x{\mathbf{q}}_1) }{({\mathbf{k}}-x{\mathbf{q}}_1)^2 + \chi} \Big)  + \Big( \frac{({\mathbf{k}}-x{\mathbf{q}}_1)^2}{(({\mathbf{k}}-x{\mathbf{q}}_1)^2 + \chi)^2}
  - \frac{{\mathbf{k}}^2}{({\mathbf{k}}^2+ \chi)^2} \Big) \Big\}.
\end{align}
\end{widetext}

\section{\label{sec:Uniform} Uniform distribution of static scattering centers}

 \begin{figure*}
\includegraphics[scale=0.5]{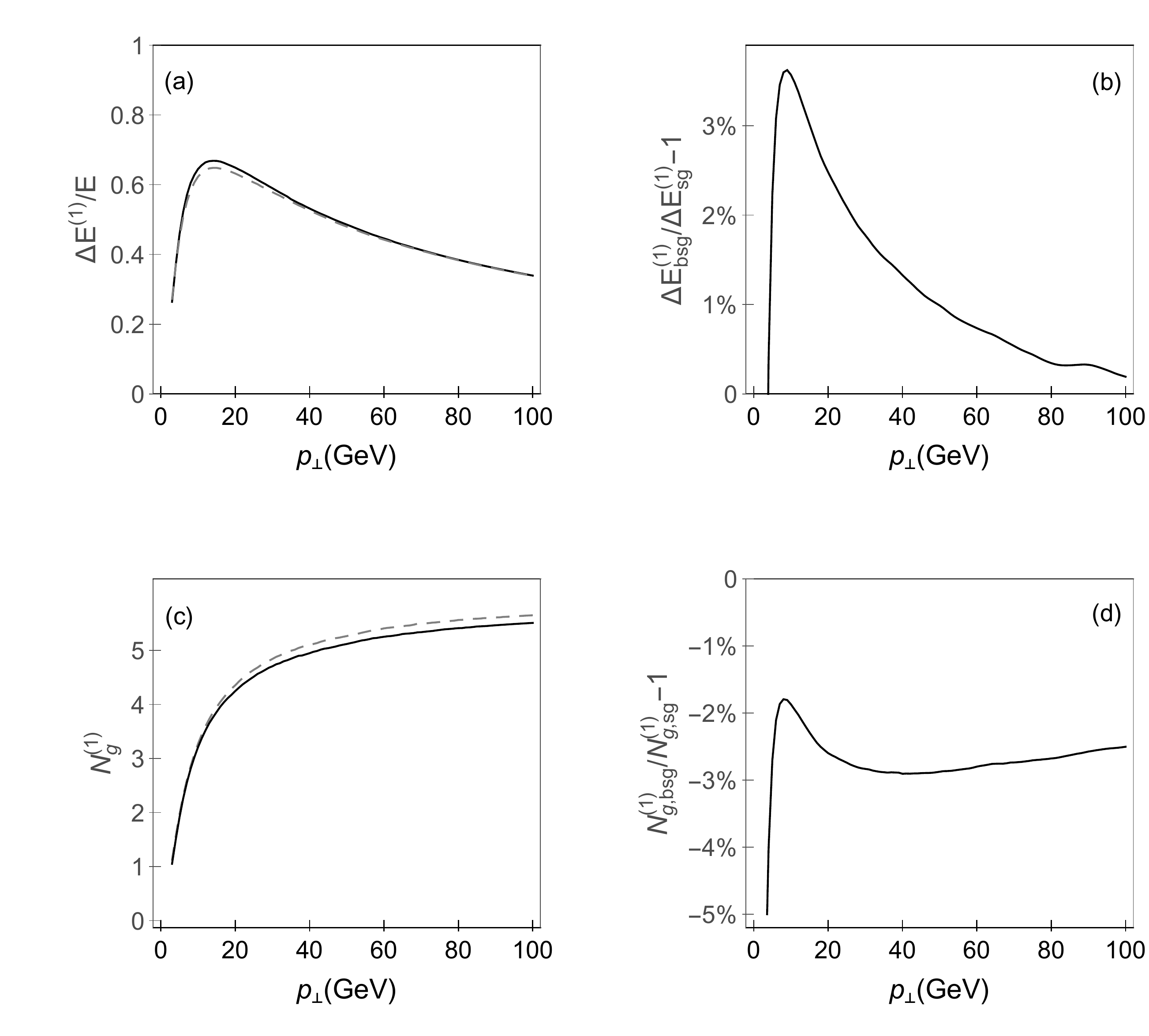}
\caption{\small The counterpart of Fig.~\ref{1}, when uniform longitudinal distance distribution is considered. The effect of relaxing the soft-gluon approximation on integrated variables to the $1^{st}$ order in opacity of DGLV formalism, as a function of $p_{\perp}$. 
(a) Comparison of gluon's fractional radiative energy loss in {\it bsg} (the solid curve) with {\it sg} (the dashed curve) case. 
(b) The percentage change of the radiative energy loss when the soft-gluon approximation is relaxed with respect to the {\it sg} case. 
(c) Comparison of number of radiated gluons in {\it bsg} (the solid curve) with {\it sg} (the dashed curve) case. 
 (d) The relative change of radiated gluon number when the soft-gluon approximation is relaxed with respect to the {\it sg} case.}
\label{11}
\end{figure*}
\begin{figure*}
\includegraphics[scale=0.5]{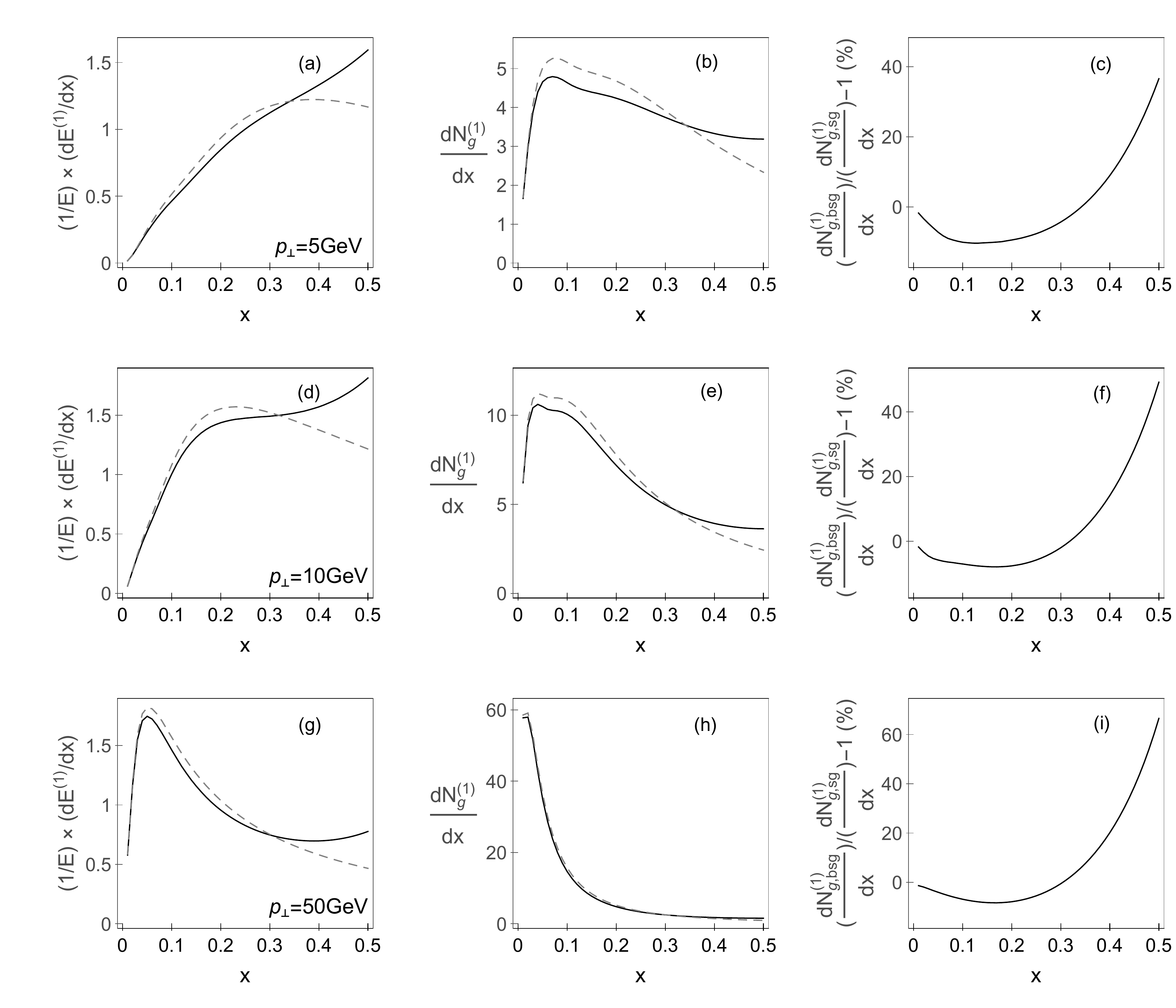}
\caption{\small The counterpart of Fig.~\ref{2}, when uniform longitudinal distance distribution is considered. The effect of relaxing the soft-gluon approximation on differential variables to the $1^{st}$ order in opacity of DGLV formalism, as a function of $x$. The comparison of $(1/E) \times (dE^{(1)}/dx)$ and $dN^{(1)}_{g}/dx$ between {\it bsg} (the solid curve) and {\it sg} (the dashed curve) case, for different values of initial jet $p_{\perp}$ (5 GeV, 10 GeV, 50 GeV, as indicated in panels) is shown in the first ((a), (d) and (g)) and second ((b), (e) and (h)) column, respectively. 
 The quantification of the effect on the single gluon radiation spectrum and its expression in percentage is shown in (c), (f) and (i).}
\label{21}
\end{figure*}
\begin{figure*}
\includegraphics[scale=0.5]{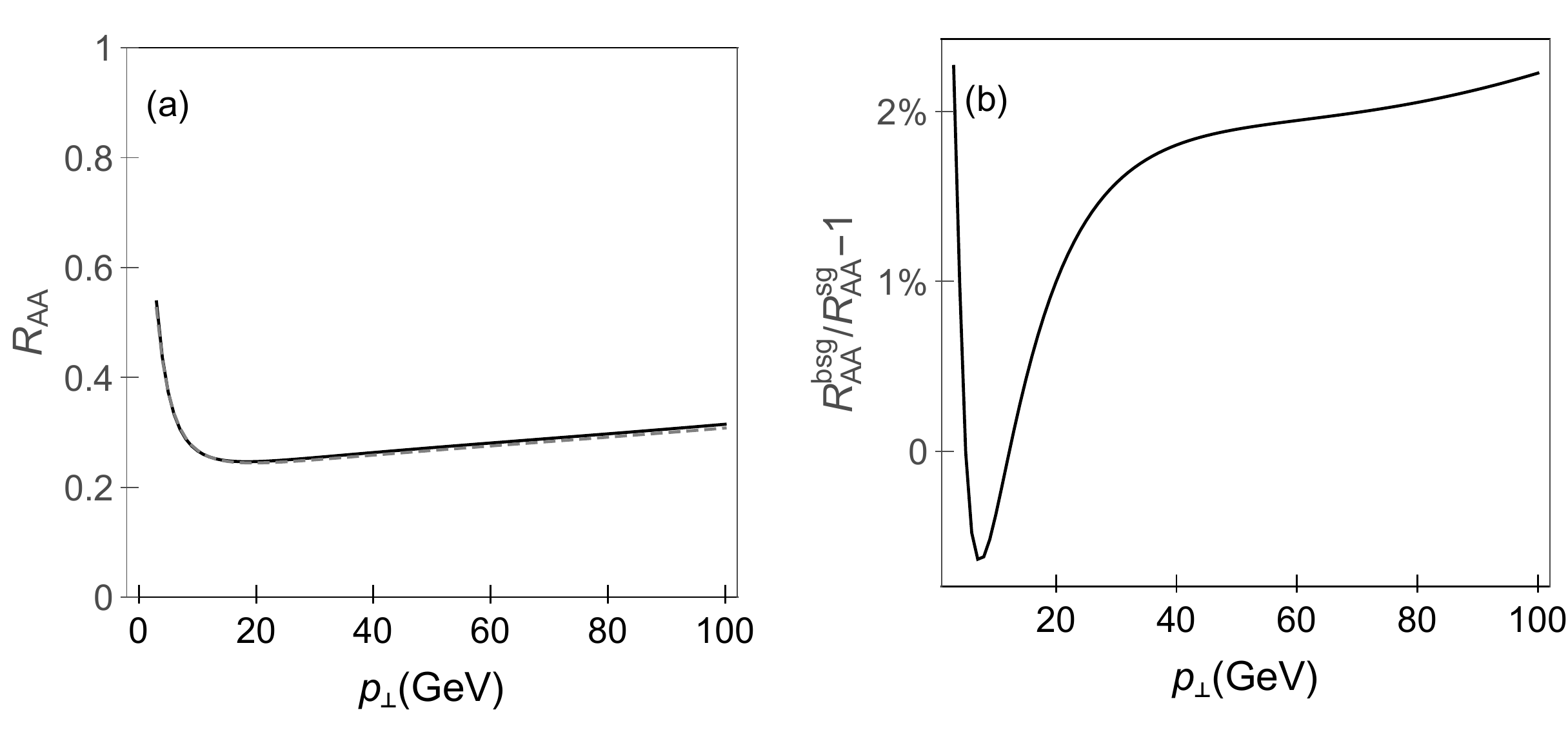}
\caption{\small The counterpart of Fig.~\ref{4}, when uniform longitudinal distance distribution is considered. The effect of relaxing the soft-gluon approximation on gluon $R_{AA}$ as a function of final $p_{\perp}$. 
(a) Comparison of gluon-jet $R_{AA}$ between {\it bsg} (the solid curve) and {\it sg} (the dashed curve) case. 
(b) The quantification of the effect and its expression in percentage.}
\label{41}
\end{figure*}  

In this section we assess how the choice of  distribution of longitudinal distance between the gluon-jet production site and target rescattering site affects our results and conclusions. To this end, we here concentrate on the  limit opposite to the exponential one (which mimics rapidly evolving medium, and which was used throughout this paper) $-$ the uniform distribution, as was done in~\citep{DynEL1,DynEL}. Thus, similarly to the procedure in Appendix~\ref{sec:E}, after incorporating Eq.~\eqref{opacity1} in  Eq.~\eqref{E_M1_2}, then substituting the  obtained expression in Eq.~\eqref{E1} (also keeping in mind that $\vec{\mathbf{p}}$ is 3D momentum of a final jet, and that we need to apply Eqs.~\eqref{dnj2} and~\eqref{cv}, we obtain the following expression for single gluon radiation spectrum in the limit of uniformly distributed static scattering centers for massless gluons:
\begin{widetext}
\begin{align}~\label{dN_dxAhom}
\frac{dN^{(1)}_{g}}{dx^{}}={} & \frac{C_2(G) \alpha_s}{\pi} \frac{L}{\lambda} \frac{(1-x+x^2)^2}{x(1-x)}\int{\frac{d^2{\mathbf{q}}_1}{\pi} \frac{\mu^2}{({\mathbf{q}}^2_1 + \mu^2)^2}}
\int{d{\mathbf{k}}^2} \nonumber \\
 \times {} & \Big\{ \Big( 1- \frac{\sin{\big(\frac{({\mathbf{k}}-{\mathbf{q}}_1)^2}{2x(1-x)E} L \big)}}{\frac{({\mathbf{k}}-{\mathbf{q}}_1)^2 }{2x(1-x)E} L} \Big) \frac{1}{({\mathbf{k}}-{\mathbf{q}}_1)^2} \Big(2 -\frac{{\mathbf{k}} \cdot ({\mathbf{k}}-{\mathbf{q}}_1)}{{\mathbf{k}}^2} - \frac{({\mathbf{k}}-{\mathbf{q}}_1) \cdot ({\mathbf{k}}-x{\mathbf{q}}_1)}{({\mathbf{k}}-x{\mathbf{q}}_1)^2} \Big)  \nonumber \\
   + {} &\Big( 1- \frac{\sin{\big(\frac{{\mathbf{k}}^2}{2x(1-x)E} L \big)}}{\frac{{\mathbf{k}}^2 }{2x(1-x)E} L} \Big) \frac{1}{{\mathbf{k}}^2}
 \Big(1 - \frac{{\mathbf{k}} \cdot ({\mathbf{k}}-x{\mathbf{q}}_1) }{({\mathbf{k}}-x{\mathbf{q}}_1)^2} \Big)  + \Big( \frac{1}{({\mathbf{k}}-x{\mathbf{q}}_1)^2}
  - \frac{1}{{\mathbf{k}}^2} \Big) \Big\},
\end{align}
\end{widetext} 
which is also symmetric to the exchange of radiated and final gluons, while the differential radiative energy loss for massless gluons in the case of the uniform distribution acquires the form:
\begin{widetext}
\begin{align}~\label{dE_dxAhom}
\frac{dE^{(1)}}{dx}={} & \frac{C_2(G) \alpha_s}{\pi} \frac{L}{\lambda} E \frac{(1-x+x^2)^2}{(1-x)}\int{\frac{d^2{\mathbf{q}}_1}{\pi} \frac{\mu^2}{({\mathbf{q}}^2_1 + \mu^2)^2}}
\int{d{\mathbf{k}}^2} \nonumber \\
 \times {} & \Big\{ \Big( 1- \frac{\sin{\big(\frac{({\mathbf{k}}-{\mathbf{q}}_1)^2}{2x(1-x)E} L \big)}}{\frac{({\mathbf{k}}-{\mathbf{q}}_1)^2 }{2x(1-x)E} L} \Big) \frac{1}{({\mathbf{k}}-{\mathbf{q}}_1)^2} \Big(2 -\frac{{\mathbf{k}} \cdot ({\mathbf{k}}-{\mathbf{q}}_1)}{{\mathbf{k}}^2} - \frac{({\mathbf{k}}-{\mathbf{q}}_1) \cdot ({\mathbf{k}}-x{\mathbf{q}}_1)}{({\mathbf{k}}-x{\mathbf{q}}_1)^2} \Big)  \nonumber \\
   + {} &\Big( 1- \frac{\sin{\big(\frac{{\mathbf{k}}^2}{2x(1-x)E} L \big)}}{\frac{{\mathbf{k}}^2 }{2x(1-x)E} L} \Big) \frac{1}{{\mathbf{k}}^2}
 \Big(1 - \frac{{\mathbf{k}} \cdot ({\mathbf{k}}-x{\mathbf{q}}_1) }{({\mathbf{k}}-x{\mathbf{q}}_1)^2} \Big)  + \Big( \frac{1}{({\mathbf{k}}-x{\mathbf{q}}_1)^2}
  - \frac{1}{{\mathbf{k}}^2} \Big) \Big\}.
\end{align}
\end{widetext}
Note that $L$ dependence in the case of uniform distribution of $\frac{dN_g^{(1)}}{dx}$ and $\frac{dE^{(1)}}{dx}$ (given by Eqs.~\eqref{dN_dxAhom} and~\eqref{dE_dxAhom}) is quite distinct to the one in the case of exponential distribution (given by Eqs.~\eqref{dN_dx_massless} and~\eqref{dE_dx2}). 

Proceeding in the same manner as in Appendix~\ref{sec:Emg}, the single gluon radiation spectrum and differential radiative energy loss for gluon-jet embedded in a finite temperature QCD medium, i.e. for gluon with effective mass $m_g$~\citep{mg}, in the case of uniformly distributed scattering centers  read:   
\begin{widetext}
\begin{align}~\label{dN_dxmassiveAhom}
\frac{dN^{(1)}_{g}}{dx^{}}={} & \frac{C_2(G) \alpha_s}{\pi} \frac{L}{\lambda} \frac{(1-x+x^2)^2}{x(1-x)}\int{\frac{d^2{\mathbf{q}}_1}{\pi} \frac{\mu^2}{({\mathbf{q}}^2_1 + \mu^2)^2}}
\int{d{\mathbf{k}}^2} \nonumber \\
 \times {} & \Big\{ \Big( 1- \frac{\sin{\big(\frac{({\mathbf{k}}-{\mathbf{q}}_1)^2 + \chi}{2x(1-x)E} L \big)}}{\frac{({\mathbf{k}}-{\mathbf{q}}_1)^2 + \chi}{2x(1-x)E} L} \Big) \frac{1}{({\mathbf{k}}-{\mathbf{q}}_1)^2+ \chi} \Big(2\frac{({\mathbf{k}}-{\mathbf{q}}_1)^2}{({\mathbf{k}}-{\mathbf{q}}_1)^2 + \chi} -\frac{{\mathbf{k}} \cdot ({\mathbf{k}}-{\mathbf{q}}_1)}{{\mathbf{k}}^2 + \chi} - \frac{({\mathbf{k}}-{\mathbf{q}}_1) \cdot ({\mathbf{k}}-x{\mathbf{q}}_1)}{({\mathbf{k}}-x{\mathbf{q}}_1)^2 + \chi} \Big)  \nonumber \\
   + {} &\Big( 1- \frac{\sin{\big(\frac{{\mathbf{k}}^2 + \chi}{2x(1-x)E} L \big)}}{\frac{{\mathbf{k}}^2 + \chi}{2x(1-x)E} L} \Big) \frac{1}{{\mathbf{k}}^2+ \chi}
 \Big(\frac{{\mathbf{k}}^2}{{\mathbf{k}}^2+ \chi} - \frac{{\mathbf{k}} \cdot ({\mathbf{k}}-x{\mathbf{q}}_1) }{({\mathbf{k}}-x{\mathbf{q}}_1)^2 + \chi} \Big)  + \Big( \frac{({\mathbf{k}}-x{\mathbf{q}}_1)^2}{(({\mathbf{k}}-x{\mathbf{q}}_1)^2 + \chi)^2}
  - \frac{{\mathbf{k}}^2}{({\mathbf{k}}^2+ \chi)^2} \Big) \Big\}
\end{align}
\end{widetext} 
and 
\begin{widetext}
\begin{align}~\label{dE_dxmassiveAhom}
\frac{dE^{(1)}}{dx^{}}={} & \frac{C_2(G) \alpha_s}{\pi} \frac{L}{\lambda} E \frac{(1-x+x^2)^2}{(1-x)}\int{\frac{d^2{\mathbf{q}}_1}{\pi} \frac{\mu^2}{({\mathbf{q}}^2_1 + \mu^2)^2}}
\int{d{\mathbf{k}}^2} \nonumber \\
 \times {} & \Big\{ \Big( 1- \frac{\sin{\big(\frac{({\mathbf{k}}-{\mathbf{q}}_1)^2 + \chi}{2x(1-x)E} L \big)}}{\frac{({\mathbf{k}}-{\mathbf{q}}_1)^2 + \chi}{2x(1-x)E} L} \Big) \frac{1}{({\mathbf{k}}-{\mathbf{q}}_1)^2+ \chi} \Big(2\frac{({\mathbf{k}}-{\mathbf{q}}_1)^2}{({\mathbf{k}}-{\mathbf{q}}_1)^2 + \chi} -\frac{{\mathbf{k}} \cdot ({\mathbf{k}}-{\mathbf{q}}_1)}{{\mathbf{k}}^2 + \chi} - \frac{({\mathbf{k}}-{\mathbf{q}}_1) \cdot ({\mathbf{k}}-x{\mathbf{q}}_1)}{({\mathbf{k}}-x{\mathbf{q}}_1)^2 + \chi} \Big)  \nonumber \\
   + {} &\Big( 1- \frac{\sin{\big(\frac{{\mathbf{k}}^2 + \chi}{2x(1-x)E} L \big)}}{\frac{{\mathbf{k}}^2 + \chi}{2x(1-x)E} L} \Big) \frac{1}{{\mathbf{k}}^2+ \chi}
 \Big(\frac{{\mathbf{k}}^2}{{\mathbf{k}}^2+ \chi} - \frac{{\mathbf{k}} \cdot ({\mathbf{k}}-x{\mathbf{q}}_1) }{({\mathbf{k}}-x{\mathbf{q}}_1)^2 + \chi} \Big)  + \Big( \frac{({\mathbf{k}}-x{\mathbf{q}}_1)^2}{(({\mathbf{k}}-x{\mathbf{q}}_1)^2 + \chi)^2}
  - \frac{{\mathbf{k}}^2}{({\mathbf{k}}^2+ \chi)^2} \Big) \Big\}.
\end{align}
\end{widetext}
Note that Eq.~\eqref{dN_dxmassiveAhom} is also symmetric to the exchange of radiated and final gluons, and in massless case reproduces Eq.~\eqref{dN_dxAhom}, whereas Eq.~\eqref{dE_dxmassiveAhom}  for $m_g \rightarrow 0$ coincides with our massless beyond soft-gluon approximation expression Eq.~\eqref{dE_dxAhom}.    
 Again, by comparing analytical expressions given by Eqs.~\eqref{dN_dxmassiveAhom} and~\eqref{dE_dxmassiveAhom} with Eqs.~\eqref{dN_dxmassiveA} and~\eqref{dE_dx2m} we observe significantly different $L$ dependence in these two opposite cases of longitudinal distance distribution. 

Finally, the number of radiated gluons and fractional radiative energy loss to the first order in opacity and beyond the soft-gluon approximation for gluons with effective mass $m_g$ in the limit of uniform longitudinal distance distribution, respectively, read: 
\begin{widetext}
\begin{align}~\label{N_massiveAhom}
N^{(1)}_{g}={} & \frac{C_2(G) \alpha_s}{\pi} \frac{L}{\lambda} \int_{0}^{\frac{1}{2}} dx \frac{(1-x+x^2)^2}{x(1-x)}\int{\frac{d^2{\mathbf{q}}_1}{\pi} \frac{\mu^2}{({\mathbf{q}}^2_1 + \mu^2)^2}}
\int{d{\mathbf{k}}^2} \nonumber \\
 \times {} & \Big\{ \Big( 1- \frac{\sin{\big(\frac{({\mathbf{k}}-{\mathbf{q}}_1)^2 + \chi}{2x(1-x)E} L \big)}}{\frac{({\mathbf{k}}-{\mathbf{q}}_1)^2 + \chi}{2x(1-x)E} L} \Big) \frac{1}{({\mathbf{k}}-{\mathbf{q}}_1)^2+ \chi} \Big(2\frac{({\mathbf{k}}-{\mathbf{q}}_1)^2}{({\mathbf{k}}-{\mathbf{q}}_1)^2 + \chi} -\frac{{\mathbf{k}} \cdot ({\mathbf{k}}-{\mathbf{q}}_1)}{{\mathbf{k}}^2 + \chi} - \frac{({\mathbf{k}}-{\mathbf{q}}_1) \cdot ({\mathbf{k}}-x{\mathbf{q}}_1)}{({\mathbf{k}}-x{\mathbf{q}}_1)^2 + \chi} \Big)  \nonumber \\
   + {} &\Big( 1- \frac{\sin{\big(\frac{{\mathbf{k}}^2 + \chi}{2x(1-x)E} L \big)}}{\frac{{\mathbf{k}}^2 + \chi}{2x(1-x)E} L} \Big) \frac{1}{{\mathbf{k}}^2+ \chi}
 \Big(\frac{{\mathbf{k}}^2}{{\mathbf{k}}^2+ \chi} - \frac{{\mathbf{k}} \cdot ({\mathbf{k}}-x{\mathbf{q}}_1) }{({\mathbf{k}}-x{\mathbf{q}}_1)^2 + \chi} \Big)  + \Big( \frac{({\mathbf{k}}-x{\mathbf{q}}_1)^2}{(({\mathbf{k}}-x{\mathbf{q}}_1)^2 + \chi)^2}
  - \frac{{\mathbf{k}}^2}{({\mathbf{k}}^2+ \chi)^2} \Big) \Big\}
\end{align}
\end{widetext}
and 
\begin{widetext}
\begin{align}~\label{E_massiveAhom}
\frac{\Delta E^{(1)}}{E} ={} & \frac{C_2(G) \alpha_s}{\pi} \frac{L}{\lambda} \int_{0}^{\frac{1}{2}} dx \frac{(1-x+x^2)^2}{(1-x)}\int{\frac{d^2{\mathbf{q}}_1}{\pi} \frac{\mu^2}{({\mathbf{q}}^2_1 + \mu^2)^2}}
\int{d{\mathbf{k}}^2} \nonumber \\
 \times {} & \Big\{ \Big( 1- \frac{\sin{\big(\frac{({\mathbf{k}}-{\mathbf{q}}_1)^2 + \chi}{2x(1-x)E} L \big)}}{\frac{({\mathbf{k}}-{\mathbf{q}}_1)^2 + \chi}{2x(1-x)E} L} \Big) \frac{1}{({\mathbf{k}}-{\mathbf{q}}_1)^2+ \chi} \Big(2\frac{({\mathbf{k}}-{\mathbf{q}}_1)^2}{({\mathbf{k}}-{\mathbf{q}}_1)^2 + \chi} -\frac{{\mathbf{k}} \cdot ({\mathbf{k}}-{\mathbf{q}}_1)}{{\mathbf{k}}^2 + \chi} - \frac{({\mathbf{k}}-{\mathbf{q}}_1) \cdot ({\mathbf{k}}-x{\mathbf{q}}_1)}{({\mathbf{k}}-x{\mathbf{q}}_1)^2 + \chi} \Big)  \nonumber \\
   + {} &\Big( 1- \frac{\sin{\big(\frac{{\mathbf{k}}^2 + \chi}{2x(1-x)E} L \big)}}{\frac{{\mathbf{k}}^2 + \chi}{2x(1-x)E} L} \Big) \frac{1}{{\mathbf{k}}^2+ \chi}
 \Big(\frac{{\mathbf{k}}^2}{{\mathbf{k}}^2+ \chi} - \frac{{\mathbf{k}} \cdot ({\mathbf{k}}-x{\mathbf{q}}_1) }{({\mathbf{k}}-x{\mathbf{q}}_1)^2 + \chi} \Big)  + \Big( \frac{({\mathbf{k}}-x{\mathbf{q}}_1)^2}{(({\mathbf{k}}-x{\mathbf{q}}_1)^2 + \chi)^2}
  - \frac{{\mathbf{k}}^2}{({\mathbf{k}}^2+ \chi)^2} \Big) \Big\}.
\end{align}
\end{widetext}
\begin{figure*}
\includegraphics[scale=0.4]{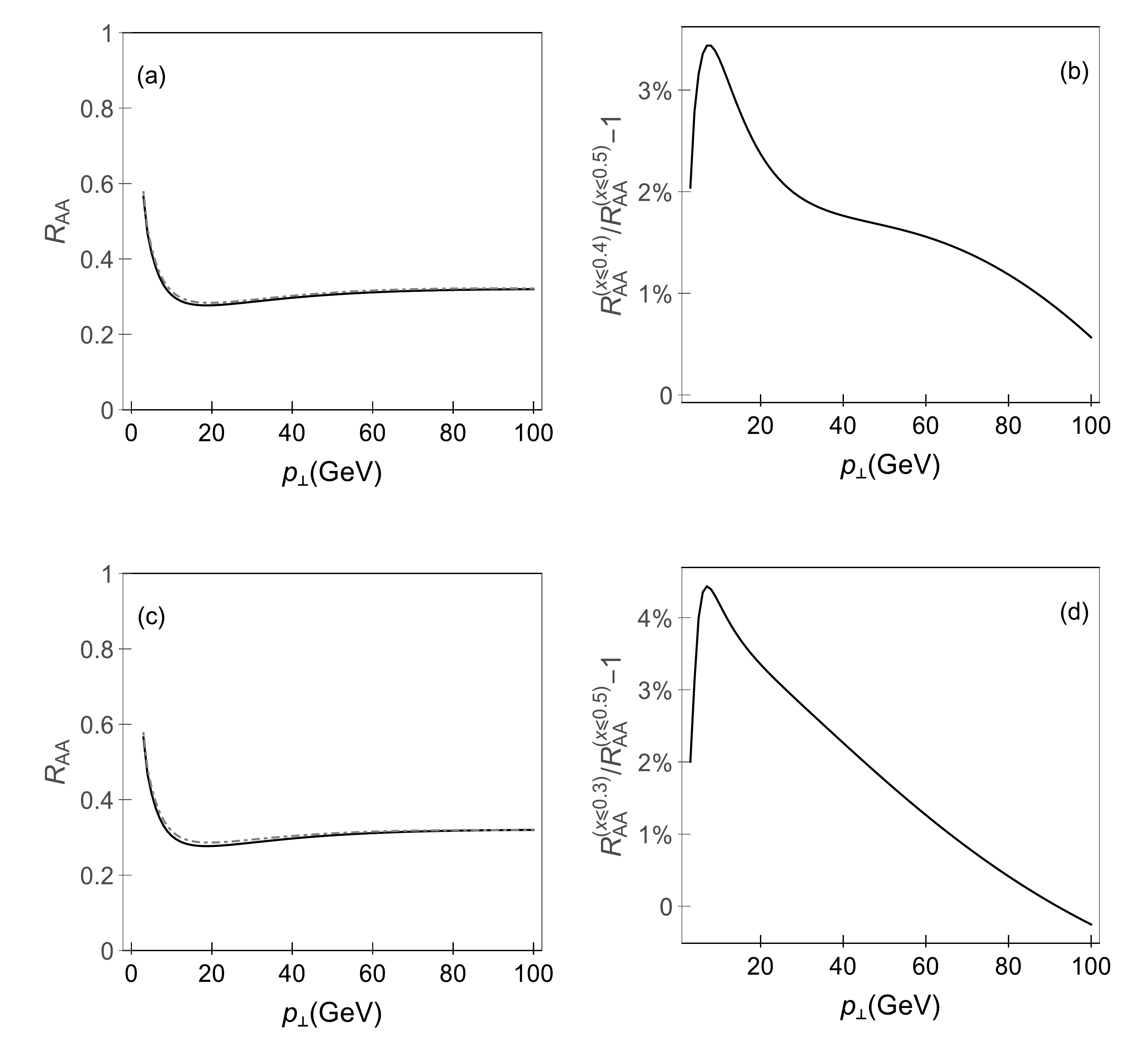}
\caption{\small The two scenarios of relevant $x$ region for the importance of the soft-gluon approximation. The effect of relaxing the soft-gluon approximation on  $R_{AA}$ as a function of $p_{\perp}$. The suppression of gluon jet beyond soft-gluon approximation (the solid curve) is compared to the combined $R_{AA}$ (the dot-dashed curve), obtained from {\it 1.} {\it bsg} expression for $x \leq 0.4$ combined with {\it sg} expression for $ x > 0.4$ in (a) ({\it 2.} {\it bsg} expression for $x \leq 0.3$ combined with {\it sg} expression for $ x > 0.3$ in (c)). 
The 
quantification of the effect and its expression  in percentage for these two scenarios is presented in (b) and (d), respectively.}
\label{43}
\end{figure*}
Note that Eq.~\eqref{E_massiveAhom} in the soft-gluon limit reduces to Eq. (2.13) from~\citep{DynEL} (or equivalently, to static case of Eq. (1) from~\citep{DynEL1}) for gluons, where likewise the uniform distribution was used.

The above obtained notably different expressions for uniform (compared to the exponential distribution of static scattering centers)  require assessing how sensitive are our conclusions on the importance of the soft-gluon approximation to the considered distribution. Therefore, in this section we also use uniform distribution case and display the effect of finite $x$ on the numerical predictions for the same variables as in section~\ref{sec:Num}. 

By comparing Figs.~\ref{11},~\ref{21} and \ref{41} with the corresponding figures from section~\ref{sec:Num} (i.e. Figs.~\ref{1},~\ref{2} and~\ref{4}) we infer that the results obtained in this section are quite similar to the ones obtained with exponential distribution in section~\ref{sec:Num}. From this, it follows that our conclusions with respect to the importance of soft-gluon approximation, presented in sections~\ref{treci} to~\ref{sec:CO}, are robust to the presumed longitudinal
distance distribution. Note that curves form this section are less smooth compared to the one from section~\ref{sec:Num}, due to oscillating sine functions in the corresponding analytical expressions (Eqs.~\eqref{dN_dxmassiveAhom} to~\eqref{E_massiveAhom}).

\section{\label{sec:RelRegion} Relevant region for the importance of the soft-gluon approximation}

 Based on the reasoning outlined in section~\ref{sec:Num} (see Fig.~\ref{Ilustracija} and the corresponding intuitive explanation), we marked $x \lesssim 0.4$ as the relevant region for the importance of the soft-gluon approximation; in this section we study this issue in more detail. We first note that claiming that certain region is not relevant for relaxing the soft-gluon approximation does not mean that this whole region can be rejected, but that beyond soft-gluon expression does not have to be applied in that region. 
Therefore, for  reliable suppression predictions, one has to take into account the entire $x$ region, while in the following lines we address the necessity of relaxing the soft-gluon approximation in a certain region. 
 
 To address this goal, we first note that  Figs.~\ref{2} and~\ref{3} highlight both: {\it I)} more "conservative" $x \approx 0.4$ and {\it II)}  $x \approx 0.3$  value as the upper limit of the relevant region. Thus, in the further text, we want to address which of the following two points is better to define as a border point of the relevant region for differentiating between {\it bsg} and {\it sg} $R_{AA}$ predictions. 
Thus, in Fig.~\ref{43} we compare suppressions obtained from {\it bsg} expression for the entire $x \leq 0.5$ region, first with {\it 1.}  results obtained from {\it bsg} expression for $x \leq 0.4$ combined with {\it sg} expression for $ x > 0.4$ (Figs.~\ref{43} (a) and (b)); and then with {\it 2.}  results obtained from {\it bsg} expression for $x \leq 0.3$ combined with {\it sg} expression for $ x > 0.3$ (Figs.~\ref{43} (c) and (d)).
 
 From Fig.~\ref{43} we observe that both $x=0.3$ and $x=0.4$ can be defined as a border point for the importance of the soft-gluon approximation.  
 However, based on the fact that difference between {\it bsg} and combined $R_{AA}$ for case {\it 1.} is smaller than for case {\it 2.}, and that taking into account $0.3 < x \leq 0.4$ region balances negative and positive $\frac{dN_g^{(1)}}{dx}$ contributions of relaxing the soft-gluon approximation (see e.g. Fig.~\ref{3}), it is safer to claim that more "conservative" region $x \lesssim 0.4$ ({\it I)}) is the relevant one. 
 

\end{document}